\documentclass[notitlepage,english,aps,prx,tightenlines,floatfix,longbibliography,10pt,longbibliography,twocolumn]{revtex4}

\usepackage{graphicx}
\usepackage{dcolumn}
\usepackage{bm}
\usepackage[dvipsnames]{xcolor}
\usepackage[colorlinks=true,urlcolor=blue,citecolor=blue, linkcolor = blue]{hyperref}
\usepackage{amssymb,ulem,amsmath}
\usepackage{physics}
\usepackage{orcidlink}
\newcommand{\qql}{\textquotedblleft}
\newcommand{\qqr}{\textquotedblright}
\newcommand{\dg}{\dagger}
\newcommand{\bo}[1]{\mathbf{#1}}
\newcommand{\ham}{\hat{\mathcal{H}}}

\newcommand{\up}{\uparrow}

\begin{document}
	
	\title{Geometry independent superfluid weight in multiorbital lattices from the generalized random phase approximation}

	\author{Minh Tam\orcidlink{https://orcid.org/0000-0003-0798-6477}} 
        \email{minh.phamnguyen@aalto.fi}
	
	\author{Sebastiano Peotta\orcidlink{https://orcid.org/0000-0002-9947-1261}}
	\email{sebastiano.peotta@aalto.fi}	
	
	\affiliation{Department of Applied Physics, Aalto University-School of Science, FI-00076 Aalto, Finland}
    \begin{abstract}
        The superfluid weight of a generic lattice model with attractive Hubbard interaction is computed analytically in the isolated band limit within the generalized random phase approximation. Time-reversal symmetry, spin rotational symmetry and the uniform pairing condition are assumed. It is found that the relation obtained in [\href{https://link.aps.org/doi/10.1103/PhysRevB.106.014518}{Huhtinen et al., Phys. Rev. B \textbf{106}, 014518 (2022)}] between the superfluid weight in the flat band limit and the so-called minimal quantum metric  is valid even at the level of the generalized random phase approximation. For an isolated, but not necessarily flat, band it is found that the correction to the superfluid weight obtained from  the generalized random phase approximation $D_{\rm s}^{(1)} = D_{\rm s,c}^{(1)}+D_{\rm s,g}^{(1)}$  is also the sum of a conventional contribution $D_{\rm s,c}^{(1)}$ and a geometric contribution $D_{\rm s,g}^{(1)}$, as in the case of the known mean-field result $D_{\rm s}^{(0)}=D_{\rm s,c}^{(0)}+D_{\rm s,g}^{(0)}$, in which the geometric term $D_{\rm s,g}^{(0)}$ is a weighted average of the quantum metric. The conventional contribution is geometry independent, that is independent of the orbital positions, while it is possible to find a preferred, or natural, set of orbital positions such that $D_{\rm s,g}^{(1)}=0$. Useful analytic expressions are derived for both the natural orbital positions and the minimal quantum metric, including its extension to bands that are not necessarily flat. Finally, using some simple examples, it is argued that the natural orbital positions may lead to a more refined classification of the topological properties of the band structure.
    \end{abstract}
	
	\maketitle
	
	\section{Introduction}
    An important result of Bardeen-Cooper-Schrieffer theory is the prediction for the superconductive critical temperature $T_c \propto e^{-\frac{1}{|U| \rho_0(E_F)}} $~\cite{Tinkham2004}, where $U$ is the interaction strength of the effective attractive interaction and $\rho_0(E_F)$ is the electronic density of state at the Fermi energy. However, the mean-field critical temperature $T_{\rm c}$ gives the energy scale for the breaking of Cooper pairs, but in many unconventional superconductors and in two-dimensional fermionic superfluids the transition to the normal state is controlled by the phase fluctuations of the order parameter rather than Cooper pair breaking~\cite{Emery1995,Carlson1999,Kosterlitz1973,Nelson1977,Kosterlitz2016}. The quantity that measures the phase stiffness of the order parameter phase with respect to perturbations or thermal fluctuations is  known as the superfluid weight $D_{\rm s}$. The superfluid weight can be obtained experimentally from the penetration depth of the magnetic field characterizing the Meissner effect in supercondutors. From a theoretical point of view, the superfluid weight is defined as a specific limit of the current-current response function~\cite{Scalapino1992,Scalapino1993}. Equivalently, it can be computed from the change of the free energy due to a twist in the boundary conditions~\cite{Peotta2022}. In this work, we adopt this second characterization since it is the most practical for our purposes. 
    
    In the rather idealized but popular model of a superconductor as a system of charged particles propagating in a homogeneous medium, the superfluid weight is simply given by $D_{\rm s} = \frac{q^2 n}{m}$, where $q$ and $m$ are the electric charge and the mass of the particles, respectively, and $n$ is the number density~\cite{Leggett1998}.
    Due to the crystalline structure, real materials possess discrete rather than continuous translational symmetry and, as consequence, the inverse mass $1/m$ in the superfluid weight has to be replaced by the inverse effective mass tensor $[\frac{1}{m_{\rm eff}}]_{i,j} = \frac{1}{\hbar^2} \partial_{k_i}\partial_{k_j}\varepsilon_{n\vb{k}}$ obtained from the energy dispersion $\varepsilon_{n\vb{k}}$ of the partially filled band (see ~\eqref{eq:D0_Sc_alt} in the following). For a long time it was assumed that only the band dispersion plays a role in determining the superfluid weight, which would then be necessarily small if the charge carriers have a large effective mass~\cite{Basov2011}. In particular the superfluid weight should be strictly vanishing in the so-called flat band limit, in which the band dispersion is constant $\varepsilon_{n\vb{k}} = \varepsilon_{n}$ as a function of quasimomentum $\vb{k}$. 
    
    The conventional wisdom was challenged by few specific examples of systems with flat bands for which the superfluid weight can be shown to be nonzero, namely exciton condensates in quantum Hall bilayers~\cite{Moon1995,Joglekar2001} and the surface states in rhombohedral graphite~\cite{Kopnin2011a,Kopnin2011}. Later on, it was shown at a general level that even in the flat band limit the superfluid weight can be nonzero and large due to an additional contribution associated to the band wave functions rather than the band dispersion~\cite{Peotta2015,Julku2016,Liang2017a}. In particular, under some symmetry assumptions, it is shown that the superfluid weight in the flat band limit is proportional to the integral over the first Brillouin zone of the quantum metric, a geometric invariant constructed out of the Bloch wave functions and their derivatives with respect to the quasimomentum~\cite{Peotta2015,Provost1980,Cheng2013}. More generally, under the same symmetry assumptions but not necessarily in the flat band limit, the superfluid weight is a sum of two contributions $D_{\rm s} = D_{\rm s,c} + D_{\rm s,g}$. The conventional contribution $D_{\rm s,c}$ depends on the inverse effective mass, while the so-called geometric contribution $D_{\rm s,g}$ is a weight averaged of the quantum metric over the Brillouin zone (see~\eqref{eq:D0_sc}-\eqref{eq:D0_sg} in the following).
    
    It is important to note that the geometric contribution to the superfluid weight can be nonzero only in multiband/multiorbital lattices, since periodic Bloch functions live in orbital space, whose dimension is the number of orbitals per unit cell. For a simple lattice with one orbital per unit cell, such as the square lattice, a periodic Bloch function  becomes a simple scalar, which cannot affect any physical property, and the quantum metric vanishes. The geometric contribution is thus an effect genuinely associated to the multiorbital character of lattice fermions. The geometric contribution is sizable for instance in multiorbital lattices with flat bands that can be realized with ultracold gases in optical lattices, such as the Lieb lattice~\cite{Julku2016}, and also in iron-based superconductors such as FeSe~\cite{Kitamura2021}, in which multiorbital effects are important~\cite{Lee2017,Sprau2017}. Another example of material with strong multiorbital character is twisted bilayer graphene~\cite{Cao2018,Cao2018b}, in which quasi-flat bands accompanied by the onset of a superconducting state are obtained by tuning the twist angle between the two graphene layers to a specific value, called the magic angle. Multiple theoretical works have reached the conclusion that the geometric contribution to the superfluid weight is comparable to the conventional one in magic angle-twisted bilayer graphene~\cite{Xie2020,Julku2020,Hu2019,Wang2020a} and, more recently, experimental evidence for this prediction has been provided~\cite{Tian2023}. The interplay between quantum geometry and superconductivity in twisted multilayer systems is the topic of two recent review articles~\cite{Rossi2021,Torma2022}. The quantum metric  has also been shown to affect the superfluid properties of lattice bosons~\cite{Julku2021,Julku2021a,Julku2023}.  
    
    The quantum metric is intimately related to the Berry curvature, another band structure invariant that plays a crucial role in the quantum Hall effect. Indeed, the two quantities are respectively the real and imaginary parts of the quantum geometry tensor, a positive semidefinite complex matrix obtained from the Bloch functions and their derivatives~\cite{Peotta2015}. Due to positive definiteness, there are relations between them expressed by inequalities. While the Berry curvature and the associated Berry phase has been extensively studied for instance in the context of the quantum Hall effect~\cite{Thouless1982}, the semiclassical theory of electronic motion~\cite{Xiao2010} and the modern theory of polarization~\cite{Resta2010}, the role of the quantum metric in determining various observable properties is currently the subject of ongoing research~\cite{Gao2014,Srivastava2015,Piechon2016,Ozawa2018,Gianfrate2020,Rhim2020,Zhu2021,Gao2023}. It is clear by now that the quantum metric is relevant for many phenomena other than superfluidity in multiorbital lattices.
    
    
    The relation between superfluid weight and quantum metric within the mean-field approximation is by now a rather established fact, nevertheless in Refs.~\cite{Huhtinen2022,Chan2022a} it was pointed out that these quantities have an unphysical dependence on the positions of the lattice sites, more specifically on the basis vectors determining the relative positions of different sublattices within the unit cell~\cite{Parravicini2013}. Intuitively, one expects that the superfluid weight should not depend on the relative positions of the lattice sites, called in the following orbital positions. The idea that certain physical observables are independent of the geometry of the lattice has been discussed in Ref.~\cite{Simon2020}. In the mathematical physics literature the related concept of unit cell consistency has also been introduced~\cite{Marcelli2019,Marcelli2021}. The quantum metric is neither geometry independent nor unit cell consistent and this leads to the unphysical result that the superfluid weight can be nonzero even in the case of a trivial flat band realized in a lattice model composed of completely disconnected unit cells. The simplest example of this unphysical phenomenon is probably the Su-Schrieffer-Heeger model discussed in Sec.~\ref{sec:SSH} of the present work.
    
    The reason behind this inconsistency has been investigated already in Ref.~\cite{Huhtinen2022} (see also the lecture notes~\cite{Peotta2023} for a more pedagogical presentation) and is due to the fact that the dependence of self-consistently calculated quantities on the change of boundary conditions   is neglected when applying mean-field theory. This is a widely used approximation, equivalent to the prescription of replacing the many-body Hamiltonian with the mean-field Hamiltonian when calculating response functions~\cite{Peotta2022,Scalapino1993}, and has the well-known drawback of breaking gauge invariance~\cite{Schrieffer1964}. More specifically, the superfluid weight is computed as the second derivative of the free energy (or thermodynamic grand potential) with respect to the phase angle parametrizing twisted boundary conditions. To explicitly preserve translational invariance, it is convenient to implement twisted boundary conditions by means of a constant electromagnetic vector potential $\vb{A}$, which cannot be gauged away in a finite system with torus geometry~\cite{Peotta2022}. The usual approximation is to replace the exact free energy with the mean-field free energy, which depends on quantities that need to be computed self-consistently, such as the Hartree-Fock potential and, in the case of superconducting systems, the pairing potential. The mean-field potentials depend on the boundary conditions, which means that they 
    need to be computed self-consistently for each different value of $\vb{A}$. It is shown in Ref.~\cite{Huhtinen2022} that the geometry independence of the superfluid weight is restored by taking into account the $\vb{A}$-dependence of the pairing potential alone. Indeed, the crucial new result of Ref.~\cite{Huhtinen2022} is that the superfluid weight in the flat band limit is proportional to the first Brillouin zone integral of the quantum metric minimized with respect to all possible orbital positions. This so-called minimal quantum metric is a geometry independent quantity and it singles out a specific choice of the orbital positions providing a physically sensible result for the superfluid weight.
    
    The purpose of the present work is to extend the results of Ref.~\cite{Huhtinen2022} in many ways. In Sec.~\ref{sec:gauge}, we prove rigorously that the superfluid weight is a geometry independent quantity as a consequence of gauge invariance. More specifically, it is shown that a shift in the orbitals positions amounts to a gauge transformation, which does not affect the value of the thermodynamic grand potential. 
    Following Ref.~\cite{Peotta2022}, in Sec.~\ref{sec:GRPA} we introduce mean-field  theory as a variational approximation for the grand potential. The mean-field grand potential depends on $\vb{A}$ either directly, through the Peierls phase, or indirectly through the variational parameters that enter in the mean-field Hamiltonian, namely the Hartree-Fock potential and the pairing potential, which are calculated self-consistently for each value of $\vb{A}$. A gauge-symmetry preserving approximation for the superfluid weight is obtained by retaining the $\vb{A}$-dependence of the pairing potential, as suggested in Ref.~\cite{Huhtinen2022}. An important difference with this latter work is that here we also retain the $\vb{A}$-dependence of the Hartree-Fock potential. As shown in Ref.~\cite{Peotta2022}, this
    is equivalent to computing the superfluid weight within the generalized random phase approximation (GRPA). The variational approach used here and in Ref.~\cite{Peotta2022} has the advantage to make it clear that the GRPA is simply mean-field theory applied in a fully conserving fashion by taking into account the fluctuations of the mean-field potentials, and is not, strictly speaking, a beyond mean-field approximation.
    In Sec.~\ref{sec:GRPA} and in Appendix~\ref{app:grpa}, we derive the  expression for the superfluid weight within the GRPA in the terms of various correlation functions between quadratic operators evaluated at $\vb{A} = 0$.
    The derivation is carried out using an alternative method compared to Ref.~\cite{Peotta2022} and more similar to Ref.~\cite{Huhtinen2022}, first in the case of a general interaction term and later specialized to the Hubbard interaction term. 
    
   	In Sec.~\ref{sec:correction}, we compute analytically the superfluid weight within the GRPA in the case of a generic lattice model with an Hubbard interaction term. We adopt the same symmetry assumptions that allow to derive the relation with the quantum metric in an isolated, but not necessary flat, band~\cite{Liang2017a}. To make this work self-contained and easier to read, we first repeat with our notation the derivation of the conventional and geometric contributions to the superfluid weight within the simplest non-gauge invariant mean-field theory. This derivation is found originally in Ref.~\cite{Liang2017a}. Then in Sec.~\ref{sec:GRPA_correction}, we extend this derivation to provide a closed-form expression for the correction term to the superfluid weight that is obtained by taking into account the dependence of both the Hartree-Fock and pairing potentials on the constant vector potential $\vb{A}$. An important result of Sec.~\ref{sec:GRPA_correction} is that this GRPA correction term is also the sum of a conventional part and a geometric part, which depend on the derivatives of the band dispersion and the periodic Bloch functions, respectively. The conventional GRPA correction is geometry independent and always leads to an increase of the superfluid weight compared to the mean-field result. 
   	The expression for the conventional part of the GRPA correction is an original result of our work. The conventional GRPA term appears because we take into account the Hartree potential in our case, contrary to Ref.~\cite{Huhtinen2022} (the Fock potential vanishes for the Hubbard interaction in the absence of magnetic order).
   	On the other hand, the geometric part of the GRPA correction is essential to cure the problem found in Ref.~\cite{Huhtinen2022} of the unphysical dependence of the superfluid weight on the orbital positions. Compared to Ref.~\cite{Huhtinen2022}, we provide an explicit expression for the geometric GRPA correction even for an isolated band that is not necessarily flat.
   	
   	In Sec.~\ref{sec:Natural_orbital} the geometric part of the GRPA correction is analyzed in more detail. 
   	More precisely, we show that the sum of the geometric GRPA correction and mean-field geometric contribution gives the minimal quantum metric  in the flat band case, as found in Ref.~\cite{Huhtinen2022}. Simple analytical expressions are also provided for both the minimal quantum metric and the associated natural orbital positions that minimize the integrated quantum metric, see Eqs.~\eqref{eq:minimal_qm}-\eqref{eq:R_integrated_def} and~\eqref{eq:linear_sys_R}. At the end of Sec.~\ref{sec:Natural_orbital}, we explain how these results can be straightforwardly extended to an isolated band that is not necessarily flat. Since we expect the minimal quantum metric and the natural orbital positions to find applications also beyond the context of superfluidity in multiorbital lattices, in Sec.~\ref{sec:examples} we provide some examples, that is we compute these band structure invariants for three different representative lattice models: the Su-Schrieffer-Heeger model, the Creutz ladder and the dice lattice. Finally, in Sec.~\ref{sec:conclusion} we summarize and discuss our results and single out possible directions for further work.
   	
   	Appendices~\ref{app:grpa},~\ref{app:corr_computation} and~\ref{app:uniform_pairing_condition} contain all the necessary computational details that should allow the interested reader to understand, reproduce and ultimately extend and apply our results to other related problems. Appendix~\ref{app:grpa} provides a derivation of the GRPA result for the superfluid weight, which in the case of the Hubbard interaction is given by Eqs.~\eqref{eq:vect_v_def}-\eqref{eq:GRPA} in Sec.~\ref{sec:GRPA}. This derivation is different but equivalent to the one of Ref.~\cite{Peotta2022}. Appendix~\ref{app:corr_computation} collects useful results for computing the correlation functions that enter in the GRPA expression for the superfluid weight. Finally, in Appendix~\ref{app:uniform_pairing_condition} the self-consistency equations of  mean-field theory are derived in the case of the Hubbard interaction.

	\section{Geometry independence of superfluid weight and gauge invariance}
    \label{sec:gauge}
	
	In this section we introduce the superfluid weight as the second derivative of the thermodynamic grand potential $\Omega(\vb{A})$ with respect to a constant electromagnetic vector potential $\vb{A}$. The vector potential parametrizes twisted boundary conditions in a finite size lattice model with a torus geometry. Moreover, we provide a simple argument based on gauge invariance showing that the superfluid weight is a geometry independent quantity. This section contains only a short summary of the concepts that are needed for the present work. For a more extensive and rigorous presentation, the reader should consult Refs.~\cite{Peotta2022} and~\cite{Huhtinen2022}, which are the basis for the results presented here. The notation used here is essentially the same as the one of Ref.~\cite{Peotta2022}.
	
	We consider a generic lattice model described by a noninteracting, or free, Hamiltonian that is quadratic in the fermionic field operators $\hat{c}_{\vb{i}\alpha\sigma}$, $\hat{c}_{\vb{i}\alpha\sigma}^\dg$, and depends parametrically on a constant electromagnetic vector potential $\vb{A}$ through the usual Peierls phase
	\begin{equation}
		\label{eq:Ham_free_A}
		\mathcal{\hat H}_{\rm free}(\vb{A}) = \sum_{\sigma = \uparrow,\downarrow}\sum_{\vb{i}\alpha,\vb{j}\beta}  \hat{c}_{\vb{i}\alpha\sigma}^\dg [H^{\sigma}_{\rm free}]_{\vb{i}\alpha,\vb{j}\beta}
		e^{i\vb{A}\cdot\vb{r}_{\vb{i}\alpha,\vb{j}\beta}} 
		\hat{c}_{\vb{j}\beta\sigma}\,.
	\end{equation} 
	Here $H^{\sigma}_{\rm free}$ is the hopping matrix, which is translationally invariant since its matrix elements $[H^{\sigma}_{\rm free}]_{\vb{i}\alpha,\vb{j}\beta} = [H^{\sigma}_{\rm free}(\vb{i}-\vb{j})]_{\alpha,\beta}$  depend only on the difference between the unit cell indices $\vb{i} = (i_1, i_2)^T$ and $\vb{j}= (j_1,j_2)^T$, and  $\vb{r}_{\vb{i}\alpha,\vb{j}\beta}$ is the displacement vector from site $\vb{j}\beta$ to site $\vb{i}\alpha$, where $\alpha,\beta = 1,2,\dots,N_{\rm orb}$ label the orbitals inside the unit cell. The number of orbitals is denoted by $N_{\rm orb}$, while $N_{\rm c}$ is the number of unit cells. Without loss of generality, only two-dimensional lattice models are considered in the present work. Note also that the free Hamiltonian commutes by construction with the $z$-axis spin component operator
	\begin{gather}
		\label{eq:Sz}
		\hat{S}^z = \frac{1}{2}\sum_{\vb{i}\alpha}(\hat{n}_{\vb{i}\alpha\uparrow}-\hat{n}_{\vb{i}\alpha\downarrow})\,,\quad \hat{n}_{\vb{i}\alpha\sigma} = \hat{c}_{\vb{i}\alpha\sigma}^\dg \hat{c}_{\vb{i}\alpha\sigma}\,.
	\end{gather}

	Periodic boundary conditions are imposed by starting with an infinitely extended and translational invariant lattice, that is a collection of lattice sites located at positions $\vb{r}_{\vb{i}\alpha}$. As a consequence of translational invariance, the site positions transform as follows under shifts of the  unit cell index
	\begin{equation}
		\vb{r}_{\vb{i}+\vb{j},\alpha} = \vb{r}_{\vb{i},\alpha}
		+ j_1\vb{a}_1 + j_2\vb{a}_2\,,\qq{with} \vb{j} = \pmqty{j_1\\j_2}\,.
	\end{equation}
	The fundamental vectors $\vb{a}_1$ and $\vb{a}_2$ generate the Bravais lattice $B = \mathrm{Span}_{\mathbb{Z}}(\vb{a}_1,\vb{a}_2)$ that encodes the translational symmetry of the model. Then, in order to obtain a finite size lattice model with a torus geometry, one fixes two noncollinear Bravais lattice vectors $\vb{R}_1,\,\vb{R}_{2}\in B$  and identifies any pair of lattice sites, labeled by  $\vb{i}\alpha$ and $\vb{j}\beta$, such that $\vb{r}_{\vb{i}\alpha} - \vb{r}_{\vb{j}\beta} = m_1\vb{R}_1 + m_2\vb{R}_2$ with $m_1\,,m_2$ arbitrary intergers. The hopping matrix elements of the finite size lattice are obtained unambiguously from the ones of the infinite lattice, provided that the hopping matrix that enters in the free Hamiltonian~\eqref{eq:Ham_free_A} has finite range, namely $[H_{\rm free}(\vb{i}-\vb{j})]_{\alpha,\beta} = 0$ for $|\vb{r}_{\vb{i}\alpha} - \vb{r}_{\vb{j}\beta}| > R$, and $R \ll \abs{\vb{R}_1},\,\abs{\vb{R}_2}$. The procedure just presented for obtaining a finite size lattice model with periodic boundary conditions from an infinite extended one has been introduced and explained in more detail in Ref.~\cite{Tovmasyan2018}.
 	
 	It is crucial to note that in a finite size lattice with periodic boundary condition, it is not possible to write the displacement vectors as the difference of the site positions, namely $\vb{r}_{\vb{i}\alpha,\vb{j}\beta} \neq \vb{r}_{\vb{i}\alpha} -\vb{r}_{\vb{j}\beta}$, otherwise it would be possible to eliminate the Peierls phase $e^{i\vb{A}\cdot\vb{r}_{\vb{i}\alpha,\vb{j}\beta}}$ in~\eqref{eq:Ham_free_A} by means of a gauge transformation and the constant vector potential $\vb{A}$ would have no observable effects. Instead, one has to define the displacement vectors more carefully, as explained in Ref.~\cite{Peotta2022}. For periodic boundary conditions, the vector potential $\vb{A}$ affects the eigenvalues and eigenfunctions of the free Hamiltonian~\eqref{eq:Ham_free_A} since it correspond to magnetic fluxes through the holes  of the torus. It can be shown using a gauge transformation, that a nonzero $\vb{A}$ is equivalent to twisted boundary conditions~\cite{Tovmasyan2018}.
	
	The free Hamiltonian~\eqref{eq:Ham_free_A} is translationally invariant, thus it is convenient to expand the field operators in their Fourier components
	\begin{equation}
		\label{eq:Fourier_exp}
		\hat{c}_{\vb{i}\alpha\sigma} = \frac{1}{\sqrt{N_{\rm c}}}\sum_{\vb{k}} e^{i\vb{k}\cdot \vb{r}_{\vb{i}\alpha}}\hat{c}_{\vb{k}\alpha\sigma}\,,
	\end{equation}
	where the wave vectors $\vb{k}$ are discretized according to the relation $\vb{k}\cdot \vb{R}_j = 2\pi n_j$ for integers $n_j$. Inserting the Fourier expansion~\eqref{eq:Fourier_exp} in~\eqref{eq:Ham_free_A} leads to
	\begin{equation}
		\label{eq:H_free_Fourier}
		\begin{split}
			&\mathcal{\hat H}_{\rm free}(\vb{A}) = \sum_\sigma \sum_{\vb{k},\alpha,\beta}
			\hat{c}_{\vb{k}\alpha\sigma}^\dg [H^\sigma_{\rm free}(\vb{k}-\vb{A})]_{\alpha,\beta}\hat{c}_{\vb{k}\beta\sigma}\,,\\
			&\quad[H_{\rm free}^\sigma(\vb{k})]_{\alpha,\beta} = \sum_{\vb{i}-\vb{j}}  [H^{\sigma}_{\rm free}(\vb{i}-\vb{j})]_{\alpha, \beta}e^{-i\vb{k}\cdot\vb{r}_{\vb{i}\alpha,\vb{j}\beta}}\,.
		\end{split}
	\end{equation}
	In the following we need the average current density operator (current operator for short) defined as
	\begin{equation}
		\label{eq:current_operator_definition}
		\begin{split}
			\hat{\vb{J}} &= -\grad_{\vb{A}}\ham_{\rm free}(\vb{A})\big|_{\vb{A} = \vb{0}}
			\\
			&= \sum_\sigma \sum_{\vb{k},\alpha,\beta}
			\hat{c}_{\vb{k}\alpha\sigma}^\dg [\grad_{\vb{k}}H^\sigma_{\rm free}(\vb{k})]_{\alpha,\beta}\hat{c}_{\vb{k}\beta\sigma}\,.
		\end{split}
	\end{equation}
	Note that here we have not normalized the current density operator by the area of the system as done in Ref.~\cite{Peotta2022}.
	
	In order to formulate the concept of geometry independence, introduced in Ref.~\cite{Simon2020}, we define an operator encoding a shift in the orbital positions
	\begin{equation}
		\hat{\vb{b}} = \sum_\sigma\sum_{\vb{i}\alpha} \vb{b}_\alpha \hat{n}_{\vb{i}\alpha\sigma}\,,\qquad \hat{b}_l = \sum_\sigma\sum_{\vb{i}\alpha} [\vb{b}_\alpha]_l \hat{n}_{\vb{i}\alpha\sigma}\,. 
	\end{equation}
	The vector $\vb{b}_\alpha = \qty\big([\vb{b}_\alpha]_x, [\vb{b}_\alpha]_y)^T$ is the shift of the position of the orbital labeled by $\alpha$ since the new orbital positions $\vb{r}'_{\vb{i\alpha}}$ and displacement vectors $\vb{r}'_{\vb{i}\alpha,\vb{j}\beta}$ are given
	\begin{equation}
		\label{eq:pos_displ_shift}
		\vb{r}'_{\vb{i}\alpha} = \vb{r}_{\vb{i}\alpha} + \vb{b}_\alpha\,,\qquad \vb{r}'_{\vb{i}\alpha,\vb{j}\beta} = \vb{r}_{\vb{i}\alpha,\vb{j}\beta}+ \vb{b}_\alpha -\vb{b}_\beta\,.
	\end{equation}
	The hopping Hamiltonian $H^\sigma_{\rm free}(\vb{k})$ in momentum space depends on the choice of the position vectors and the displacement vectors, in fact one has
	\begin{gather}
		\label{eq:H_free_orbital_transf}
		\begin{split}
		[H_{\rm free}^{\sigma\prime}(\vb{k})]_{\alpha,\beta} &= \sum_{\vb{i}-\vb{j}}  [H^{\sigma}_{\rm free}(\vb{i}-\vb{j})]_{\alpha, \beta}e^{-i\vb{k}\cdot\vb{r}'_{\vb{i}\alpha,\vb{j}\beta}}
		\\
		&=[e^{-i\vb{k}\cdot\vb{b}}H_{\rm free}^{\sigma}(\vb{k})e^{i\vb{k}\cdot\vb{b}}]_{\alpha,\beta}\,,
		\end{split}
	\end{gather}
	where in the last equation $\vb{b}$ is a single-particle operator in orbital space, whose eigenvalues are the orbital position shifts $\vb{b}_\alpha$, namely $\vb{b}\ket{\alpha} = \vb{b}_\alpha\ket{\alpha}$.
	The operator $\hat{\vb{b}}$ generates gauge transformations parameterized by $\vb{A}$
	\begin{equation}
		\label{eq:gauge_trans}
		\hat{U}(\vb{A}) = e^{i\vb{A}\cdot\hat{\vb{b}}}\,,\qquad \hat{U}(\vb{A}) \hat{c}_{\vb{i}\alpha} \hat{U}^\dg(\vb{A}) = e^{-i\vb{A}\cdot \vb{b}_\alpha}\hat{c}_{\vb{i}\alpha}\,.	
	\end{equation}
	Using the shifted displacement vectors, one obtains a new noninteracting Hamiltonian
	\begin{equation}
		\begin{split}
		&\mathcal{\hat H}'_{\rm free}(\vb{A})= \hat{U}(\vb{A}) \mathcal{\hat H}_{\rm free}(\vb{A}) \hat{U}^\dg(\vb{A}) \\
		&\quad= \sum_\sigma\sum_{\vb{i}\alpha,\vb{j}\beta}  \hat{c}_{\vb{i}\alpha\sigma}^\dg [H^{\sigma}_{\rm free}(\vb{i}-\vb{j})]_{\alpha,\beta}
		e^{i\vb{A}\cdot\vb{r}'_{\vb{i}\alpha,\vb{j}\beta}}\hat{c}_{\vb{j}\beta\sigma} \,.
		\end{split}
	\end{equation}
	The current operator also changes accordingly
	\begin{equation}
		\begin{split}
			&\hat{\vb{J}}' = -\grad_{\vb{A}}\ham'_{\rm free}(\vb{A})\big|_{\vb{A} = \vb{0}} = 
			\hat{\vb{J}} - i[\hat{\vb{b}},\ham_{\rm free}]\,,\\
			&\qq*{with} \ham_{\rm free} = \ham_{\rm free}(\vb{A} = \vb{0}) = \ham'_{\rm free}(\vb{A} = \vb{0})\,.
		\end{split}
	\end{equation}
	
	In this work, we consider the case of an attractive Hubbard interaction term
	\begin{equation}
		\label{eq:Hubbard_int}
		\ham_{\rm int} = -\sum_{\vb{i}\alpha}U_\alpha\hat{n}_{\vb{i}\alpha\uparrow}\hat{n}_{\vb{i}\alpha\downarrow}\,,\qq{with} U_\alpha \geq 0\,.
	\end{equation}
	Thus, the full many-body Hamiltonian is
	\begin{equation}
		\ham(\vb{A}) = \ham_{\rm free}(\vb{A}) -\mu\hat{N}+ \ham_{\rm int}\,, 
	\end{equation}
	with $\hat{N} = \sum_{\sigma}\sum_{\vb{i}\alpha} \hat{n}_{\vb{i}\alpha\sigma}$ the particle number operator.
	Note that the interaction term is invariant under the gauge transformations~\eqref{eq:gauge_trans}, namely $[\ham_{\rm int},\hat{U}(\vb{A})] = 0$, as a consequence only the noninteracting part of the many-body Hamiltonian is modified by the gauge transformation~\eqref{eq:gauge_trans}
	\begin{equation}
		\begin{split}
		\ham'(\vb{A})  &= \hat{U}(\vb{A}) \ham(\vb{A}) \hat{U}^\dg(\vb{A}) \\ &= \ham_{\rm free}'(\vb{A})-\mu\hat{N} + \ham_{\rm int}\,.
		\end{split}
	\end{equation}
	Due to the property of similarity-invariance of the trace $\Tr[H] = \Tr[UHU^{-1}]$, it follows that the thermodynamic grand potential is invariant under gauge transformations 
	\begin{equation}
		\label{eq:grand_pot_invariance}
		\begin{split}
		\Omega(\vb{A}) &= -\beta^{-1}\ln\Tr\qty\big[e^{-\beta \ham(\vb{A})}] \\
		&= -\beta^{-1}\ln\Tr\qty\big[e^{-\beta \ham'(\vb{A})}]\,,
		\end{split}
	\end{equation}
	($\beta = 1/(k_{\rm B}T)$ is the inverse temperature).
	The superfluid weight is defined as the second derivative of the grand potential with respect to the uniform vector potential  $\vb{A}$
	\begin{equation}
		\label{eq:d2F_dA2}
		D_{{\rm s},lm} = \frac{1}{\mathcal{A}}\pdv{\Omega(\vb{A})}{A_l}{A_m}\bigg|_{\vb{A}=\vb{0}}\,,
	\end{equation}
	where $\mathcal{A} = |\vb{R}_1\times \vb{R}_2|$ is the area of the finite size system with periodic boundary conditions. From~\eqref{eq:grand_pot_invariance}, it is evident that the superfluid weight is a geometry independent quantity~\cite{Simon2020}, i.e. it is invariant under shifts of the orbital positions inside the unit cell. Notice how this result is a consequence of the fact that, under an orbital position shift, the displacement vectors in~\eqref{eq:pos_displ_shift} are modified by a difference term $\vb{b}_\alpha-\vb{b}_{\beta}$, while the displacement vector $\vb{r}_{\vb{i}\alpha,\vb{j}\beta}$ cannot be written as a difference, as discussed above.
	
	 The concept of geometry independence, or dependence, has been discussed in Ref.~\cite{Simon2020} and its utility has been pointed out in relation to several observable quantities. However, the superfluid weight is the first observable that has been shown to be geometry independent as a consequence of gauge invariance. The same approach may be useful to prove the geometry independence of other quantities in the future.
	
	\section{Generalized random phase approximation}
    \label{sec:GRPA}
	
	An attractive interaction, such as the one in~\eqref{eq:Hubbard_int}, generally leads to the emergence of superfluid phases characterized by a nonzero superfluid weight~\cite{Scalapino1992,Scalapino1993}. In the context of superfluidity and superconductivity, mean-field theory is the simplest and most common approximation. This is a variational approximation for the grand potential based on the Bogoliubov inequality~\cite{Feynman1972,J.J.Binney1992,Peotta2022}
	\begin{equation}
		\label{eq:Bogololiubv_inequality}
		\Omega \leq \Omega_{\rm m.f.} = \Omega_0 + \ev*{\ham -\ham_0}\,.
	\end{equation}
	The auxiliary grand potential $\Omega_0 = -\beta^{-1}\ln\Tr\qty\big[e^{-\beta \ham_0}]$ is obtained from a quadratic variational Hamiltonian $\ham_0$, which, in the case of the Hubbard interaction~\eqref{eq:Hubbard_int}, takes the form
	\begin{gather}
		\begin{split}
		\label{eq:H0}
		\ham_0(\vb{A}) &= \ham_{\rm free}(\vb{A}) -\mu\hat{N} + \sum_{\sigma,\alpha} \Gamma_{\alpha}^\sigma \hat{N}_{\alpha\sigma} \\
		&\quad + \sum_\alpha\qty(\Delta_\alpha D_\alpha^\dg + \Delta^*_\alpha D_\alpha)\,, 
		\end{split}	\\
        \label{eq:operator_N}
		\hat{N}_{\alpha\sigma} = \sum_{\vb{i}} \hat{c}^\dg_{\vb{i}\alpha\sigma}\hat{c}_{\vb{i}\alpha\sigma}
		 =  \sum_{\vb{k}} \hat{c}^\dg_{\vb{k}\alpha\sigma}\hat{c}_{\vb{k}\alpha\sigma}\,,
		\\
        \label{eq:operator_D}
		\hat{D}_\alpha =  \sum_{\vb{i}}\hat{c}_{\vb{i}\alpha\downarrow}\hat{c}_{\vb{i}\alpha\uparrow} =  \sum_{\vb{k}}\hat{c}_{-\vb{k}\alpha\downarrow}\hat{c}_{\vb{k}\alpha\uparrow}\,.
	\end{gather}
	The expectation value on the right hand side of~\eqref{eq:Bogololiubv_inequality} is evaluated with respect to the statistical ensemble associated to the variational Hamiltonian $\ham_0$, namely
	\begin{equation}
		\label{eq:H-H0}
		\ev*{\ham -\ham_0} = \frac{1}{\Tr\qty\big[e^{-\beta\ham_0}]}\Tr\qty\big[(\ham -\ham_0)e^{-\beta\ham_0}]\,.
	\end{equation}
	The coefficients $\Gamma_{\alpha}^\sigma$ and $\Delta_\alpha$ that appear in $\ham_0$ are variational coefficients that are chosen so as to minimize the mean-field grand potential $\Omega_{\rm m.f.}$ on the right hand side of~\eqref{eq:Bogololiubv_inequality}. When the minimum of $\Omega_{\rm m.f.}$ is attained, the Hartree potential $\Gamma^\sigma_\alpha$ and the pairing potential $\Delta_\alpha$ satisfy the self-consistency conditions of mean-field theory
	\begin{gather}
        \label{eq:Gamma_self_eq}
		\Gamma_\alpha^\sigma = -U_\alpha\ev*{\hat{c}^\dg_{\vb{i}\alpha\bar\sigma}\hat{c}_{\vb{i}\alpha\bar\sigma}} = -\frac{U_\alpha}{N_{\rm c}}\ev*{\hat{N}_{\alpha\bar\sigma}} \,,\\
        \label{eq:Delta_self_eq}
		\Delta_\alpha = -U_\alpha\ev*{\hat{c}_{\vb{i}\alpha\downarrow}\hat{c}_{\vb{i}\alpha\uparrow}} = -\frac{U_\alpha}{N_{\rm c}}\ev*{\hat{D}_\alpha}\,.
	\end{gather}
	where the expectation values are evaluated as in~\eqref{eq:H-H0}. In~\eqref{eq:Gamma_self_eq} we define $\bar{\sigma} = \downarrow$ if $\sigma = \uparrow$ and viceversa.
 In the equations above, we have assumed that translational symmetry is not broken, therefore the expectation values $\ev*{\hat{c}^\dg_{\vb{i}\alpha\sigma}\hat{c}_{\vb{i}\alpha\sigma}}$ and $\ev*{\hat{c}_{\vb{i}\alpha\downarrow}\hat{c}_{\vb{i}\alpha\uparrow}}$ do not depend on the unit cell index $\vb{i}$. It is also assumed that spin rotational symmetry around the $z$-axis is preserved by the mean-field solution, indeed the variational Hamiltonian~\eqref{eq:H0} commutes with the spin operator~\eqref{eq:Sz}. This implies that the Fock mean-field potential term, proportional to $\hat{c}_{\vb{i}\alpha\uparrow}^\dg \hat{c}_{\vb{i}\alpha\downarrow}$ in the case of the Hubbard interaction, does not appear in $\ham_0$. This assumption is justified since spin rotational symmetry breaking is associated to magnetic order, which generally occurs for repulsive interactions.
	
	The mean-field grand potential is minimized with respect to $\Delta_\alpha$ and $\Gamma_\alpha^\sigma$ for each separate value of the vector potential $\vb{A}$, therefore the Hartree and pairing potentials become themselves function of $\vb{A}$. It has been shown in Ref.~\cite{Huhtinen2022} that one cannot ignore this dependence in the pairing potential $\Delta(\vb{A})$ when the mean-field approximation for the superfluid weight is evaluated as
	\begin{equation}
		\label{eq:D_mf}
		D_{{\rm s},lm} \approx 
		\frac{1}{\mathcal{A}}
		\frac{d^2\Omega_{\rm m.f.}\big(\vb{A},\Gamma(\vb{A}),\Delta(\vb{A})\big)}{dA_l dA_m}
		\bigg|_{\vb{A}=\vb{0}}\,.
	\end{equation}
	We denote by $d\Omega_{\rm m.f.}\big(\vb{A},\Delta(\vb{A}),\Gamma(\vb{A}))/dA_l$ the full derivative  of the mean-field grand potential, including also the $\vb{A}$-dependence of the mean-field potentials $\Gamma_\alpha^\sigma(\vb{A})$ and $\Delta_\alpha(\vb{A})$, while the partial derivative $\partial \Omega_{\rm m.f.}\big(\vb{A},\Gamma(\vb{A}),\Delta(\vb{A}))/\partial A_l$ denotes the derivative with respect to the first argument only.
	Replacing the full derivatives with the partial derivatives in~\eqref{eq:D_mf} is a commonly used approximation~\cite{Scalapino1993,Peotta2015,Liang2017a}. However, it has the disadvantage of breaking gauge invariance and thus geometry independence~\cite{Huhtinen2022}. On the other hand, it has been shown~\cite{Peotta2022} that the superfluid weight computed from~\eqref{eq:D_mf} (with the full derivatives) is in fact equivalent to the generalized random phase approximation~\cite{Anderson1958,Rickayzen1959}, which is an approximation that preserves gauge invariance. 
	
	As shown in Appendix~\ref{app:grpa}, it is possible to express the full second derivatives of the mean-field grand potential in~\eqref{eq:D_mf} in terms of correlation functions evaluated on the mean-field statistical ensemble for $\vb{A} = \vb{0}$ only. This is advantageous since it is not necessary to solve the mean-field problem for several different values of $\vb{A}$ in order to evaluate the superfluid weight. To present this result, we introduce the following convenient notation
	\begin{equation}
		\label{eq:corr_func_def}
		(\hat{A},\hat{B}) = -\int_0^\beta\dd{\tau} \big\langle 
		\qty\big(\hat{A}(\tau)-\ev*{\hat{A}})
		\qty\big(\hat{B}(\tau)-\ev*{\hat{B}})
		\big\rangle\,,
	\end{equation} 
	where $\hat{A}$ and $\hat{B}$ are arbitrary operators and the notation $\hat{A}(\tau) = e^{\tau\ham_0} \hat{A} e^{-\tau \ham_0}$ for the time evolution of an operator in imaginary time $\tau$ has been used. In fact, the symbol introduced in ~\eqref{eq:corr_func_def} is simply the imaginary time Green's function evaluated at the Matsubara frequency $i\omega_n = 0$~\cite{Fetter2003} and has some useful properties that are easy to prove
	\begin{gather}
	\label{eq:corr_func_prop_1}
	(\hat{A},\hat{B}) = (\hat{B},\hat{A})\,,\\
	\label{eq:corr_func_prop_2}
	(\hat{A},\hat{B})^* = (\hat{B}^\dg,\hat{A}^\dg)\,,\\
	\label{eq:corr_func_prop_3}
	(\hat{A}^\dg,\hat{A}) \leq 0\,.
	\end{gather}
    In Appendix \ref{app:corr_computation}, we introduce a compact method to evaluate the correlation function in  \eqref{eq:corr_func_def} between two translationally invariant quadratic operators $\hat A$ and $\hat B$. 

	To express the result for the superfluid weight in the generalized random phase approximation, we need correlations functions of the form~\eqref{eq:corr_func_def} of pairs of operators taken from the set $\{\hat{J}_l,\,\hat{N}_{\alpha\sigma},\,\hat{D}_\alpha\}$. The correlation functions that involve the components of the current operator $\hat{J}_l$~\eqref{eq:current_operator_definition} are organized into a vector
	\begin{equation}
		\label{eq:vect_v_def}
		\vb{v}_l = \pmqty{\vb{v}_{l,\alpha = 1} \\ 
			\vb{v}_{l,\alpha = 2} \\ \vdots \\ \vb{v}_{l,\alpha = N_{\rm orb}}}\,,\quad 
		\vb{v}_{l\alpha} =
		\pmqty{(\hat{J}_l, \hat{N}_{\alpha\uparrow}) \\ 
			(\hat{J}_l, \hat{N}_{\alpha\downarrow}) \\
			(\hat{J}_l, \hat{D}_{\alpha}) \\ (\hat{J}_l, \hat{D}^\dg_{\alpha})}\,.
	\end{equation}
	Instead, all the remaining correlation functions are collected into a matrix
	\begin{gather}
		\label{eq:A_alphabeta}
		\begin{split}
		&A_{\alpha,\beta} = \\&\pmqty{(\hat{N}_{\alpha \uparrow}, \hat{N}_{\beta \uparrow}) & (\hat{N}_{\alpha \uparrow}, \hat{N}_{\beta \downarrow}) & (\hat{N}_{\alpha \uparrow}, \hat{D}_{\beta}) &  (\hat{N}_{\alpha \uparrow}, \hat{D}^\dg_{\beta}) \\
			(\hat{N}_{\alpha \downarrow}, \hat{N}_{\beta \uparrow}) & (\hat{N}_{\alpha \downarrow}, \hat{N}_{\beta \downarrow}) & (\hat{N}_{\alpha \downarrow}, \hat{D}_{\beta}) &  (\hat{N}_{\alpha \downarrow}, \hat{D}^\dg_{\beta}) \\
			(\hat{D}_{\alpha}, \hat{N}_{\beta \uparrow}) & (\hat{D}_{\alpha}, \hat{N}_{\beta \downarrow}) & (\hat{D}_{\alpha}, \hat{D}_{\beta}) &  (\hat{D}_{\alpha}, \hat{D}^\dg_{\beta}) \\
			(\hat{D}_{\alpha}^\dg, \hat{N}_{\beta \uparrow}) & (\hat{D}_{\alpha}^\dg, \hat{N}_{\beta \downarrow}) & (\hat{D}_{\alpha}^\dg, \hat{D}_{\beta}) &  (\hat{D}^\dg_{\alpha}, \hat{D}^\dg_{\beta})}\,,
		\end{split}
		\\
		A = \pmqty{A_{1,1} & A_{1,2} & \dots & A_{1,N_{\rm orb}}\\
			A_{2,1} & A_{2,2}  & & \\
			\vdots &  & \ddots &  \\
			A_{N_{\rm orb},1} & & & A_{N_{\rm orb},N_{\rm orb}}}\,.	
	\end{gather}
	Finally, we need also the following matrix
	\begin{gather}
		\label{eq:B_alpha}
		B_{\alpha} = -\frac{U_\alpha}{N_{\rm c}}\pmqty{0 & 1 & 0 & 0 \\ 
			1 & 0 & 0 & 0 \\
			0 & 0 & 0 & 1 \\
			0 & 0 & 1 & 0} \,,
		\\
		B = \pmqty{\dmat{B_{\alpha = 1}, B_{\alpha = 2},\ddots,B_{\alpha = N_{\rm orb}}}}\,.
	\end{gather} 
	As shown in Appendix~\ref{app:grpa}, the full derivative of the mean-field grand potential can then be expressed as
	\begin{equation}
		\label{eq:GRPA}
		\begin{split}
			&\left.\frac{d^2}{dA_m dA_l} \qty[\Omega_{\rm m.f.}\qty\big(\vb{A},\Delta(\vb{A}),\Gamma(\vb{A}))]\right|_{\vb{A} = \vb{0}} \\
			&= \left.\pdv{\Omega_0}{A_l}{A_m}\right|_{\vb{A} = \vb{0}} +\vb{v}_l^TB\vb{v}_m + \vb{v}_l^TB\frac{1}{A^{-1}-B}B\vb{v}_m \\
			&=\left.\pdv{\Omega_0}{A_l}{A_m}\right|_{\vb{A} = \vb{0}} +\vb{v}_l^TB\vb{v}_m +  
			\vb{v}_l^TB AB\vb{v}_m \\
			&\qquad +  
			\vb{v}_l^TB AB AB\vb{v}_m+\vb{v}_l^TB AB ABAB\vb{v}_m+\dots \\ 
		\end{split}	
	\end{equation}
	In the last equality we have used the geometric series expansion
	\begin{equation}
			\frac{1}{A^{-1}-B} = A + ABA+ABABA+\dots
	\end{equation}
	As mentioned above, the superfluid weight is often computed by retaining only the second partial derivatives of the mean-field grand potential in~\eqref{eq:GRPA}, given by
	\begin{equation}
		\label{eq:zero_order}
		\left.\pdv{\Omega_0}{A_l}{A_m}\right|_{\vb{A} = \vb{0}}
		= \left.\ev{\pdv{\ham_{\rm free}(\vb{A})}{A_l}{A_m}}\right|_{\vb{A} = \vb{0}} + \qty\big(\hat{J}_l,\hat{J}_m)\,.
	\end{equation}
	The first term on the right hand side is known as the diamagnetic part of the current-current response function, while the second term is the paramagnetic one~\cite{Scalapino1993,Peotta2022}. It can be shown~\cite{Liang2017a} that the diamagnetic part is equal to a correlation function of the form~\eqref{eq:corr_func_def} (see~\eqref{eq:diagmagnetic_resp_corr_func}).

    \section{Superfluid weight in the isolated band limit}
    \label{sec:correction}
    
    The aim of this section is to compute analytically the superfluid weight in the isolated band limit within the generalized random phase approximation, which means that~\eqref{eq:GRPA} is reduced to integrals over the first Brillouin zone of certain combinations of the band dispersions and band wave functions, and of their derivatives with respect to quasimomentum. Ultimately, these integrals have to be evaluated numerically, however, in Sec.~\ref{sec:examples}, we provide also some examples in which fully analytical results can be obtained. The final expression presented below is valid for generic lattice models under few assumptions, the most important being time-reversal symmetry and the uniform pairing condition to be introduced in the following.
    For completeness, we first evaluate the terms corresponding to the second partial derivatives of $\Omega_0$~\eqref{eq:zero_order} and reobtain the known result that the superfluid weight in a multiband/multiorbital lattice can be separated into two contributions, called conventional and the geometric, respectively~\cite{Peotta2015,Liang2017a}. In particular, the quantum metric enters in the geometric contribution to the superfluid weight. However, as pointed in Refs.~\cite{Chan2022a,Huhtinen2022}, the quantum metric depends on the orbital positions, therefore it is not a geometry independent quantity. In order to restore gauge invariance and thus geometry independence, we evaluate the remaining terms in ~\eqref{eq:GRPA}. This amounts to computing the full derivatives with respect to the vector potential $\vb{A}$ rather than just the partial derivatives. The same approach has been used in Ref.~\cite{Huhtinen2022}, with the only difference that the Hartree potential $\Gamma_\alpha^\sigma$ in the variational Hamiltonian $\ham_0(\vb{A})$ is neglected in this latter work. By slightly modifying the derivation in Appendix~\ref{app:grpa}, it is shown that neglecting the Hartree potential amounts to setting to zero in~\eqref{eq:GRPA} all the correlation functions in which the number operators $\hat {N}_{\alpha\sigma}$ appear. 
    One of the main result of this work is to take into account the Hartree potential and show that in this way one obtains a new correction to the superfluid weight that is geometry independent and is proportional to the derivatives of the band dispersion. Therefore, in the language of Refs.~\cite{Julku2016,Liang2017a}, this is a conventional contribution since it vanishes in the flat band limit.
    

    We start by rewriting the variational quadratic Hamiltonian $\mathcal{\hat H}_0$ in Nambu form
    \begin{equation}
	\begin{split}
		\mathcal{\hat H}_0 &=  \hat{\vb{c}}_{\vb{k}}^\dg H_0(\vb{k}) \hat{\vb{c}}_{\vb{k}} + \text{const.}
	\end{split} 
	\end{equation}
	where the column (row) vector $\hat{\vb{c}}_{\vb{k}}$ ($\hat{\vb{c}}_{\vb{k}}^\dg$) is defined in ~\eqref{eq:Nambu_spinor} and $H_{0}(\bo k)$ is the single-particle Bogoliubov-de Gennes (BdG) Hamiltonian given by
    \begin{gather}
    	\label{eq:Bdg_Hamiltonian}
        H_0(\vb{k}) = 
		\pmqty{H_{\rm free}^\uparrow(\vb{k})+\Gamma^\uparrow -\mu & \Delta  \\ \Delta^* & -[H_{\rm free}^\downarrow(-\vb{k})]^*-\Gamma^\downarrow +\mu} \,,\\
		\Gamma^\sigma = \mathrm{diag}(\Gamma^\sigma_1,\Gamma^\sigma_2,\dots,\Gamma^\sigma_{N_{\rm orb}})\,,\\
		\Delta = \mathrm{diag}(\Delta_1,\Delta_2,\dots,\Delta_{N_{\rm orb}})\,. \label{eq:Delta_matrix}
    \end{gather}
    The Nambu form for translational invariant quadratic operators is discussed in Appendix~\ref{app:corr_computation}.
    From now on time-reversal symmetry is assumed, which implies $[H_{\rm free}^\downarrow(-\vb{k})]^*  = H_{\rm free}^\uparrow(\vb{k})$ and $\Gamma^\uparrow = \Gamma^\downarrow$. The second assumption, which enables the analytic evaluation of the superfluid weight for generic lattices, is called the uniform pairing condition, expressed by
    \begin{equation}
    	\label{eq:uniform_pairing}
    	\Delta_\alpha  = \Delta_\beta = \Delta\,,\qq{for all} \alpha,\beta\,.
    \end{equation}
    The pairing potential $\Delta$ can also be taken real and positive.
    With a slight abuse of notation, we indicate with $\Delta$ both the scalar value of the uniform pairing potential in~\eqref{eq:uniform_pairing} and the matrix in~\eqref{eq:Delta_matrix}, which becomes proportional to the identity. The uniform pairing conditions is justified in Appendix~\ref{app:uniform_pairing_condition} starting from the self-consistency equations of mean-field theory~\eqref{eq:Gamma_self_eq}-\eqref{eq:Delta_self_eq}.
    
    Under the above assumptions, the BdG Hamiltonian can be diagonalized as follows
    \begin{equation}
    \label{eq:BdG_Ham_diagonalized}
    \begin{split}
	H_0(\vb{k}) & = 
	\pmqty{U_{\vb{k}} & 0 \\ 0 & U_{\vb{k}}}
	\begin{pmatrix}
		\varepsilon_{\vb{k}} -\mu & \Delta \\
		\Delta & -(\varepsilon_{\vb{k}} -\mu)
	\end{pmatrix}
	\pmqty{U_{\vb{k}}^\dagger & 0 \\ 0 & U_{\vb{k}}^\dagger} \\
    & = (U_{\vb{k}}\oplus U_{\vb{k}})W_{\vb{k}}E_{\vb{k}}W_{\vb{k}}^\dagger
	(U_{\vb{k}}^\dagger\oplus U_{\vb{k}}^\dagger)\,.
    \end{split}
    \end{equation}
Here, $E_{\vb{k}} = E_{\vb{k}}^>\oplus (-E_{\vb{k}}^>)$ is the diagonal matrix of quasiparticle excitations with $E_{\vb{k}}^> = \mathrm{diag}(E_{n\vb{k}}) > 0$ the block of positive excitation energies, $[U_{\vb{k}}]_{\alpha,n} = g_{n\vb{k}}(\alpha) = \braket{\alpha}{g_{n\vb{k}}}$ the unitary matrix of Bloch functions that diagonalizes $H_{\rm free}^\uparrow(\vb{k}) + \Gamma^{\up} = U_{\vb{k}} \varepsilon_{\vb{k}} U_{\vb{k}}^\dagger$ and $\varepsilon_{\vb{k}} = \mathrm{diag}(\varepsilon_{n\vb{k}})$ the diagonal matrix of the band dispersions. Finally,
the BdG wave functions are given by
\begin{gather}
	W_{\vb{k}} = \begin{pmatrix}
		\mathrm{diag}(u_{n\vb{k}}) & -\mathrm{diag}(v_{n\vb{k}}) \\
		\mathrm{diag}(v_{n\vb{k}}) & \mathrm{diag}(u_{n\vb{k}})
	\end{pmatrix}\,, \\
    \label{eq:def_unk}
	u_{n\vb{k}} = \frac{1}{\sqrt{2}}\left(1+\frac{\varepsilon_{n\vb{k}}-\mu}{E_{n\vb{k}}}\right)^{\frac{1}{2}}\,,\\
    \label{eq:def_vnk}
	v_{n\vb{k}} = \frac{1}{\sqrt{2}}\left(1-\frac{\varepsilon_{n\vb{k}}-\mu}{E_{n\vb{k}}}\right)^{\frac{1}{2}}\,,\\
    \label{eq:def_Enk}
	E_{n\vb{k}} = \sqrt{(\varepsilon_{n\vb{k}}-\mu)^2+\Delta^2}\,.
\end{gather} 
The quantities $u_{n\vb{k}}$ and $v_{n\vb{k}}$ are the usual BCS coherence factors~\cite{Schrieffer1964,deGennes1966}. Note that in a multiband lattice model there is a pair $(u_{n\vb{k}},v_{n\vb{k}})$ of coherence factors for each band.

The total superfluid weight is separated into two contributions
\begin{equation}
	\label{eq:D0-D1}
	D_{\rm s} = D_{\rm s}^{(0)} + D_{\rm s}^{(1)}\,.
\end{equation}
The first contribution~$D_{\rm s}^{(0)}$ is the one given by the second partial derivative of $\Omega_0$, that is~\eqref{eq:zero_order}, while $D_{\rm s}^{(1)}$ includes all the remaining terms in~\eqref{eq:GRPA}. In the following, $D_{\rm s}^{(0)}$ is called the \qql mean-field theory (MFT) superfluid weight\qqr, because in all previous works, with the  exception of~Refs.~\cite{Huhtinen2022,Chan2022a}, this is the only term that is evaluated when the superfluid weight is computed within the mean-field approximation. Instead, $D_{\rm s}^{(1)}$ is referred to as the \qql GRPA correction\qqr to the superfluid weight. In fact, it would not be incorrect to consider the sum of the two contributions in~\eqref{eq:D0-D1} the actual mean-field theory result for the superfluid weight since it is obtained by taking the full derivative of the mean-field free energy $\Omega_{\rm m.f.}$, see~\eqref{eq:GRPA}. However, we avoid this nomenclature in order not to create confusion when referring to previous works.

As shown in the following, both the MFT superfluid weight and the GRPA correction can be written as the sums of a conventional contribution and a geometric contribution, which we indicate as $D_{\rm s}^{(j)} = D_{\rm s,c}^{(j)}+ D_{\rm s,g}^{(j)}$, with $j = 0,1$. As explained in Appendix~\ref{app:corr_computation}, the conventional contribution depends only on the intraband matrix elements of the current operator. According to~\eqref{eq:matr_elem_curr_op_identity}, this means that only the derivative of the band dispersions $\pdv*{\varepsilon_{n\vb{k}}}{k_l}$ with respect to the quasimomentum $\vb{k}$, that is the group velocity, enters into $D_{\rm s,c}^{(j)}$ and not the derivatives of the Bloch functions $\ket{\partial_lg_{n\vb{k}}}$. On the other hand, the geometric contribution  $D_{\rm s,g}^{(j)}$ is associated to the interband matrix elements of the current operator, therefore it depends only on scalar products of the form $\braket{g_{m\vb{k}}}{\partial_{l}g_{n\vb{k}}}$, but not on the group velocity. It was pointed out for the first time in Refs.~\cite{Peotta2015,Liang2017a} that the derivatives of the Bloch functions affect the mean-field superfluid weight in the form of an additional geometric contribution, indicated by $D^{(0)}_{\rm s,g}$ in our notation. A major result of the present work is to show that, under the same assumptions, the GRPA correction $D_{\rm s}^{(1)}$ can also be separated into a conventional $D_{\rm s,c}^{(1)}$ and a geometric part $D_{\rm s,g}^{(1)}$ to be presented below.

\subsection{MFT superfluid weight}
\label{sec:superfluid_weight_MF}

Using the results in Appendix~\ref{app:corr_computation}, more specifically by taking the isolated band limit in~\eqref{eq:Ds_0_full}, we obtain for the MFT superfluid weight
\begin{gather}
	D_{\rm s}^{(0)} = D_{\rm s,c}^{(0)}+ D_{\rm s,g}^{(0)}\,, \\
	\label{eq:D0_sc}
	\begin{split}
	&[D_{\rm s,c}^{(0)}]_{l,m} = \int \frac{\dd[2]{\vb{k}}}{(2\pi)^2}\frac{\Delta^2}{E_{\bar{n}\vb{k}}^2}\partial_{l}\varepsilon_{\bar{n}\vb{k}}\partial_{m}\varepsilon_{\bar{n}\vb{k}} \\ &\hspace{1cm}\times\qty[\frac{\tanh(\frac{\beta E_{\bar{n}\vb{k}}}{2})}{E_{\bar{n}\vb{k}}}-\frac{\beta}{2\cosh[2](\frac{\beta E_{\bar{n}\vb{k}}}{2})}]\,,
	\end{split}
	\\ 
	\label{eq:D0_sg}
	\begin{split}
	[D_{\rm s,g}^{(0)}]_{l,m} = \int\frac{\dd[2]{\vb{k}}}{(2\pi)^2} \frac{2\Delta^2}{E_{\bar{n}\vb{k}}}\tanh(\frac{\beta E_{\bar{n}\vb{k}}}{2})\mathcal{G}_{l,m}(\vb{k})\,,
	\end{split}
\end{gather}
where we have introduced the quantum metric
\begin{equation}
\label{eq:quantum_metric}
 \mathcal{G}_{l,m}(\vb{k}) = \Tr[\partial_lP(\vb{k})\partial_mP(\vb{k})]
\end{equation}
defined in terms of the projector $P(\vb{k}) = \dyad{g_{\bar{n}\vb{k}}}$ on the only partially filled band labeled by $\bar{n}$. The partially filled band in the noninteracting limit is determined by the condition $\min_{\vb{k}}\varepsilon_{\bar{n}\vb{k}} \leq \mu \leq  \max_{\vb{k}}\varepsilon_{\bar{n}\vb{k}}$ on the chemical potential $\mu$.  Isolated band limit means that the partially filled band is separated from all other bands by a large band gap $E_{\rm gap}$
\begin{equation}
 |\varepsilon_{m\vb{k}} - \varepsilon_{\bar{n}\vb{k}}| \geq  E_{\rm gap}	\qq{for} m\neq \bar{n}\,.
\end{equation}
Moreover, it is assumed that the interaction strength is much smaller than the band gap
\begin{equation}
	U_\alpha \ll E_{\rm gap}\,.
\end{equation}
In this limit, the superfluid weight and all of the observable properties of the system are determined by the band dispersion $\varepsilon_{\bar{n}\vb{k}}$ and the Bloch functions $\ket{g_{\bar{n}\vb{k}}}$ of the partially filled band. Rather counterintuitively, in order to obtain the correct expression for the geometric contribution to the MFT superfluid weight it is essential to take into account the interband matrix elements of the current operator, as explained in Appendix~\ref{app:corr_computation}. Na\"ively, one may neglect the interband matrix elements of the current operator in the isolated band limit, but the result is that the geometric contribution is lost. In general, the geometric contribution is considerably smaller than the conventional one for a dispersive band with a bandwidth much larger than the interaction strength, so the former can be safely neglected~\cite{Liang2017a,Torma2022}. On the other hand, when the partially dispersive band is quasi-flat, this not anymore an acceptable approximation, since the conventional contribution vanishes in the flat band limit ($\partial_{l}\varepsilon_{\bar{n}\vb{k}} = 0$) while the geometric one does not.

It is possible to rewrite the conventional contribution in a more suggestive form by using~\eqref{eq:Mm_11_nn} in the form given by the last term and performing an integration by parts. In this way, one obtains the following expression
\begin{gather}
	\label{eq:D0_Sc_alt}
	[D_{\rm s,c}^{(0)}]_{l,m} = \int \frac{\dd[2]{\vb{k}}}{(2\pi)^2}  \qty(n_{\vb{k}}\partial_l\partial_m\varepsilon_{\bar{n}\vb{k}} -\frac{\beta\partial_l\varepsilon_{\bar{n}\vb{k}}\partial_m\varepsilon_{\bar{n}\vb{k}}}{2\cosh[2](\frac{\beta E_{\bar{n}\vb{k}}}{2})} )\,,\\
	\qq*{with} n_{\vb{k}} = 1-\frac{\varepsilon_{\bar{n}\vb{k}}-\mu}{E_{\bar{n}\vb{k}}}\tanh(\frac{\beta E_{\bar{n}\vb{k}}}{2})\,.
\end{gather}
The first term inside the parenthesis is the inverse effective mass tensor $\partial_l\partial_m\varepsilon_{\bar{n}\vb{k}}  = [\frac{1}{m_{\rm eff}}]_{l,m}$ weighted by the occupation factor $n_{\vb{k}}$. It is not difficult to show that $n_{\vb{k}}$ is precisely the occupation number (including both spins) of the state with energy $\varepsilon_{\bar{n}\vb{k}}$ at thermal equilibrium. Therefore,  the interpretation of the first term is that all of the particles participate to the superfluid flow at zero temperature and the superfluid weight is an average measure of the effective mass of the carriers of the superfluid current. The second term vanishes at zero temperature and can be intepreted as a depletion of the superfluid component due to the thermal excitations of quasiparticles~\cite{Landau2006}.
However, this interpretation does not provide the full picture since the geometric contribution~\eqref{eq:D0_sg} is not taken into account. 

\subsection{GRPA correction to the superfluid weight}
\label{sec:GRPA_correction}

In the previous section we have reobtained the general result for the MFT superfluid weight in multiband/multiorbital lattices in the presence of an Hubbard interaction term, which has been discussed in a number previous works.
In this section, we move on to consider the GRPA correction to the superfluid weight, consisting of the terms in~\eqref{eq:GRPA} which depends on the matrix $A$ and the vector $\vb{v}_l$. Thus, we need to evaluate their these objects in the isolated band approximation. The details of the calculation are again provided in Appendix~\ref{app:corr_computation}.
For the matrix elements of the matrix $A_{\alpha,\beta}$ defined in~\eqref{eq:A_alphabeta}, we obtain
\begin{gather}
\label{eq:A_alphabeta_isolated_band}
\begin{split}
	A_{\alpha,\beta}  =&-\sum_{\vb{k}}
	\frac{\beta \abs{\mel{\alpha}{P(\vb{k})}{\beta}}^2}{4\cosh[2](\frac{\beta E_{\bar{n} \vb{k} }}{2})} C^{\rm c}_{\vb{k}}
	\\
	&+ \sum_{\vb{k}}
	\frac{\abs{\mel{\alpha}{P(\vb{k})}{\beta}}^2}{2E_{\bar{n} \vb{k}}} \tanh(\frac{\beta E_{\bar{n} \vb{k}}}{2}) C^{\rm g}_{\vb{k}}\,,
\end{split}
\end{gather}
with the two $4\times 4$ matrices $C^{\rm c}_{\vb{k}}$ and $C^{\rm g}_{\vb{k}}$ given by
\begin{gather}
C^{\rm c,g}_{\vb{k}} = \pmqty{[C^{\rm c,g}_{\vb{k}}]^{1,1} & [C^{\rm c,g}_{\vb{k}}]^{1,2} \\[0.4em]
[C^{\rm c,g}_{\vb{k}}]^{2,1} & [C^{\rm c,g}_{\vb{k}}]^{2,2}}\,,
\\
[C^{\rm c}_{\vb{k}}]^{1,1} = \pmqty{u_{\bar{n} \vb{k}}^4+v_{\bar{n} \vb{k}}^4 & -2u_{\bar{n} \vb{k}}^2v_{\bar{n}\vb{k}}^2 \\[0.4em]
-2u_{\bar{n} \vb{k} }^2v_{\bar{n}\vb{k}}^2 & 
u_{\bar{n}\vb{k}}^4+v_{\bar{n}\vb{k}}^4
}\,,
\\
[C^{\rm c}_{\vb{k}}]^{2,2} = -[C^{\rm g}_{\vb{k}}]^{1,1} = 
2u_{\bar{n} \vb{k}}^2v_{\bar{n} \vb{k}}^2 \pmqty{1 & 1 \\ 1 & 1}\,,
\\
[C^{\rm g}_{\vb{k}}]^{2,2} = \pmqty{ 2u_{\bar{n} \vb{k}}^2v_{\bar{n} \vb{k}}^2 &  -(u_{\bar{n} \vb{k}}^4+v_{\bar{n} \vb{k}}^4) \\-(u_{\bar{n} \vb{k}}^4+v_{\bar{n} \vb{k}}^4) &  2u_{\bar{n} \vb{k}}^2v_{\bar{n} \vb{k}}^2 }\,,
\\
[C^{\rm c,g}_{\vb{k}}]^{i,j} = 
(u_{\bar{n} \vb{k}}^2-v_{\bar{n}\vb{k}}^2)u_{\bar{n}\vb{k}}v_{\bar{n}}\pmqty{1 & 1 \\ 1 & 1}\quad (i\neq j)\,.
\label{eq:C_matrix_last}
\end{gather}
For reason that will become clear in a moment, we distinguish two contributions to the matrix $A$, a conventional one proportional to $C^{\rm c}_{\vb{k}}$ and a geometric one proportional to $C_{\vb{k}}^{\rm g}$. The same separation  applies to the vector $\vb{v}_l$~\eqref{eq:vect_v_def}
\begin{gather}
\vb{v}_{l\alpha} = \vb{v}_{l\alpha}^{\rm c} + \vb{v}_{l\alpha}^{\rm g}\,,\\
\label{eq:vc_vg_def}
\vb{v}_{l\alpha}^{\rm c} = (\hat{J}_l,\hat{N}_\alpha)\pmqty{1 \\ -1 \\ 0 \\ 0}\,,\hspace{0.5cm}
\vb{v}_{l\alpha}^{\rm g} = (\hat{J}_l,\hat{D}_\alpha)\pmqty{0 \\ 0 \\ 1 \\ -1}\,,
\\
\label{eq:JN_final_main}
\begin{split}
	&(\hat{J}_l, \hat{N}_{\alpha\uparrow}) = -(\hat{J}_l, \hat{N}_{\alpha\downarrow}) =\sum_{\vb{k}}\ev{P(\vb{k})}{\alpha}\frac{\Delta^2}{2E_{\bar{n}\vb{k}}^2}\\
	& \hspace{1cm}
	\times\qty[\frac{\tanh(\frac{\beta E_{\bar{n}\vb{k}}}{2})}{E_{\bar{n}\vb{k}}}
	-\frac{\beta}{2\cosh[2](\frac{\beta E_{\bar{n}\vb{k}}}{2})}]\partial_l \varepsilon_{\bar{n}\vb{k}}\,, 
\end{split}
\\
\label{eq:JD_final_main}
\begin{split}
	&\qty\big(\hat{J}_l,\hat{D}_\alpha) = -\qty\big(\hat{J}_l,\hat{D}_\alpha^\dg) \\ &\hspace{0.1cm}=\sum_{\vb{k}}\frac{\Delta}{2E_{\bar{n}\vb{k}}}\tanh(\frac{\beta E_{\bar{n}\vb{k}}}{2})\ev{\qty[P(\vb{k}),\partial_l P(\vb{k})]}{\alpha}\,.
\end{split} 
\end{gather}
These results are obtained from~\eqref{eq:JN_corr_M}-\eqref{eq:JD_corr_M},~\eqref{eq:JN_final} and~\eqref{eq:JD_final} in Appendix~\ref{app:corr_computation}. It is clear from~\eqref{eq:JN_final_main} that $\vb{v}_{l\alpha}^{\rm c}$ is a conventional term since it depends only on the group velocity $\partial_l \varepsilon_{\bar{n}\vb{k}}$. In fact, expressing $\vb{v}_{l\alpha}^{\rm c}$ in terms of the group velocity alone is a nontrivial result that involves a fair amount of algebra, see~\eqref{eq:M+_ii_neq_comp}. On the other hand, one can see from~\eqref{eq:JD_final_main} that $\vb{v}_{l\alpha}^{\rm g}$ is purely geometrical, since it depends only on the derivatives of the Bloch functions, or equivalently, only on the derivatives of the flat band projector $\partial_l P(\vb{k})$.

The key observation that allows to considerably simplify the calculation is to notice that the vectors $\vb{v}_{l\alpha}^{\rm c,g}$  are eigenvectors of the matrices $C_{\vb{k}}^{\rm (c,g)}$ and $B_\alpha$~\eqref{eq:B_alpha}
\begin{gather}
	C^{\rm c}_{\vb{k}} \vb{v}_{l\alpha}^{\rm c} = \vb{v}_{l\alpha}^{\rm c}\,,\qquad C^{\rm c}_{\vb{k}} \vb{v}_{l\alpha}^{\rm g} = 0\,,\\
	C^{\rm g}_{\vb{k}} \vb{v}_{l\alpha}^{\rm c} = 0\,,\qquad C^{\rm g}_{\vb{k}} \vb{v}_{l\alpha}^{\rm g} = \vb{v}_{l\alpha}^{\rm g}\,,\\
	B_\alpha \vb{v}_{l\alpha}^{\rm c,g} = \frac{U_\alpha}{N_{\rm c}} \vb{v}_{l\alpha}^{\rm c,g}\,.
\end{gather}
As a consequence, we have also
\begin{gather}
A_{\alpha,\alpha'} B_{\alpha'} \vb{v}_{l\alpha'}^{\rm (c,g)} = \lambda^{\rm (c,g)}_{\alpha,\alpha'} U_{\alpha'} \vb{v}_{l\alpha'}^{\rm (c,g)}\,,\\
\label{eq:lambda_c_alphaalpha1}
\qq*{with}\lambda^{\rm c}_{\alpha,\alpha'} = -\frac{1}{N_{\rm c}}\sum_{\vb{k}}  \frac{\beta\abs{\mel{\alpha}{P(\vb{k})}{\alpha'}}^2}{4\cosh^2(\frac{\beta E_{\bar{n}\vb{k}}}{2})}\,,\\
\label{eq:lambda_g_alphaalpha1}
\lambda^{\rm g}_{\alpha,\alpha'} = \frac{1}{N_{\rm c}}\sum_{\vb{k}}  \frac{\abs{\mel{\alpha}{P(\vb{k})}{\alpha'}}^2}{2E_{\bar{n}\vb{k}}}\tanh(\frac{\beta E_{\bar{n}\vb{k}}}{2})\,.
\end{gather}
It is then clear that GRPA correction to the superfluid weight can be written as the sum of a conventional and geometric components since the vectors introduced in~\eqref{eq:vc_vg_def} are orthogonal $\vb{v}_{l\alpha}^{\rm c}\cdot \vb{v}_{m\beta}^{\rm g} = 0$.
The conventional component takes the form
\begin{equation}
\label{eq:GRPA_Ds_conv}
\begin{split}
[D^{(1)}_{\rm s,c}]_{l,m} &= \frac{1}{\mathcal{A}}(\vb{v}^{\rm c}_l)^T \qty(B +  B\frac{1}{A^{-1}-B}B )\vb{v}^{\rm c}_m \\
&=2\mathcal{A}_c \vb{c}^T_l\qty(U+U\frac{1}{(\lambda^{\rm c})^{-1}-U}U)\vb{c}_m\,.
\end{split}
\end{equation}
Here, $\mathcal{A}_{\rm c} = \mathcal{A}/N_{\rm c}$  denotes the unit cell area, $U = \mathrm{diag}(U_1,U_2,\dots,U_{\alpha=N_{\rm orb}})$ is a diagonal matrix with the coupling constants $U_\alpha$ on the main diagonal, $\lambda^{\rm c}$ is a matrix with components given by~\eqref{eq:lambda_c_alphaalpha1} and $\vb{c}_l = (c_{l,1},c_{l,2},\dots,c_{l,\alpha=N_{\rm orb}})^T$ are vectors whose components are given by  $c_{l\alpha} = \mathcal{A}^{-1}(\hat{J}_l, \hat{N}_{\alpha\uparrow})$. Recall that, in the thermodynamic limit, the summation over wave vectors in the expression for $(\hat{J}_l, \hat{N}_{\alpha\uparrow})$ in~\eqref{eq:JN_final_main} is replaced by the Brillouin zone integral $\sum_{\vb{k}} \to \frac{\mathcal{A}}{(2\pi)^2}\int \dd[2]\vb{k}$.

At zero temperature $\lambda^{\rm c} = 0$ and only the first term in~\eqref{eq:GRPA_Ds_conv} survives. Since the components of the vector $\vb{c}_l$ are purely real and $U_\alpha >0$, the matrix $M$ with components $[M]_{l,m} = \vb{c}_l^TU\vb{c}_m$ is positive semidefinite, which means that the conventional part of the GRPA correction leads to an enhancement of the superfluid weight compared to its mean-field value at zero temperature. 

It can be also shown that the second term in~\eqref{eq:GRPA_Ds_conv} is always negative semidefinite, which is consistent with the discussion below~\eqref{eq:Omega_mf_dAldAm} (see also Ref.~\cite{Huhtinen2022}).  To prove this, one notes that the matrix $Q(\vb{k})$ with components $[Q(\vb{k})]_{\alpha,\alpha'} = \abs{\mel{\alpha}{P(\vb{k})}{\alpha'}}^2$ is positive semidefinite since
\begin{equation}
	\label{eq:Q_positivity}
	\vb{c}^\dagger Q\vb{c} = \Tr[c^\dagger PcP] = \Tr[(Pc P)^\dagger(PcP)] \geq0\,,
\end{equation}
with $c = \mathrm{diag}(c_1,c_2,\dots,c_{N_{\rm orb}})$ the diagonal matrix obtained from the components of the generic complex vector $\vb{c} = (c_1,c_2,\dots,c_{N_{\rm orb}})^T$. In fact, this is just a special case of the Schur product theorem according to which the Hadamard product (entry-wise product) $C = A \circ B$ of two positive (semi)-definite matrices $A$ and $B$ is positive (semi)-definite~\cite{Horn2012}. In our specific case, we have $Q(\vb{k}) = P(\vb{k}) \circ P^*(\vb{k})$ with both $P(\vb{k})$ and $P^*(\vb{k})$ positive semidefinite. Thus $\lambda^c$ is negative semidefinite and as a consequence also $((\lambda^{\rm c})^{-1}-U)^{-1}$. This concludes the proof.
By the same token,
\begin{equation}
	\label{eq:lambda_U_identity}
	\frac{1}{U^{-1}-\lambda^{\rm c}} = U+U\frac{1}{(\lambda^{\rm c})^{-1}-U}U
\end{equation}
is a positive semidefinite matrix, showing that the conventional GRPA correction $D_{\rm s,c}^{(1)}$ always leads to an increase of the mean-field superfluid weight even at finite temperature. It will become clear in the following that the conventional GRPA correction $D_{\rm s,c}^{(1)}$ is also geometry independent.

Using~\eqref{eq:lambda_U_identity} with $\lambda^{\rm g}$ in place of $\lambda^{\rm c}$, we can write the geometric component of the GRPA correction to the superfluid weight in the following form
\begin{equation}
\label{eq:GRPA_Ds_geom}
\begin{split}
	[D^{(1)}_{\rm s,g}]_{l,m}
	&=2\mathcal{A}_c \vb{d}^T_l\frac{1}{U^{-1}-\lambda^{\rm g}}\vb{d}_m\,.
\end{split}
\end{equation}
Similarly to~\eqref{eq:GRPA_Ds_conv}, the matrix $\lambda^{\rm g}$ has matrix elements $\lambda^{\rm g}_{\alpha,\alpha'}$ given by~\eqref{eq:lambda_g_alphaalpha1} and $\vb{d}_l = (d_{l,1},d_{l,2},\dots,d_{l,\alpha=N_{\rm orb}})^T$ is a vector with components $d_{l\alpha} = \mathcal{A}^{-1}\qty\big(\hat{J}_l,\hat{D}_\alpha)$, whose expression in terms of the periodic Bloch functions and their derivatives is given in~\eqref{eq:JD_final_main}. Note that the components of the $\vb{d}_{l}$ vector just introduced are purely imaginary since $\qty\big(\hat{J}_l,\hat{D}_\alpha) = -\qty\big(\hat{J}_l,\hat{D}_\alpha^\dg) = -\qty\big(\hat{J}_l,\hat{D}_\alpha)^*$, see~\eqref{eq:corr_func_prop_1}-\eqref{eq:corr_func_prop_2}.
Therefore, if it is shown that the matrix $U^{-1} - \lambda^{\rm g}$ is positive definite and so is its inverse, then $D^{(1)}_{\rm s,g}$ is negative semidefinite, meaning that superfluid weight is decreased by the geometric GRPA correction compared to the mean-field result $D^{(0)}_{\rm s}$. This is in contrast to the conventional GRPA correction $D^{(1)}_{\rm s,c}$ discussed previously. 
A subtle point here is that $U^{-1}-\lambda^{\rm g}$ is  not invertible, as shown below. This is not a problem since the formula in~\eqref{eq:GRPA_Ds_geom} makes sense and gives the correct result if one denotes by $1/(U^{-1}-\lambda^{\rm g})$  the Moore-Penrose inverse (pseudoinverse).

To prove that $U^{-1} - \lambda^{\rm g}$ is positive semidefinite, we need the self-consistency equation of mean-field theory for the parameters $\Delta_\alpha$. By assuming the uniform pairing condition~\eqref{eq:uniform_pairing} and taking the isolated band limit, one obtains from~\eqref{eq:self_const_exp_Delta} 
\begin{equation}
	\frac{1}{U_\alpha} = \frac{1}{N_{\rm c}} \sum_{\vb{k}}\frac{\tanh\qty(\frac{\beta E_{\bar{n}\vb{k}}}{2})}{2E_{\bar{n}\vb{k}}}\ev{P(\vb{k})}{\alpha}\,.
\end{equation}
Therefore, we can write
\begin{gather}
U^{-1} - \lambda^{\rm g} = 	\frac{1}{N_{\rm c}} \sum_{\vb{k}}\frac{\tanh\qty(\frac{\beta E_{\bar{n}\vb{k}}}{2})}{2E_{\bar{n}\vb{k}}}R(\vb{k})\,, \\
\label{eq:R_matrix_def}
\qq*{with} R(\vb{k}) = \mathrm{diag}\,P(\vb{k}) -P(\vb{k}) \circ P^*(\vb{k})\,.
\end{gather}
The matrix $R(\vb{k})$ is defined in terms of the Hadamard product $Q(\vb{k}) = P(\vb{k}) \circ P^*(\vb{k})$ introduced in~\eqref{eq:Q_positivity}, while $\mathrm{diag}\,P(\vb{k})$ is the diagonal part of $P(\vb{k})$, that is the matrix obtained by setting to zero all of the matrix elements away from the main diagonal. Since the linear combination of positive semidefinite matrices with positive coefficients is again positive semidefinite, we just need to prove that $R(\vb{k})$ is positive semidefinite. This is done as follows
\begin{equation}
\label{eq:R_positivity}
\begin{split}
2\vb{c}^\dg R \vb{c} &= 2\Tr[c^\dg Pc] - 2\Tr[Pc^\dg P c]\\
&= \Tr\qty\big[[P,c]^\dg[P,c]] \geq 0\,,
\end{split}
\end{equation}
where the vector $\vb{c}$ and the diagonal matrix $c$ are as in~\eqref{eq:Q_positivity}. This concludes the proof that $D_{\rm s,g}^{(1)}$ is negative semidefinite. By taking $c$ as the identity matrix, one obtains that the right-hand side of~\eqref{eq:R_positivity} is zero, showing that $R(\vb{k})$ and $U^{-1} -\lambda^{\rm g}$ are never invertible. This can be traced back to the fact that the mean-field free energy does not change if all the variational parameters $\Delta_\alpha$ are multiplied by the same constant phase $e^{i\phi}$. The invariance under global phase rotations of the order parameter is a general property of superconducting systems, which is a consequence of gauge invariance. 
 
The expressions~\eqref{eq:GRPA_Ds_conv} and~\eqref{eq:GRPA_Ds_geom} for the conventional and geometric components of the GRPA correction are the main results of this section, together with the statements regarding their positive or negative semidefiniteness, respectively $D_{\rm s,c}^{(1)} \geq 0$ and $D_{\rm s,g}^{(1)} \leq 0$.
The formulas~\eqref{eq:GRPA_Ds_conv} and~\eqref{eq:GRPA_Ds_geom} allow to compute the full GRPA correction $D^{(1)}_{\rm s}$ of generic lattice models in the isolated flat band limit in terms of the energy dispersion $\varepsilon_{\bar{n}\vb{k}}$ and the Bloch functions $\ket{g_{\bar{n}\vb{k}}}$ of the only partially filled band. An important difference between the conventional and the geometric components is that the latter is not geometry independent, meaning that it depends on the orbital position vectors $\vb{r}_{\vb{i}\alpha}$ that enter in the Fourier transforms of the field operators~\eqref{eq:Fourier_exp} and of the free Hamiltonian~\eqref{eq:H_free_Fourier}. In the next section, it is explained how one can take advantage of this fact and set the geometric GRPA correction to zero by a suitable choice of the orbital positions.


\section{Minimal quantum metric and natural orbital positions}
\label{sec:Natural_orbital}

In this section, we first assume that the partially filled band is not only isolated, but also flat in order to simplify the presentation. All of the results can be straightforwardly extended to the case of an isolated, but not necessarily flat, band as explained towards the end. In the isolated flat band limit, only the geometric contribution to the superfluid weight $D_{\rm s} = D_{\rm s,g}^{(0)} + D_{\rm s,g}^{(1)}$ survives, while the conventional one vanishes $D_{\rm s,c}^{(0)} = D_{\rm s,c}^{(1)}=0$. Since the energy dispersion of the band is just a constant $\varepsilon_{\bar{n}\vb{k}}  = \varepsilon_{\bar{n}}$, the superfluid weight depends only on an invariant built out of the flat band Bloch functions.  
The quantum metric~\eqref{eq:quantum_metric}, which appears in the expression~\eqref{eq:D0_sg} for the geometric contribution to the MF superfluid weight, is invariant under multiplication of the Bloch functions by an arbitrary $\vb{k}$-dependent phase factor $\ket{g_{\bar{n}\vb{k}}}\to e^{i\phi(\vb{k})}\ket{g_{\bar{n}\vb{k}}}$, because the projector $P(\vb{k}) = \dyad{g_{n\vb{k}}}$ is unaffected by this transformation. In this sense the quantum metric is a  band structure invariant. However, if we perform a shift of the orbital positions as in~\eqref{eq:pos_displ_shift}, the flat band projector transforms as (see~\eqref{eq:H_free_orbital_transf})
\begin{gather}
\label{eq:P_transformation}
P(\vb{k}) \to P'(\vb{k}) =  e^{-i\vb{k}\cdot\vb{b}}P(\vb{k})e^{i\vb{k}\cdot\vb{b}}\,,\\
\label{eq:d_P_transformation}
\begin{split}
&\partial_l P(\vb{k}) \to \partial_l P'(\vb{k})
\\
& =  e^{-i\vb{k}\cdot\vb{b}}\qty\big(\partial_l P(\vb{k})-i[b_{l},P(\vb{k})])e^{i\vb{k}\cdot\vb{b}}\,. 
\end{split}	
\end{gather}
The components $b_l$ of the vector $\vb{b}$ are operators acting in orbital space and encode the position shifts $\vb{b}_\alpha$ for each orbital, that is $b_l\ket{\alpha} = [\vb{b}_\alpha]_l\ket{\alpha}$. The presence of the commutator term in~\eqref{eq:d_P_transformation} implies that the quantum metric is not geometry independent. It follows that it is essential to include the GRPA correction in order to restore the geometry independence of the total superfluid weight, which has been established in Sec.~\ref{sec:gauge} using the principle of gauge invariance. On the other hand, it is clear that the quantities $\ev{P(\vb{k})}{\alpha}$ and~$\abs{\mel{\alpha}{P(\vb{k})}{\alpha'}}^2$, appearing in~\eqref{eq:JN_final_main} and~\eqref{eq:lambda_c_alphaalpha1}, respectively, are invariant under orbital position shifts~\eqref{eq:P_transformation}, therefore the conventional component of the GRPA correction~\eqref{eq:GRPA_Ds_conv} is geometry independent. 

The purpose of this section is to show that  it is possible to set to zero the geometric component of the GRPA correction, that is $D_{\rm s,g}^{(1)} = 0$, by a suitable choice of the orbital positions. It is also shown that these orbital positions, called in the following the natural orbital positions, are the ones that minimize the trace of the  quantum metric integrated over the Brillouin zone, in agreement with the results of Ref.~\cite{Huhtinen2022}. Using the nomenclature of this last reference, the integrated quantum metric computed using the natural orbital positions is called the minimal quantum metric.

Using~\eqref{eq:D0_sg} and~\eqref{eq:GRPA_Ds_geom}, we can write the superfluid weight in the isolated flat band limit in the following way
\begin{equation}
\label{eq:Ds_flat_band}
\begin{split}
[D_{\rm s}]_{l,m} &= [D_{\rm s,g}^{0}]_{l,m}  + [D_{\rm s,g}^{(1)}]_{l,m} \\
&=
\frac{2\Delta^2}{E_{\bar{n}}}\tanh(\frac{\beta E_{\bar{n}}}{2})
\widetilde{\mathcal{M}}_{l,m}\,,
\end{split}
\end{equation}
where $E_{\bar{n}} = \sqrt{(\varepsilon_{\bar{n}}-\mu)^2+\Delta^2} = E_{\bar{n}\vb{k}}$ is the quasiparticle dispersion, which is also flat, and the minimal quantum metric $\widetilde{\mathcal{M}}_{l,m}$  is defined as
\begin{gather}
\label{eq:minimal_qm}
\widetilde{\mathcal{M}}_{l,m} = \mathcal{M}_{l,m} -\frac{1}{2}\vb{s}^T_lR^{-1}\vb{s}_m\,,
\\
\label{eq:integrated_qm}
\mathcal{M}_{l,m} = \int\frac{\dd[2]{\vb{k}}}{(2\pi)^2} \mathcal{G}_{l,m}(\vb{k})\,,
\\
\label{eq:s_vector_def_1}
\vb{s}_l = \pmqty{s_{l,1}, s_{l,2}, \dots, s_{l,N_{\rm orb}}}^T\,,
\\
\label{eq:s_vector_def_2}
s_{l,\alpha} = i\int\frac{\dd[2]{\vb{k}}}{(2\pi)^2}\ev{\qty[P(\vb{k}),\partial_l P(\vb{k})]}{\alpha}\,,\\
\label{eq:R_integrated_def}
R = \int\frac{\dd[2]{\vb{k}}}{(2\pi)^2}\qty[\mathrm{diag}\,P(\vb{k}) -P(\vb{k}) \circ P^*(\vb{k})] \,.
\end{gather}
We call $\widetilde{\mathcal{M}}_{l,m}$ the minimal quantum metric as in Ref.~\cite{Huhtinen2022}, but a more proper name would be minimal \textit{integrated} quantum metric, since $\mathcal{M}_{l,m}$ is the quantum metric~\eqref{eq:quantum_metric} integrated over the whole Brillouin zone. The vector $\vb{s}_l$ is related to the vector $\vb{d}_l$ introduced in~\eqref{eq:GRPA_Ds_geom} and has purely real components. 
As before, in~\eqref{eq:R_integrated_def} we have used the Hadamard product $Q(\vb{k}) = P(\vb{k}) \circ P^*(\vb{k})$ whose matrix elements are $[Q(\vb{k})]_{\alpha,\alpha'} = \abs{\mel{\alpha}{P(\vb{k})}{\alpha'}}^2$.
As in~\eqref{eq:GRPA_Ds_geom}, $R^{-1}$ denotes the pseudoinverse since $R$ is not an invertible matrix. 

It is possible to show that the result in~\eqref{eq:Ds_flat_band}-\eqref{eq:R_integrated_def} for the superfluid weight in the isolated flat band limit is applicable also in the case of several degenerate flat bands. The only modification is that the projector reads $P(\vb{k}) = \sum_{n\in  \mathcal{F}}\dyad{g_{n\vb{k}}}$ in this case, where the sum runs over the set $\mathcal{F}$ of degenerate partially filled flat bands. An example of a lattice with degenerate flat bands is the dice lattice presented in Sec.~\ref{sec:dice_lattice}.

We now show that, if the orbital positions are chosen so as to minimize the trace of the integrated quantum metric $\Tr \mathcal{M} = \sum_l\mathcal{M}_{l,l}$, then the vectors $\vb{s}_l$ vanish and so does the geometric GRPA correction $D_{\rm s,g}^{(1)}$ proportional to the quantity $\vb{s}_l^{T}R^{-1}\vb{s}_m/2$ in~\eqref{eq:minimal_qm}. 
To this end, we need to compute how the integrated quantum metric changes under a shift of the orbital positions.
From~\eqref{eq:P_transformation}, we obtain
\begin{equation}
	\label{eq:S_transf}
	\begin{split}
		&\mathcal{M}_{l,m} \to \mathcal{M}'_{l,m}= \mathcal{M}_{l,m} \\
		&\hspace{3mm}-i\int\frac{\dd[2]{\vb{k}}}{(2\pi)^2}\qty(\Tr\qty\big[[P(\vb{k}),\partial_lP(\vb{k})]b_m]+ l\leftrightarrow m)\\
		&\hspace{3mm}+ 2\int\frac{\dd[2]{\vb{k}}}{(2\pi)^2}\Tr[P(\vb{k})b_l(1-P(\vb{k}))b_m]\\
		&=  \mathcal{M}_{l,m}-\vb{s}_l^T\widetilde{\vb{b}}_m -\vb{s}_m^T\widetilde{\vb{b}}_l  + 2\widetilde{\vb{b}}_l^TR\widetilde{\vb{b}}_m\,,\\
		&\qq*{with} \widetilde{\vb{b}}_l = \qty\big([\vb{b}_{\alpha = 1}]_l, [\vb{b}_{\alpha = 2}]_l,\dots, [\vb{b}_{\alpha = N_{\rm orb}}]_l)^T\,.
	\end{split}
\end{equation}
By setting to zero the derivatives of the trace $\Tr \mathcal{M}' = \sum_l \mathcal{M}'_{l,l}$ with respect to the shift components $[\vb{b}_\alpha]_l$, it is found that the integrated quantum metric is minimized when the following condition is satisfied
 \begin{equation}
 	\label{eq:linear_sys_R}
 	R\widetilde{\vb{b}}_l = \frac{\vb{s}_l}{2}\,.
 \end{equation}
 The linear system~\eqref{eq:linear_sys_R} necessarily has a solution because $\Tr \mathcal{M}'$ is a positive quantity and has a minimum as a function of the shifts $[\vb{b}_\alpha]_l$. However, the solution is not unique since $\Tr \mathcal{M}'$ is invariant if all orbitals are shifted by the same amount ($[\vb{b}_\alpha]_l = [\vb{b}_\beta]_l$ for all $\alpha,\,\beta$) as one expect from translational invariance. Indeed, if $\vb{e} = (1,1,...,1)^T$ is the vector with all the components equal to one, it is easy to verify that $\vb{s}\cdot\vb{e} = 0$ and $R\vb{e} = 0$. Thus, a solution of ~\eqref{eq:linear_sys_R}  can be expressed in terms of the pseudoinverse $R^{-1}$ of the matrix $R$, namely $\widetilde{\vb{b}}_l = R^{-1}\vb{s}_l/2$. According to the properties of the pseudoinverse, this solution has minimum norm $\norm*{\widetilde{\vb{b}}_l} = \sqrt{\sum_\alpha[\vb{b}_\alpha]^2_l}$, therefore it satisfies the condition $\vb{e}\cdot\widetilde{\vb{b}}_l=\sum_\alpha[\vb{b}_\alpha]_l = 0$. 
 
 On the other hand, we have from~\eqref{eq:d_P_transformation} and~\eqref{eq:s_vector_def_2}
 \begin{equation}
 	\label{eq:s_vector_transf}
 	\begin{split}
 	&s_{l,\alpha} \to s_{l,\alpha}' = i\int\frac{\dd[2]{\vb{k}}}{(2\pi)^2}\ev{\qty[P'(\vb{k}),\partial_l P'(\vb{k})]}{\alpha}\\
 	&= s_{l,\alpha} + \int\frac{\dd[2]{\vb{k}}}{(2\pi)^2} \ev{\qty[P(\vb{k}),[b_l, P(\vb{k})]]}{\alpha} 
 	\\
 	&= s_{l,\alpha} -2 \sum_\beta[R]_{\alpha,\beta}[\vb{b}_\beta]_l\,, \qq{or} \vb{s}_l' = \vb{s}_l -2 R\widetilde{\vb{b}}_l\,. 
 	\end{split}
 \end{equation} 	
 This result together with~\eqref{eq:S_transf} implies that the minimal quantum metric $\widetilde{\mathcal{M}}$~\eqref{eq:minimal_qm} is geometry independent, as expected from the discussion in Sec.~\ref{sec:gauge}. 
 Another consequence is that, if the orbital shifts are chosen so as to satisfy~\eqref{eq:linear_sys_R}, then $\vb{s}'_l = 0$ and the minimal quantum metric $\widetilde{\mathcal{M}}$~\eqref{eq:minimal_qm} coincides with the integrated quantum metric $\mathcal{M}$, thus it is a positive semidefinite matrix. This also means that the correlation functions $(J_l,D_\alpha)$~\eqref{eq:JD_final_main} and the GRPA correction in~\eqref{eq:Ds_flat_band}-\eqref{eq:minimal_qm} vanish if calculated with the orbitals positions defined by~\eqref{eq:linear_sys_R}.
Importantly, it is always possible to find such a preferred set of orbital positions, which we call natural orbital positions. In general, we have observed that the natural orbital positions are always unique up to arbitrary translations. This is indeed the case of the examples presented in Sec.~\ref{sec:Natural_orbital}, however a general proof of this fact is lacking at present.
 
Even though only the flat band case has been considered so far in this section, it is easy to show that natural orbital positions, for which $D_{\rm s,g}^{(1)}=0$, exist also in the case of an isolated band that is not necessarily flat. In order to do this, one simply needs to repeat the arguments presented in this section with the inclusion of the weight factor $E_{\bar{n}\vb{k}}^{-1}\tanh(\beta E_{\bar{n}\vb{k}}/2)$ under the integral sign $\int\dd[2]{\vb{k}}$ in~\eqref{eq:integrated_qm},~\eqref{eq:s_vector_def_2} and~\eqref{eq:R_integrated_def}. Indeed, it is immediate to see that the derivations of the transformation properties in~\eqref{eq:S_transf} and~\eqref{eq:s_vector_transf} are unaffected by the weight factor. 
 
It is a remarkable fact that the expression~\eqref{eq:minimal_qm}-\eqref{eq:R_integrated_def} for the minimal quantum metric and the linear system~\eqref{eq:linear_sys_R} defining the natural orbital positions are obtained also from the analysis of the two-body excitation spectrum. Here, we quickly recall how this has been done in Ref.~\cite{Huhtinen2022}. The dispersion of propagating two-body bound states is given by the effective Hamiltonian
\begin{equation}
\label{eq:h_effective}
\mel{\alpha}{h(\vb{q})}{\beta} = -  \int\frac{\dd[2]{\vb{k}}}{(2\pi)^2} \mel{\alpha}{P(\vb{k}+\vb{q})}{\beta}\hspace{-1mm} \mel{\beta}{P(\vb{k})}{\alpha}\,,
\end{equation}
first introduced in Ref.~\cite{Herzog-Arbeitman2022}, under the assumption that
\begin{equation}
	\label{eq:uniform_pairing_3}
	\int\frac{\dd[2]{\vb{k}}}{(2\pi)^2} \mel{\alpha}{P(\vb{k})}{\alpha} = \int\frac{\dd[2]{\vb{k}}}{(2\pi)^2} \mel{\beta}{P(\vb{k})}{\beta} 
\end{equation}
for all $\alpha$ and $\beta$. This condition guarantees that the uniform pairing condition is satisfied for $U_\alpha = U_\beta = U$ (see~\eqref{eq:uniform_pairing_2}) and implies that the two-body bound state with lowest energy for $\vb{q}=0$ is represented by the effective state vector $\ket{\Psi_0}$ with $\braket{\alpha}{\Psi_0} = \braket{\beta}{\Psi_0} = N_{\rm orb}^{-1/2}$ for all $\alpha,\beta$. The effective Hamiltonian transforms as $h(\vb{q}) \to h'(\vb{q}) =  e^{-i\vb{q}\cdot\vb{b}}h(\vb{q})e^{i\vb{q}\cdot\vb{b}}$ under an orbital position shift, therefore its eigenvalues are geometry independent quantities. Using second order perturbation theory, one finds that the inverse effective mass of the bound state represented by $\ket{\Psi_0}$ is proportional to the minimal quantum metric~\eqref{eq:minimal_qm}, which coincides with the integrated quantum metric $\mathcal{M}_{l,m} = N_{\rm orb}\mel{\Psi_0}{\partial_l\partial_m h(\vb{q})}{\Psi_0}$, if the natural orbital positions are used in~\eqref{eq:h_effective}. In particular, the linear system determining the natural orbital positions obtained in Ref.~\cite{Huhtinen2022} (see Eq.~(38) in this reference) is a special case of~\eqref{eq:linear_sys_R} when~\eqref{eq:uniform_pairing_3} holds. Indeed, the components of $\vb{s}_l$ can be written in the following equivalent form
\begin{equation}
\begin{split}
\frac{s_{l,\alpha}}{2} &= \frac{i}{2}\int\frac{\dd[2]{\vb{k}}}{(2\pi)^2}\ev{\qty[P(\vb{k}),\partial_l P(\vb{k})]}{\alpha} \\
&= -i\int\frac{\dd[2]{\vb{k}}}{(2\pi)^2} \ev{\partial_l P(\vb{k}) P(\vb{k})}{\alpha} \\
&= i\sum_\beta \mel{\alpha}{\partial_l h(0)}{\beta}\,,
\end{split}
\end{equation}
which can be identified with the right-hand side of Eq.~(38) in Ref.~\cite{Huhtinen2022}, while on the left-hand side it is also easy to identify the matrix $R$~\eqref{eq:R_integrated_def}. The second equality in the above equation is a consequence of the fact that the projector $P(\vb{k})$ is in general not periodic, but rather $\varrho$-equivariant~\cite{Marcelli2021}, namely $P(\vb{k}+\vb{g}_j) = e^{-i\vb{g}_j\cdot \vb{b}}P(\vb{k})e^{i\vb{g}_j\cdot \vb{b}}$ for some given shift operator $\vb{b}$ [see~\eqref{eq:H_free_orbital_transf}], while $\vb{g}_j$ are reciprocal lattice vectors defined by $\vb{a}_i\cdot \vb{g}_j = 2\pi \delta_{i,j}$, therefore  $\ev{P^2(\vb{k})}{\alpha} = \ev{P(\vb{k})}{\alpha}$ is periodic.

Our derivation of ~\eqref{eq:minimal_qm}-\eqref{eq:R_integrated_def} and~\eqref{eq:linear_sys_R} is equivalent to the one of Ref.~\cite{Huhtinen2022} but is based on the direct evaluation of the GRPA expression for the superfluid weight. It is rather instructive to see how the same results can be obtained with completely different methods. This provides strong evidence that the relation between minimal quantum metric and superfluid weight in a flat band is an accurate, or even exact, result.  Our approach has the advantage that it is straightforward to relax the flat band assumption, moreover, we have shown that the condition~\eqref{eq:uniform_pairing_3} is unnecessary. This means that the minimal quantum metric and the associated natural orbital positions can be defined for arbitrary bands or composite bands, therefore they are likely to find applications in other contexts. With this perspective in mind, we illustrate these band structure invariants by considering some representative examples in the following section.
  
 \section{Examples}
 \label{sec:examples}
 
 In this section, we illustrate the ideas introduced in the previous section by considering three examples: Su-Schrieffer-Heeger (SSH) model, Creutz ladder and dice lattice. In  all of them, the Brillouin zone integrals that enter in the definitions of the minimal quantum metric and natural orbital positions can be worked out analytically. This gives us the opportunity to understand these concepts in the simplest possible setting. See also Ref.~\cite{Huhtinen2022} for an analogous discussion of the Lieb lattice.

\subsection{Su-Schrieffer-Heeger model}
\label{sec:SSH}
 
 \begin{figure}
     \centering
     \includegraphics[scale=1.4]{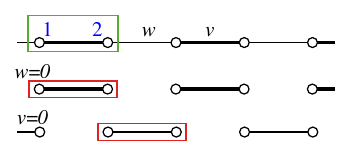}
     \caption{The SSH model is a simple linear chain with alternating hoppings. The unit cell, shown in the top row as a green rectangle, consists of two orbitals labeled by $\alpha=1,2$,  while $v$ and $w$ denote the hopping amplitudes within a unit cell (thick line) and between neighboring unit cells (thin line), respectively. By convention, the two orbitals in the unit cell have the same position, see~\eqref{eq:SSH_orb_pos}. For illustration purposes, they are displaced from each other in the figure.
     The two energy bands of the SSH model are both flat when either $w = 0$ (middle row) or $v=0$ (bottom row). In both cases, the lattice model reduces to a collection of disconnected dimers, therefore transport is not possible in any form. However, the quantum metric and thus the MF superfluid weight is zero in the first case ($w = 0$) and nonzero if $w\neq 0$. To resolve this inconsistency, it is necessary to take into account the geometric GRPA correction to the superfluid weight, which amounts to computing the quantum metric using the natural orbital positions. The red rectangles in the middle and bottom rows contain orbitals whose natural positions are identical. 
     }
     \label{fig:SSH_lattice}
 \end{figure}
 
    The SSH model is a one-dimensional lattice model widely used to illustrate the topological properties of the band structure and the concept of bulk-edge correspondence~\cite{Asboth2016}. As illustrated in Fig.~\ref{fig:SSH_lattice}, the unit cell of the SSH model contains two orbitals, labeled by $\alpha = 1,2$. The single-particle free Hamiltonian $H_{\rm free}$~\eqref{eq:Ham_free_A} for the SSH model is given by
    \begin{gather}
    	H^\sigma_{\rm free}(i-j) = 0\qq{for} |i-j| > 1\,,\\
    	H^\sigma_{\rm free}(0) = \pmqty{0 & v\\ v & 0}\,,\\
    	H^\sigma_{\rm free}(1) = [H_{\rm free}(-1)]^\dg =  \pmqty{0 & w\\ 0 & 0}\,.
    \end{gather} 
    The two real and positive parameters $v$ and $w$ denote the intra-cell and inter-cell hopping matrix elements, respectively. 
    The lattice constant is fixed to $a = 1$, thus the crystal momentum $k$ takes values in the range $[-\pi,\pi]$.
    The orbital positions are chosen to be the same for the two orbitals inside the same unit cell (see top row in Fig.~\ref{fig:SSH_lattice}), that is
    \begin{equation}
    	\label{eq:SSH_orb_pos}
    	r_{j,(\alpha = 1)} = r_{j,(\alpha = 2)} = j\,. 
    \end{equation}
    With this convention for the orbital positions, the Fourier transform of the free Hamiltonian~\eqref{eq:H_free_Fourier} reads
    \begin{equation}
    \begin{split}
    &H^\sigma_{\rm free}(k) = \sum_{j}H^\sigma_{\rm free}(j)e^{-ikj} \\ &\hspace{6mm}= \pmqty{0 & v+ we^{-ik} \\ v+ we^{ik} & 0}\,.
    \end{split}
    \end{equation} 
    It is straightforward to obtain the energy dispersions $\varepsilon_{n}(k)$ and projection operators $P_n(k)$ for the upper ($n = +$) and lower ($n = -$) band
    \begin{gather}
    \varepsilon_{\pm}(k) = \pm |v+we^{ik}| =  \pm \sqrt{v^2+w^2+2vw\cos k}\,,\\
    \label{eq:proj_SSH}
    \begin{split}
    &P_{\pm}(k) = P^\sigma_{\pm}(k)= \frac{1}{2}\qty(1+\frac{H_{\rm free}(k)}{\varepsilon_{\pm}(k)}) \\
    &\hspace{2mm}= \frac{1}{2}\pmqty{1 & \pm f^*(k) \\
    \pm f(k) & 1}\,,\quad f(k) = \frac{v+we^{ik}}{|v+we^{ik}|}\,.
	\end{split}	
    \end{gather} 
    The spin index $\sigma$ has been dropped since since the projector operator $P^\sigma_{\pm}(k)$ does not depend on the spin. 
	We need also the derivative of the projector given by
	\begin{equation}
	\label{eq:d_proj_SSH}
	\begin{split}
	&\partial_k P_{\pm}(k) = \frac{1}{2}\pmqty{0 & \pm[f'(k)]^*\\
	\pm f'(k) & 0}\,, \\
	&\qq*{with} f'(k) = \frac{iw(w+v\cos k)}{\abs{v+we^{ik}}(v+we^{-ik})}\,.
	\end{split} 
	\end{equation}
	Since the diagonal matrix elements  of the projector~\eqref{eq:proj_SSH} are constant and equal to $\ev{P_{\rm \pm}(k)}{\alpha}=1/2$, the SSH model satisfies the uniform pairing condition if $U_1 = U_2 = U$. Under this condition, the geometric contribution to the MF superfluid weight is given by the formula in~\eqref{eq:D0_sg} in the isolated band limit (adapted to one-dimension), which depends on the quantum metric~$\mathcal{G}(k)$. We obtain from~\eqref{eq:d_proj_SSH} that the quantum metric is
	\begin{equation}
		\begin{split}
		\mathcal{G}(k) &= \Tr[\partial_k P_{\pm}(k)\partial_kP_{\pm}(k)] = \frac{\abs{f'(k)}^2}{2} \\
		&=\frac{w^2(w+v\cos k)^2}{2(v^2+w^2+2vw\cos k)^2}\,.
		\end{split}
	\end{equation} 
	The quantum metric is the same for the two bands and diverges at $k = \pi$ when $|w-v|\to 0$ concomitantly with the closure of the energy gap between the two bands at the same $k$ point. The integral of the quantum metric over the Brillouin zone $\mathcal{M} = \int_{-\pi}^{\pi}\frac{\dd{k}}{2\pi}\mathcal{G}(k)$ can be performed by using the change of variable $e^{ik} \to z$ and applying the residue theorem to the resulting contour integral
	\begin{equation}
		\label{eq:M_SSH_contour}
		\begin{split}
		\mathcal{M} &=
		\frac{1}{2\pi i}\oint_{\abs{z}=1}\dd{z} \frac{w^2(2wz + vz^2 + v)^2}{8z(v+wz)^2(vz+w)^2} \\
		&=\begin{cases}
			\frac{w^2}{4(v^2-w^2)} & \qq*{for} v > w\,,\\
			\frac{2w^2-v^2}{4(w^2-v^2)} & \qq*{for} v < w\,.
		\end{cases}
		\end{split}
	\end{equation}
	We have computed the integrated quantum metric even away from the flat band limit, which in the SSH model is obtained when $v = 0$ or $w = 0$, since it can be of interest also for other applications besides the computation of the superfluid weight.
	In the case $w=0$, the unit cells are completely disconnected, as shown in Fig.~\ref{fig:SSH_lattice}, and one expects the absence of all forms of transport. In fact, the superfluid weight vanishes since $\mathcal{M} = 0$. On the other hand, if $v = 0$, then the quantum metric is nonzero ($\mathcal{M} = 1/2$) even though the lattice is again composed of disconnected two-orbital dimers  and one would again expect vanishing superfluid weight. As explained below, this apparent paradox is due to the fact that we have ignored the geometric GRPA correction $D_{\rm s,g}^{(1)}$ or, in other words, we have calculated the quantum metric using orbitals positions that are different from the natural ones. 
	
	To compute the natural orbital positions and the GRPA correction, we need the vector $\vb{s} = (s_{\alpha=1},s_{\alpha = 2})^T$~\eqref{eq:s_vector_def_1}-\eqref{eq:s_vector_def_2}, in which the spatial index $l$ has been dropped since we are dealing with a one dimensional model, and the pseudoinverse of the matrix $R$~\eqref{eq:R_integrated_def}.
	It is straightforward to obtain the latter from~\eqref{eq:proj_SSH}
	\begin{equation}
	\label{eq:R_SSH model}
	R^{-1} = \pmqty{1 & -1 \\ -1 & 1} = 4R\,.	
	\end{equation}
	Instead, the components of the vector $\vb{s}$ are calculated again using a contour integral as in~\eqref{eq:M_SSH_contour}
	\begin{equation}
	\label{eq:s_alpha_SSH_contour}
	\begin{split}
	s_\alpha &= 	i\int_{-\pi}^{\pi} \frac{\dd{k}}{2\pi} \ev{[P_{\pm}(k),\partial_k P_{\pm}(k)]}{\alpha} \\
	&= \frac{(-1)^\alpha}{2\pi i} \int_{\abs{z}^2 = 1} \dd{z} \frac{w(2w z + vz^2 + v)}{4z(v+wz)(vz+w)} \\
	&=\begin{cases}
		0 & \qq*{for} v > w \\
		\frac{(-1)^\alpha}{2} & \qq*{for} v < w \,.
	\end{cases}
	\end{split}
	\end{equation}
	The position shifts, given by $\widetilde{\vb{b}} = (b_{\alpha=1}, b_{\alpha=2})^T = R^{-1}\vb{s}/2$ vanish in the case $v > w$, which confirms that the choice in~\eqref{eq:SSH_orb_pos} corresponds to the natural orbital positions. On the other hand, for $v < w$ the natural orbital positions are given by
	\begin{equation}
	r'_{j,\alpha} = r_{j,\alpha} + b_\alpha	= j + \frac{(-1)^\alpha}{2}\,.
	\end{equation} 
	As a consequence, we have $r_{j,2} = r_{j+1,1}$. This means that in the SSH model the natural orbital positions are given by assigning the same positions to the orbitals connected by the hopping with largest magnitude, as illustrated in Fig.~\ref{fig:SSH_lattice}. We can also compute the minimal quantum metric for arbitrary value of the hopping amplitudes. From~\eqref{eq:R_SSH model} and~\eqref{eq:s_alpha_SSH_contour}, one obtains $\vb{s}^TR^{-1}\vb{s} = 1$ in the case $v < w$, therefore the minimal quantum metric is given by
	\begin{equation}
		\widetilde{\mathcal{M}} = \mathcal{M}-\frac{1}{2}\vb{s}^TR^{-1}\vb{s} =
		\begin{cases}
			\frac{w^2}{4(v^2-w^2)} & \qq*{for} v > w\,,\\
			\frac{v^2}{4(w^2-v^2)} & \qq*{for} v < w\,.
		\end{cases}
	\end{equation}
	In particular $\widetilde{\mathcal{M}} = 0$ for $v = 0$ or $w=0$, thus the superfluid weight vanishes in the flat band limit in the case of the SSH model after taking into account the geometric GRPA correction, as expected. 
	Note that the expressions of the minimal quantum metric in the two cases are related by to each other through the interchange $v \leftrightarrow w$. 
	This is consistent with the fact that one can interchange the two hopping hopping amplitudes $v$ and $w$ by a simple redefinition of the unit cell, which amounts to a different specification of the orbitals positions. Thus, the minimal quantum metric captures an intrinsic property of the band wave functions, which is not affected by the unit cell choice or the specific assignment of the orbital positions. Note also that the natural orbital positions respect the symmetries of the lattice, more specifically the reflection symmetry with respect to the middle point of the line connecting two nearest-neighbor orbitals. We will see more examples of this general phenomenon in the following.
	
	Besides the superfluid weight, another potentially interesting application of the natural orbital positions is related to topological invariants.  
	The SSH model possess a topological invariant, the winding number $\mathcal{W}$~\cite{Bernevig2015,Chiu2016,Asboth2016}, which takes value $\mathcal{W} = 0$ for $v > w$ and $\abs{\mathcal{W}}=1$ for $v < w$ when computed using the projector operator given in~\eqref{eq:proj_SSH}. In fact, one can show that the quantum metric is bounded from below by the winding number~\cite{Tovmasyan2016}, namely $\mathcal{M} > \mathcal{W}^2/2$. However, it is well-known that the winding number depends on the choice of unit cell used to compute it, which amounts to a choice of the orbitals positions~\cite{Cayssol2021,Fuchs2021}. Interestingly, our results show that the winding number of the SSH model computed with the natural orbital positions is always zero. Our next example, the Creutz ladder, possesses bands with nonzero winding number, when computed using the natural orbital positions. This suggests that the natural orbital positions can be used to provide a more refined classifications of the topological properties of the band structure. 
	 
\subsection{Creutz ladder}

\begin{figure}
	\includegraphics[scale=1.4]{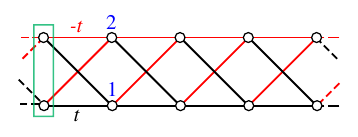}
	\caption{\label{fig:creutz_ladder} The Creutz ladder is a one-dimensional lattice model composed of two simple linear chains. The two chains are distinguished by the orbitals index $\alpha = 1,2$ and coupled by inter-chain hopping matrix elements, shown as the red and black diagonal lines in the figure. The horizontal and diagonal black lines correspond to the hopping amplitude $t>0$ in the free Hamiltonian~\eqref{eq:Creutz_Ham} and the red lines to $-t$. With this choice of the hopping matrix elements, the two bands of the Creutz ladder are perfectly flat and geometrically non-trivial since the minimal quantum metric is nonzero. As in the case of the SSH, orbitals in the same unit cell (green rectangle) have the same position. As shown in the main text, this choice of the orbital position is natural.
	The two chains are displaced from each other in the transverse direction only for illustration purposes.}
\end{figure}

The Creutz ladder shown in Fig.~\ref{fig:creutz_ladder} is a one-dimensional lattice model with two orbitals per unit cell introduced in Ref.~\cite{Creutz1999}. It has the peculiar property that both of its two bands are perfectly flat for a specific choice of the hopping matrix elements. This model has been studied extensively both in the bosonic and the fermionic case with the inclusion of different types of interaction terms~\cite{Tovmasyan2013,Takayoshi2013,Tovmasyan2016,Junemann2017,Mondaini2018,Tovmasyan2018,Kang2020,Kuno2020,Orito2021a,Chan2022a}. 

Adopting the convention for the phases of hopping matrix elements introduced in Ref.~\cite{Tovmasyan2018}, the free Hamiltonian of the Creutz ladder reads
\begin{gather}
\label{eq:Creutz_Ham}
H^\sigma_{\rm free}(1) = [H^\sigma_{\rm free}(-1)]^\dg = t\pmqty{1 & 1 \\
-1 & -1}\,,	\\
H^\sigma_{\rm free}(j) = 0 \qq{for} j \neq \pm 1\,.
\end{gather}
with $t >0$ the energy scale of the hopping amplitudes. This Hamiltonian is represented graphically in Fig.~\ref{fig:creutz_ladder}.
Choosing the orbital positions as for the SSH model~\eqref{eq:SSH_orb_pos}, i.e. same position for orbitals in the same unit cell, the Fourier transform of the free Hamiltonian of the Creutz ladder takes the form
\begin{equation}
H^\sigma_{\rm free}(k) = 2t\pmqty{\cos k & -i\sin k \\
i\sin k & -\cos k}\,.	
\end{equation} 
The dispersions of the two bands, labeled by $n = \pm$, and the associated projection operators are given by
\begin{gather}
\varepsilon_{\pm}(k)= \pm 2t\,,\\
\label{eq:comm_proj_creutz}
\begin{split}
&P_{\pm}(k) = \frac{1}{2}\pmqty{1\pm\cos k & \mp i\sin k \\ \pm i\sin k & 1\mp \cos k
}\,.
\end{split}
\end{gather}
As in the case of the SSH model, the spin index has been dropped since the projection operator is the same for spin up and spin down. Again, the uniform pairing conditions is satisfied for $U_1 = U_2 = U$ since
$\int_{-\pi}^\pi\frac{\dd{k}}{2\pi}\ev{P_{\pm}(k)}{\alpha} = 	\frac{1}{2}$.
From~\eqref{eq:comm_proj_creutz} it is immediate to obtain the result
\begin{equation}
	[P_{\pm}(k),\partial_k P_{\pm}(k)] = -\frac{1}{2}\pmqty{0 & i \\ i & 0}\,,  
\end{equation}
thus the components of the vector $\vb{s}$ in~\eqref{eq:s_vector_def_2} are all zero. It follows that the orbital positions given by~\eqref{eq:SSH_orb_pos} are in fact natural ones for the Creutz ladder and the superfluid weight is obtained simply from the integrated quantum metric  $\mathcal{M}$, while the GRPA correction vanishes. As noted in Ref.~\cite{Tovmasyan2016}, the quantum metric of the Creutz ladder is constant, independent of $k$,
\begin{equation}
\mathcal{G}(k)=	\Tr[\partial_k P(k)\partial_k P(k)] = \frac{1}{2}\,.
\end{equation} 
For completeness, we compute also the matrix $R$~\eqref{eq:R_integrated_def} and its pseudoinverse
\begin{equation}
	\label{eq:R_Creutz}
	R^{-1} = 2\pmqty{1 & -1 \\ -1 & 1} = 16R\,.	
\end{equation}%

Note that the natural orbital positions respect the symmetries of the Creutz ladder, in this specific case they are preserved under the interchange of the two orbitals inside the unit cell. The unitary operator $\mathcal{\hat{R}}$ implementing this transformation on the field operators is
\begin{equation}
\mathcal{\hat{R}}\hat{c}_{j,1} \mathcal{\hat{R}}^\dg = (-1)^j\hat{c}_{j,2}\,,\quad \mathcal{\hat{R}}	\hat{c}_{j,2} \mathcal{\hat{R}}^\dg = (-1)^j\hat{c}_{j,1}\,.
\end{equation}
The sign factors $(-1)^j$ in the definition of $\mathcal{\hat{R}}$ correspond a gauge transformation and are necessary in order to preserve the signs of the hopping matrix elements of the Creutz ladder Hamiltonian (see Fig.~\ref{fig:creutz_ladder}). 

As mentioned in the previous section, the winding number computed with the natural orbital positions is $\mathcal{W}=1$~\cite{Tovmasyan2016}, in contrast to the SSH model, for which it is always zero. It is an interesting open question for the future is to understand whether the different winding numbers of the two models manifest in some observable properties, for instance in the edge states that occur in topologically nontrivial lattice models due to the bulk-edge correspondence.

\subsection{Dice lattice}
\label{sec:dice_lattice}

\begin{figure}
	\includegraphics[scale=1.1]{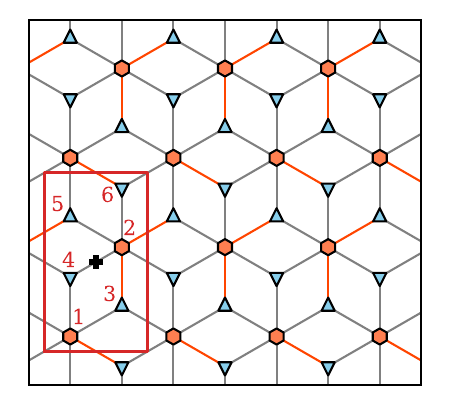}
	\caption{\label{fig:dice_lattice} Schematics of the dice lattice. The orange hexagons and the blue triangles denotes lattice sites. The hexagons are called hub sites and are six-fold coordinated while the triangles are called rim sites and are three-fold coordinated. The bonds between sites represent Hamiltonian matrix elements, which are all equal up to a sign encoded in the color of the bond ($+$ for black and $-$ for orange). With this choice of the hopping matrix elements, the Bravais lattice is rectangular with the fundamental vectors given in~\eqref{eq:fund_vec_rect}. The red rectangle denotes the associated unit cell, which contains six orbitals, labelled by the orbital index $\alpha=1,\dots,6$ as shown in the figure. The black cross denotes the baricenter of the orbitals within the chosen unit cell. The natural orbital positions are specified by the vectors $\widetilde{\vb{b}}_x$~\eqref{eq:b_tilda_dice_x} and $\widetilde{\vb{b}}_y$~\eqref{eq:b_tilda_dice_y} with the baricenter taken as the origin of the coordinate system. The positions of the lattice sites in the figure coincide with the natural orbital positions.}
\end{figure}

In this final example, we explicitly calculate the set of natural orbital positions and the minimal quantum metric for the two-dimensional dice lattice, also known as $\mathcal{T}_3$ lattice \cite{Vidal1998,Moller2012,Tovmasyan2018}. The graphical representation of the free Hamiltonian of the dice lattice is given in Fig.~\ref{fig:dice_lattice}, while its Fourier transform is provided in Ref.~\cite{Tovmasyan2018} and is not repeated here for brevity. The labeling of the orbitals shown in the figure is the same as in Ref.~\cite{Swaminathan2023} and different from Ref.~\cite{Tovmasyan2018}. Note that in Ref.~\cite{Tovmasyan2018} a more general model is considered, whereas here we specialize to the case in which all the hopping matrix elements are equal up to a sign. The hopping sign is denoted by the color of the bonds in Fig.~\ref{fig:dice_lattice}. As discussed in Ref.~\cite{Vidal1998}, with this specific choice of hopping signs, the band structure of the dice lattice is composed of six doubly degenerated flat bands. We focus on the lowest pair of degenerate flat bands $(n = 1,2)$ with energy 
\begin{equation}
	\label{eq:epsilon_minus}
	\bar{\varepsilon} = \varepsilon_{n=1,2} = \frac{1}{2}\qty(\varepsilon_{\rm h} - \sqrt{\varepsilon_{\rm h}^2+24})\,.
\end{equation}
The parameter $\varepsilon_{\rm h}$ is the on-site energy of the hub sites, while the rim sites have zero on-site energy  (see Fig.~\ref{fig:dice_lattice} for the definition of hub and rim sites). The unit of energy in the above expression is the absolute value of the hopping amplitude between any two sites.
The periodic Bloch functions for this pair of flat bands are
\begin{gather}
	\ket{g_{1,\vb{k}}} = c\qty\big(\bar{\varepsilon}, 0, 1+e^{ik_1},1,  e^{ik_2},e^{ik_2}(e^{ik_1}-1))^T\,,\label{eq:g_1_dice}\\
	\ket{g_{2,\vb{k}}} = c\qty\big(0,\bar{\varepsilon},-1,1+e^{-ik_1},1-e^{-ik_1},1)^T\,.\label{eq:g_2_dice}
\end{gather}
where the normalizing constant is given by $c = (\bar{\varepsilon}^2 + 6)^{-1/2}$ and we have introduced the quantities $k_i = \vb{k}\cdot \vb{a}_i$, with $\vb{a}_{i=1,2}$ the fundamental vectors of the Bravais lattice
\begin{equation}
	\label{eq:fund_vec_rect}
	\vb{a}_1 =  \pmqty{1\\0}\,,\qquad \vb{a}_2 = \pmqty{0 \\ \sqrt{3}}\,.
\end{equation} 
The periodic Bloch functions in~\eqref{eq:g_1_dice} and~\eqref{eq:g_2_dice} are obtained with the following choice of the orbital positions
\begin{equation}
\vb{r}_{\vb{i}\alpha} = i_1\vb{a}_1+i_2\vb{a}_2\,,\qquad \vb{i} = \pmqty{i_1\\i_2}\,,	
\end{equation}
i.e. the same position vector is assigned to all of the orbitals inside the rectangular unit cell shown in Fig.~\ref{fig:dice_lattice} since $\vb{r}_{\vb{i}\alpha}$ is independent of the orbital index $\alpha$. Note that these are different from the positions of the lattice sites actually used in Fig.~\ref{fig:dice_lattice}.

A convenient feature of the dice lattice is that the components of the periodic Bloch functions and therefore also of the projection operator $P(\vb{k}) = \dyad{g_{\vb{k},1}}{g_{\vb{k},1}}+\dyad{g_{\vb{k},2}}{g_{\vb{k},2}}$ are polynomials in $e^{ik_1}$ and $e^{ik_2}$, which allows to obtaining analytical results. Indeed, the calculation of the matrix $R$~\eqref{eq:R_integrated_def} and the vector $\vb{s}_l$~\eqref{eq:s_vector_def_1}-\eqref{eq:s_vector_def_2} is straightfoward but tedious and is best done with the help of a computer algebra system. The explicit expressions of these quantities are not particularly illuminating and are not provided here. We simply note that the following vectors
\begin{gather}
\label{eq:b_tilda_dice_x}
\widetilde{\vb{b}}_x = \frac{1}{4}\qty(-1,1,1,-1,-1,1)^T\,,\\
\label{eq:b_tilda_dice_y}
\widetilde{\vb{b}}_y = \frac{1}{4\sqrt{3}}\qty(-5,1,-3,-1,3,5)^T\,,
\end{gather}
give a solution of the linear system~\eqref{eq:linear_sys_R} for arbitrary values of the parameter $\varepsilon_{\rm h}$. In fact, they give the unique solution satisfying the constrain $\vb{e}\cdot\widetilde{\vb{b}}_l=0$, thus this is the solution obtained by using the pseudoinverse of the matrix $R$, as explained in Sec.~\ref{sec:Natural_orbital}. Due to this constrain, the vectors $\widetilde{\vb{b}}_l$ give the natural orbital positions for the dice lattice in a coordinate system in which the origin is the baricenter of the orbitals inside a unit cell. The positions of the lattice sites shown in Fig.~\ref{fig:dice_lattice} coincide with the natural orbital positions given by~\eqref{eq:b_tilda_dice_x}-\eqref{eq:b_tilda_dice_y}. One can see once again that the natural orbital positions provide a maximally symmetric arrangement of the lattice sites.    

It also straightforward to compute the minimal quantum metric~\eqref{eq:minimal_qm}, which is proportional to the identity and whose diagonal elements are ($l=x,y$)
\begin{equation}
	\widetilde{\mathcal{M}}_{l,l} = \frac{1}{18} \qty(5+\frac{12}{\varepsilon_{\rm h}^2+24}+\frac{5 \varepsilon_{\rm h}}{\sqrt{\varepsilon_{\rm h}^2+24}})\,.
\end{equation}
The superfluid weight is proportional to the minimal quantum metric according to~\eqref{eq:Ds_flat_band}, provided the uniform pairing condition~\eqref{eq:uniform_pairing}, or equivalently~\eqref{eq:uniform_pairing_2}, is satisfied. For a discussion of the uniform pairing condition  in the case of the dice lattice see Ref.~\cite{Swaminathan2023}.
Note that the integrated quantum metric $\mathcal{M}$~\eqref{eq:integrated_qm} of the dice lattice is in general not proportional to the identity matrix when computed using orbitals positions that are different from the natural ones. This means that using the integrated quantum metric alone in~\eqref{eq:Ds_flat_band} can lead to an unphysical anisotropy of the superfluid weight. Thus, it is important to include the geometric GRPA correction, which amounts to using the natural orbitals positions, in order to restore the proper symmetry of the superfluid weight tensor.  
 
\section{Discussion and conclusion}
\label{sec:conclusion}

In this work, we have provided the analytic expression for the superfluid weight within the GRPA in the isolated band limit. By analytic we mean that the evaluation of the superfluid weight is reduced to performing Brillouin zone integrals of certain combinations of the band dispersions $\varepsilon_{n\vb{k}}$ and the band wave functions $\ket{g_{n\vb{k}}}$ of a generic lattice model and their derivatives with respect to quasimomentum. This allows to relate the superfluid weight, which is an important observable for superfluid systems, to the properties of the band structure, in particular the effective mass and invariants such as the quantum metric~\eqref{eq:quantum_metric}. Our results hold under specific assumptions: the interaction is of the Hubbard form~\eqref{eq:Hubbard_int} with negative coupling constants $U_\alpha <0$ (attractive) that can depend on the orbital $\alpha$, and the free Hamiltonian~\eqref{eq:Ham_free_A} is invariant under spin rotations along a given axis and also time-reversal symmetric for $\vb{A} = 0$. Moreover, it is assumed that the uniform pairing condition is satisfied, namely that the pairing potential $\Delta_\alpha$ is independent of the orbital index $\alpha$~\eqref{eq:uniform_pairing}. Under the same assumptions, it was shown in Refs.~\cite{Peotta2015} and~\cite{Liang2017a} that the superfluid weight computed within the mean-field approximation (BCS theory), denoted here by $D_{\rm s}^{(0)}$, can be split into two contributions $D_{\rm s}^{(0)} = D_{\rm s,c}^{(0)} + D_{\rm s,g}^{(0)}$. The conventional contribution $D_{\rm s,c}^{(0)}$ can be written as in~\eqref{eq:D0_sc} or, equivalently,  as in~\eqref{eq:D0_Sc_alt} in terms of the effective mass, while the geometric contribution $D_{\rm s,g}^{(0)}$ is a weighted averaged of the quantum metric over the Brillouin zone~\eqref{eq:D0_sg}. The geometric contribution becomes important for bands with small bandwidth compared to the interaction coupling constants $U_\alpha$ since the conventional contribution vanishes in the flat band limit.

A major result of this work is that the same separation holds  also at the level of the GRPA. Indeed, we find that the correction term to the superfluid obtained from the GRPA $D_{\rm s}^{(1)} = D_{\rm s,c}^{(1)} + D_{\rm s,g}^{(1)}$ is also the sum of a conventional contribution $D_{\rm s,c}^{(1)}$ and a geometric contribution $D_{\rm s,g}^{(1)}$.
The expression for the conventional part of the GRPA correction is given by combining~\eqref{eq:JN_final_main},~\eqref{eq:lambda_c_alphaalpha1} and~\eqref{eq:GRPA_Ds_conv}. One can see from the expression~\eqref{eq:JN_final_main} for correlation functions of the form $(\hat{J}_l, \hat{N}_{\alpha\sigma})$  that $D_{\rm s,c}^{(1)}$ vanishes in the flat band limit, as in the case of $D_{\rm s,c}^{(0)}$. We also show that the conventional GRPA correction is a positive semidefinite tensor, meaning that this correction term leads to an increase of the superfluid weight compared to the mean-field result. 

On the other hand, the geometric part of the GRPA correction, given by~\eqref{eq:JD_final_main},~\eqref{eq:lambda_g_alphaalpha1} and~\eqref{eq:GRPA_Ds_geom}, is not necessarily zero in the flat band limit and, as discussed in Sec.~\ref{sec:Natural_orbital} and through examples in Sec.~\ref{sec:examples}, it is important to include it in order to restore the geometry independence of the superfluid weight. Indeed, we find that, within the GRPA, the superfluid weight in the flat band limit is proportional to the minimal quantum metric $\widetilde{\mathcal{M}}$~\eqref{eq:minimal_qm}-\eqref{eq:R_integrated_def}, the integral of the quantum metric over the Brillouin zone minimized with respect to the orbital positions (see Sec.~\ref{sec:Natural_orbital}). The relation between minimal quantum metric and superfluid weight was pointed out in Ref.~\cite{Huhtinen2022} for the first time, where it was shown that the geometry independence of the superfluid weight is restored by not neglecting the $\vb{A}$-dependence of the pairing potential $\Delta(\vb{A})$ when taking derivatives of the mean-field free energy~\eqref{eq:D_mf}. 

In this work we extend the results of Ref.~\cite{Huhtinen2022} in several ways. First, in~\eqref{eq:D_mf} we take also into account the $\vb{A}$-dependence of the Hartree-Fock potential $\Gamma(\vb{A})$, which amounts to computing the superfluid weight within the full GRPA~\cite{Peotta2022}. By doing so, we obtain the conventional part of the GRPA correction to the superfluid weight, which is a new result. This does not affect the results of Ref.~\cite{Huhtinen2022} since $D_{\rm s,c}^{(1)} = 0$ in the flat band limit, which means that the relation between superfluid weight and minimal quantum metric holds at the level of the GRPA. This is consistent with numerical investigations performed using quantum Monte Carlo methods~\cite{Hofmann2020,Peri2021}, which find that~\eqref{eq:Ds_flat_band} gives a rather accurate estimate of the superfluid weight in the flat band limit. 

Another useful result of the present work is the derivation of the simple analytical expressions for both the minimal quantum metric~\eqref{eq:minimal_qm}-\eqref{eq:R_integrated_def} and the natural orbitals positions, the latter given as the solution of the linear system~\eqref{eq:linear_sys_R}. 
Our approach is based on the direct evaluation of the GRPA formula for the superfluid weight. Compared to the one of Ref.~\cite{Huhtinen2022}, where these band structure invariants were introduced, we are able to remove unnecessary assumptions, namely the flat band requirement and the condition~\eqref{eq:uniform_pairing_3}. This is very important in view of potential future applications in other contexts. While the quantum metric has found by now many important applications, the issue of its geometry dependence is almost never addressed, therefore we expect the analytical formulation of the minimal quantum metric and the natural orbital positions to become very useful in this sense. Also, our gauge symmetry-based argument for establishing the geometry independence of the superfluid weight (Sec.~\ref{sec:gauge}) may  be extended to other observable quantities.

In Sec.~\ref{sec:examples}, an interesting application is  already provided, since we find that the SSH model is topologically trivial when the winding number is computed using the natural orbital positions, while the Creutz ladder is not. For the future, it would be interesting to better understand in what sense these two models, and other ones as well, are topologically distinct. Note that the SSH model is often used as a toy model to illustrate the concepts of bulk-edge correspondence and topological phase transition~\cite{Asboth2016}, while at the same time it is debated whether it is a truly topologically lattice model given that its winding number is not unit cell consistent~\cite{Fuchs2021}. In view of these examples, we speculate that the concept of natural orbital positions may ultimately lead to a more refined classification of the topological properties of the band structure. 

As we have seen in Sec.~\ref{sec:examples} (see also Ref.~\cite{Huhtinen2022}), another useful feature of the natural orbital positions is to provide a set of positions that are maximally symmetric without requiring any input other than the band projector $P(\vb{k})$. Thus, they could find applications in the field of electronic structure theory, for instance. In this sense, more work is needed in order to better understand the physical meaning of the natural orbital positions since symmetry alone is not sufficient to determine them. For instance, this is the case of the Lieb lattice with staggered hopping~\cite{Huhtinen2022}, in which many orbital positions are compatible with the lattice symmetries, but the natural ones are uniquely determined up to translations.

Finally, an interesting direction for future work is to extend our results to bosonic superfluids. Some work as already been done in this direction~\cite{Julku2021,Julku2021a,Julku2023,Lukin2023}, however the superfluid weight has not been computed in the full GRPA approximation since the $\vb{A}$-dependence of the Hartree-Fock potential is not taken into account. This might be especially important in cases where translational  symmetry is spontaneously broken by interactions, for instance the supersolid phase observed in ultracold gases with dipolar interactions~\cite{Chomaz2019,Tanzi2019,Bottcher2019,Donner2019}.

\acknowledgments

This work has been supported by the Academy of Finland under Grants No. 330384 and No. 336369. We acknowledge the computational resources provided by the Aalto Science-IT project. We thank Kukka-Emilia Huhtinen, P\"aivi T\"orm\"a, Ville Pyykk\"onen and Jami Kinnunen for useful suggestions and discussions.
 	
\appendix
	
	\section{Derivation of the generalized random phase approximation for the superfluid weight}
	\label{app:grpa}
	
	In this Appendix, we derive the expression for the superfluid weight in the generalized random phase approximation~\eqref{eq:GRPA} using a method different from the one presented in Ref.~\cite{Peotta2022}. While in this latter reference an approach based on the one-particle density matrix has been used, here we work directly with the mean-field potentials $\Gamma$ and $\Delta$. \\
	To keep the derivation general, we consider a more general type of interaction term of the form
	\begin{gather}
	\label{eq:H_int_alt}
	\ham_{\rm int} = \frac{1}{2}\sum_{i,j}V_{i\sigma,j\sigma'}\hat{n}_{i\sigma}\hat{n}_{j\sigma'}\,,
	\end{gather}
	with $V_{i\sigma,j\sigma'} = V_{j\sigma',i\sigma}$ and $V_{i\sigma,i\sigma} = 0$. Moreover, translational invariance is not assumed, therefore the unit cell index and the orbital index are grouped together into collective indices $\vb{i}\alpha \to i $,  $\vb{j}\beta\to j$, as done in Ref.~\cite{Peotta2022}. Contrary to Ref.~\cite{Peotta2022} however,  spin rotational symmetry along the $z$-axis is assumed from the start, that is the many-body Hamiltonian commutes with the spin operator~\eqref{eq:Sz} and so does the variational Hamiltonian
	\begin{equation}
		\label{eq:H_BdG}
		\begin{split}
		\mathcal{\hat H}_0 &= \mathcal{\hat H}_{\rm free}-\mu\hat{N} + \sum_\sigma\sum_{j,j'} \Gamma^\sigma_{j,j'} \hat{c}_{j\sigma}^\dg \hat{c}_{j'\sigma} \\
        &\quad+\sum_{j,j'} \qty(\Delta_{j,j'}\hat{c}_{j\uparrow}^\dg \hat{c}_{j'\downarrow}^\dg + \Delta_{j,j'}^*\hat{c}_{j'\downarrow}\hat{c}_{j\uparrow})
		\end{split}
	\end{equation}
	that enters in the right hand side of the Bogoliubov inequality~\eqref{eq:Bogololiubv_inequality}.
	This is the most general quadratic Hamiltonian constrained only by spin rotational symmetry along the $z$-axis. The expectation value in~\eqref{eq:Bogololiubv_inequality} can be evaluated using Wick's theorem and, for an interaction term of form~\eqref{eq:H_int_alt}, reads
	%
	%
	\begin{equation}
		\label{eq:inter_Bog}
		\begin{split}
			&\expval*{\mathcal{\hat H}-\mathcal{\hat H}_0} = 
			\frac{1}{2}\sum_{\sigma,\sigma'}\sum_{i,j} V_{i\sigma,j\sigma'}\expval{\hat{n}_{i\sigma}}\expval{\hat{n}_{j\sigma'}} \\
			&-
			\frac{1}{2}\sum_{\sigma}\sum_{i,j} V_{i\sigma,j\sigma}\expval*{\hat{c}_{i\sigma}^\dg\hat{c}_{j\sigma}}\expval*{\hat{c}_{j\sigma}^\dg\hat{c}_{i\sigma}}
			\\
			&+\sum_{i,j}V_{i\uparrow,j\downarrow}\expval*{\hat{c}_{i\uparrow}^\dg\hat{c}_{j\downarrow}^\dg}\expval*{\hat{c}_{j\downarrow}\hat{c}_{i\uparrow}}
			- \sum_{i,j}\sum_\sigma\Gamma^\sigma_{i,j} \expval*{\hat{c}_{i\sigma}^\dg \hat{c}_{j\sigma}} \\
			&- \sum_{i,j}\qty(\Delta_{i,j} \expval*{\hat{c}_{i\uparrow}^\dg \hat{c}_{j\downarrow}^\dg} + \Delta_{i,j}^*  \expval*{\hat{c}_{j\downarrow}\hat{c}_{i\uparrow}})\,.
		\end{split}
	\end{equation}
	%
	The expectation values that appear in the above expression are obtained as the derivatives of $\Omega_0 = -\beta^{-1}\Tr\qty\big[e^{-\beta\ham_0}]$ with respect to the mean-field potentials
	\begin{gather}
		\pdv{\Omega_0}{\Gamma^\sigma_{i,j}} = \expval*{\hat{c}_{i\sigma}^\dg\hat{c}_{j\sigma}}\,,
		\label{eq:omega_deriv_1}
		\\
		\pdv{\Omega_0}{\Delta_{i,j}} = \expval*{\hat{c}_{i\uparrow}^\dg \hat{c}_{j\downarrow}^\dg}\,,\quad \pdv{\Omega_0}{\Delta^*_{i,j}} = \expval*{\hat{c}_{j\downarrow}\hat{c}_{i\uparrow}}\,.
		\label{eq:omega_deriv_2}
	\end{gather}
	Then, by minimizing the mean-field grand potential~\cite{Peotta2022}, one obtains the self-consistency equations of mean-field theory
	\begin{gather}
		\label{eq:BCS_self_cons_1}
		\Gamma^\sigma_{j,j} =  \sum_{j'} V_{j\sigma,j'\sigma'}\expval{\hat{n}_{j'\sigma'}}\,,\\
		\label{eq:BCS_self_cons_2}
		\Gamma_{i,j}^\sigma = - V_{i\sigma,j\sigma}\expval*{\hat{c}_{j\sigma}^\dg\hat{c}_{i\sigma}}\,, \qq{for} i \neq j\,,\\
		\label{eq:BCS_self_cons_3}
		\Delta_{i,j} = V_{i\uparrow,j\downarrow}\expval*{\hat{c}_{j\downarrow}\hat{c}_{i\uparrow}}\,,\qq{for} i \neq j\,.
	\end{gather}
	The variational coefficients $\Gamma^\sigma_{j,j}$ corresponds to the Hartree potential,  $\Gamma_{i,j}^\sigma$ for $i\neq j$ to the non-local Fock potential and $\Delta_{i,j}$ to the pairing potential.
	
	Following the notation of Ref.~\cite{Peotta2022}, we define the vector of mean-field potentials as
	\begin{equation}
		\Delta^a = (\Gamma_{i,j}^\sigma,\Delta_{i,j},\Delta_{i,j}^*)\,, 	
	\end{equation}
	and we use the following variation of the Einstein summation convention
	\begin{equation}
		\label{eq:Einstein2}
		\begin{split}
		\pdv{f}{\Delta^a}\pdv{\Delta^a}{g} &\overset{\rm def}{=} \sum_\sigma\sum_{i,j}\pdv{f}{\Gamma^\sigma_{i,j}} \pdv{\Gamma^\sigma_{i,j}}{g} \\
        & + \sum_{i,j}\qty\bigg(\pdv{f}{\Delta_{i,j}} \pdv{\Delta_{i,j}}{g}+
		\pdv{f}{\Delta^*_{i,j}} \pdv{\Delta^*_{i,j}}{g})\,.
		\end{split}
	\end{equation}  
	The vector of expectation values of operators quadratic in $\hat{c}_{j\sigma},\,\hat{c}_{j\sigma}^\dg$ is then written as (see~\eqref{eq:omega_deriv_1}-\eqref{eq:omega_deriv_2})
	\begin{equation}
		\begin{split}
		\pdv{\Omega_0}{\Delta^a} &= \qty(\pdv{\Omega_0}{\Gamma_{i,j}^\sigma},\pdv{\Omega_0}{\Delta_{i,j}},\pdv{\Omega_0}{\Delta_{i,j}^*}) ^T \\
        & = \qty(\expval*{\hat{c}_{i\sigma}^\dg\hat{c}_{j\sigma}},\expval*{\hat{c}_{i\uparrow}^\dg \hat{c}_{j\downarrow}^\dg},\expval*{\hat{c}_{j\downarrow}\hat{c}_{i\uparrow}})^T\,.
		\end{split}
	\end{equation}
	Using this notation we can write the expectation value of the interaction term in a concise form
    \begin{widetext}
	\begin{equation}
		\label{eq:Hint_ev}
		\begin{split}
			&\ev*{\mathcal{\hat{H}}_{\rm int}}	=  \frac{1}{2}\pdv{\Omega_0}{\Delta^a}V^{ab}\pdv{\Omega_0}{\Delta^b} 
			= \frac{1}{2}\sum_{i,j,i',j'} \qty(\expval*{\hat{c}_{i\uparrow }^\dg\hat{c}_{j\uparrow}},\expval*{\hat{c}_{i\downarrow }^\dg\hat{c}_{j\downarrow}},\expval*{\hat{c}_{i\uparrow}^\dg \hat{c}_{j\downarrow}^\dg},\expval*{\hat{c}_{j\downarrow}\hat{c}_{i\uparrow}}) \times\\ &\pmqty{\mqty{V_{i\uparrow,i'\uparrow}\delta_{i,j}\delta_{i',j'} -V_{i\uparrow,j\uparrow} \delta_{i,j'}\delta_{j,i'} & V_{i\uparrow,i'\downarrow}\delta_{i,j}\delta_{i',j'}\\ 
					V_{i\downarrow,i'\uparrow}\delta_{i,j}\delta_{i',j'} & V_{i\downarrow,i'\downarrow}\delta_{i,j}\delta_{i',j'} -V_{i\downarrow,j\downarrow} \delta_{i,j'}\delta_{j,i'} } & 0 \\
				0 & \mqty{0 & V_{i\uparrow,j\downarrow}\delta_{i,i'}\delta_{j,j'}\\ V_{j\downarrow,i\uparrow}\delta_{i,i'}\delta_{j,j'} & 0}}
			\pmqty{\expval*{\hat{c}_{i'\uparrow }^\dg\hat{c}_{j'\uparrow}}\\\expval*{\hat{c}_{i'\downarrow }^\dg\hat{c}_{j'\downarrow}}\\\expval*{\hat{c}_{i'\uparrow}^\dg \hat{c}_{j'\downarrow}^\dg}\\\expval*{\hat{c}_{j'\downarrow}\hat{c}_{i'\uparrow}}}\,.
		\end{split}
	\end{equation}
    \end{widetext}
	Thus, from this last equation and~\eqref{eq:inter_Bog}, we have for the the mean-field grand potential 
	\begin{equation}
		\Omega_{\rm m.f.} = \Omega_0 -\Delta^a \pdv{\Omega_0}{\Delta^a}	+ 
		\frac{1}{2}\pdv{\Omega_0}{\Delta^a}V^{ab}\pdv{\Omega_0}{\Delta^b}\,.
	\end{equation}
	Taking the partial derivative with respect to $\Delta^c$ gives
	\begin{equation}
		\label{eq:Omega_mf_first}
		\pdv{\Omega_{\rm m.f.}}{\Delta^c} = \qty( V^{ab}\pdv{\Omega_0}{\Delta^b} -\Delta^a)\pdv{\Omega_0}{\Delta^c}{\Delta^a}\,,
	\end{equation}
	The first partial derivatives of the mean-field grand potential vanish when
	\begin{equation}
		\label{eq:self-consistent_Einstein}
		V^{ab}\pdv{\Omega_0}{\Delta^b} -\Delta^a = 0\,.
	\end{equation}
	These are just the self-consistency equations of mean-field theory~\eqref{eq:BCS_self_cons_1}-\eqref{eq:BCS_self_cons_3} written using our Einstein summation convention~\eqref{eq:Einstein2}-\eqref{eq:Hint_ev}. After taking another partial derivative of~\eqref{eq:Omega_mf_first} and imposing that the mean-field potentials are self-consistent~\eqref{eq:self-consistent_Einstein}, we have
	\begin{equation}
		\label{eq:Omega_DeDe}
		\pdv{\Omega_{\rm m.f.}}{\Delta^c}{\Delta^d} = -\pdv{\Omega_0}{\Delta^c}{\Delta^d} +
		\pdv{\Omega_0}{\Delta^c}{\Delta^a}V^{ab}\pdv{\Omega_0}{\Delta^b}{\Delta^d}\,.	
	\end{equation}
	Recall that both the mean-field grand potential $\Omega_{\rm m.f.}$ and $\Omega_0$ depend  either directly or indirectly on the vector potential $\vb{A}$, namely $\Omega_{\rm m.f.}\qty\big(\vb{A},\Delta^a(\vb{A}))$ and $\Omega_0\qty\big(\vb{A},\Delta^a(\vb{A}))$ (see~\eqref{eq:D_mf}). The self-consistency equations are satisfied for any $\vb{A}$, which means that 
	\begin{equation}
		\label{eq:Omega_mf_dAl}
		\begin{split}
		&\dv{}{A_l}\bigg(\pdv{\Omega_{\rm m.f.}}{\Delta^a} \big(\vb{A},\Delta^a(\vb{A})\big)\bigg) \\
		&= \pdv{\Omega_{\rm m.f.}}{A_l}{\Delta^a} + \pdv{\Omega_{\rm m.f.}}{\Delta^a}{\Delta^b}\pdv{\Delta^b}{A_l} = 0\,.
		\end{split}
	\end{equation}
	This last result gives the derivatives of the mean-field potentials with respect to $\vb{A}$
	\begin{equation}
		\label{eq:Delta_dAl}
		\pdv{\Delta^a}{A_l} = - \qty[\pdv{\Omega_{\rm m.f.}}{\Delta^a}{\Delta^b}]^{-1}\pdv{\Omega_{\rm m.f.}}{A_l}{\Delta^b}\,.
	\end{equation}
	Here and in the following, we use the notation $\qty[M_{a,b}]^{-1}$ to denote the inverse of the matrix with elements $M_{a,b}$ rather than the inverse of a single matrix element. We ignore here the subtleties occurring when the inverse of the Hessian matrix $\pdv*{\Omega_{\rm m.f.}}{\Delta^a}{\Delta^b}$ does not exists. In this case one should express the solution of the linear system~\eqref{eq:Omega_mf_dAl} using the pseudoinverse (Moore-Penrose inverse). 
	
	We can now compute the full derivatives of the mean-field grand potential and express them in terms of the mean-field solution for $\vb{A} = \vb{0}$.
	We start with the first full  derivative
	\begin{equation}
		\label{eq:Omega_dAl}
		\begin{split}
		&\dv{}{A_l}	\qty[\Omega_{\rm m.f.}\qty\big(\vb{A},\Delta^a(\vb{A}))] =  \\
		&= \pdv{\Omega_{\rm m.f.}}{A_l} + \pdv{\Omega_{\rm m.f.}}{\Delta^a}\pdv{\Delta^a}{A_l} = \pdv{\Omega_{\rm m.f.}}{A_l}\,,
		\end{split}
	\end{equation}
	where in the last equality we have used~\eqref{eq:Omega_mf_first}-\eqref{eq:self-consistent_Einstein}.
	The second full derivative is then
	\begin{equation}
		\label{eq:Omega_mf_dAldAm}
		\begin{split}
			&\frac{d^2}{dA_m dA_l} \qty[\Omega_{\rm m.f.}\qty\big(\vb{A},\Delta^a(\vb{A}))] \\ &=			
			\dv{}{A_m}\bigg(\pdv{\Omega_{\rm m.f.}}{A_l} \big(\vb{A},\Delta^a(\vb{A})\big)\bigg) \\
            & = \pdv{\Omega_{\rm m.f.}}{A_m}{A_l} + \pdv{\Delta^a}{A_m} \pdv{\Omega_{\rm m.f.}}{\Delta^a}{A_l} \\
			&= \pdv{\Omega_{\rm m.f.}}{A_m}{A_l} - \pdv{\Delta^a}{A_m} 
			\pdv{\Omega_{\rm m.f.}}{\Delta^a}{\Delta^b}\pdv{\Delta^b}{A_l}\ \\
            & = \pdv{\Omega_{\rm m.f.}}{A_m}{A_l} - \pdv{\Omega_{\rm m.f.}}{A_m}{\Delta^b}\qty[\pdv{\Omega_{\rm m.f.}}{\Delta^b}{\Delta^a}]^{-1}\pdv{\Omega_{\rm m.f.}}{\Delta^a}{A_l}\,.
		\end{split}	
	\end{equation}
	where in the last two equalities we have used~\eqref{eq:Omega_mf_dAl} and~\eqref{eq:Delta_dAl}.
	Since the self-consistent solution minimizes $\Omega_{\rm m.f.}$, the Hessian matrix $\pdv*{\Omega_{\rm m.f.}}{\Delta^a}{\Delta^b}$ (and also its inverse) is positive semidefinite. This fact together with~\eqref{eq:Omega_mf_dAldAm} implies that the superfluid weight is bounded from above (in the sense of matrix inequalities) by the matrix $\pdv*{\Omega_{\rm m.f.}}{A_m}{A_l}$. 
	
	It is convenient to express the result in~\eqref{eq:Omega_mf_dAldAm} only in terms of the quantities $\pdv*{\Omega_0}{\Delta^a}{\Delta^b}$, which are correlations function relative to the mean-field statistical ensemble, as shown below. We have already done this is in the case of $\pdv*{\Omega_{\rm m.f.}}{\Delta^a}{\Delta^b}$ in ~\eqref{eq:Omega_DeDe}, thus we only need to do the same for $\pdv*{\Omega_{\rm m.f.}}{A_m}{A_l}$ and $\pdv*{\Omega_{\rm m.f.}}{\Delta^a}{A_l}$. Since $\Omega_{\rm m.f.} = \Omega_0 + \ev*{\ham-\ham_0}$ we consider the partial derivatives of the expectation value
	\begin{equation}
		\label{eq:H-H_dAl}
		\begin{split}
		\pdv{\ev*{\ham-\ham_0}}{A_l} &= \pdv{}{A_l}\qty(-\Delta^a \pdv{\Omega_0}{\Delta^a}	+ 
		\frac{1}{2}\pdv{\Omega_0}{\Delta^a}V^{ab}\pdv{\Omega_0}{\Delta^b}) \\ & = \pdv{\Omega_0}{A_l}{\Delta^a}\qty(V^{ab}\pdv{\Omega_0}{\Delta^b} - \Delta^a)\,.
		\end{split}
	\end{equation}
	Again this quantity vanishes if the self-consistency equations~\eqref{eq:self-consistent_Einstein} are satisfied, therefore from~\eqref{eq:Omega_dAl},  we have 
	\begin{equation}
		\begin{split}
		&\dv{}{A_l}	\qty[\Omega_{\rm m.f.}\qty\big(\vb{A},\Delta^a(\vb{A}))] = \pdv{\Omega_{\rm m.f.}}{A_l} = \pdv{\Omega_0}{A_l} + \pdv{\ev*{\ham-\ham_0}}{A_l} \\
		&\quad= \pdv{\Omega_0}{A_l} = \ev{\pdv{\ham_{\rm free}(\vb{A})}{A_l}} = - \ev*{\hat{J}_l}\,,
		\end{split}
	\end{equation} 
	where~\eqref{eq:current_operator_definition} has been used in the last equality.
	Thus, we find the important result that the first full derivative of the grand potential with respect to $\vb{A}$ is proportional to the current. Taking another partial derivative with respect to $A_m$ in~\eqref{eq:H-H_dAl} and~\eqref{eq:Omega_mf_first} and imposing self-consistency gives
	\begin{gather}
		\pdv{\ev*{\ham-\ham_0}}{A_l}{A_m} = \pdv{\Omega_0}{A_l}{\Delta^a}V^{ab}\pdv{\Omega_0}{\Delta^b}{A_m}\,,\\
		\pdv{\Omega_{\rm m.f.}}{A_l}{\Delta^c} = \pdv{\Omega_0}{A_l}{\Delta^a} V^{ab} \pdv{\Omega_0}{\Delta^b}{\Delta^c}\,.
	\end{gather}
	Using these results together with~\eqref{eq:Omega_DeDe} and the last line of~\eqref{eq:Omega_mf_dAldAm} leads to 
    \begin{equation}
		\label{eq:D_final_1}
		\begin{split}
			&\frac{d^2}{dA_m dA_l} \qty[\Omega_{\rm m.f.}\qty\big(\vb{A},\Delta^a(\vb{A}))] =
			\pdv{\Omega_0}{A_m}{A_l} \\ &+\pdv{\Omega_0}{A_l}{\Delta^a}V^{ab}\pdv{\Omega_0}{\Delta^b}{A_m} -v_{l,c}\qty[M_{c,d}]^{-1} v_{m,d}\,, \\
			&\qq*{with} M_{c,d} = -\pdv{\Omega_0}{\Delta^c}{\Delta^d} +
			\pdv{\Omega_0}{\Delta^c}{\Delta^g}V^{gh}\pdv{\Omega_0}{\Delta^h}{\Delta^d}\\
			&\qq*{and} v_{l,c} = \pdv{\Omega_0}{A_l}{\Delta^a} V^{ab} \pdv{\Omega_0}{\Delta^b}{\Delta^c}\,. 
		\end{split}
	\end{equation}
	To simplify this expression we use the identity
	\begin{equation}
        \begin{split}
		&A(A-ABA)^{-1}A = (A^{-1}-B)^{-1} \\
        &=A+ABA+ABABA+\dots\,,
        \end{split}
	\end{equation}
	where $A$ and $B$ are matrices defined by
	\begin{equation}
		[A]_{a,b} = \pdv{\Omega_0}{\Delta^a}{\Delta^b}\qq{and} [B]_{a,b} = V^{ab}\,.
	\end{equation}
	Thus, our final result is
	\begin{equation}
		\label{eq:D_final_2}
		\begin{split}
			& \frac{d^2}{dA_m dA_l} \qty[\Omega_{\rm m.f.}\qty\big(\vb{A},\Delta^a(\vb{A}))] =
			\pdv{\Omega_0}{A_m}{A_l} \\
   &+ \pdv{\Omega_0}{A_l}{\Delta^a}V^{ab}\pdv{\Omega_0}{\Delta^b}{A_m} \\
			&+ \pdv{\Omega_0}{A_l}{\Delta^a} V^{ab} \qty[\qty[\pdv{\Omega_0}{\Delta^b}{\Delta^c}]^{-1} - V^{bc}]^{-1} V^{cd}\pdv{\Omega_0}{\Delta^d}{A_m}\,.
		\end{split}
	\end{equation}
	With further simple manipulations, it is possible to prove the equivalence between this and the result for the superfluid weight in the generalized random phase approximation provided in Ref.~\cite{Peotta2022}. 
	
	We can specialize~\eqref{eq:Hint_ev} to the case of the attractive Hubbard interaction given by
	\begin{equation}
		\label{eq:Hubbard_non_trans_symm}
		\ham_{\rm int} = -\sum_{j}U_j\hat{n}_{j\uparrow}\hat{n}_{j\downarrow},\,\qq{with} U_j> 0\,. 
	\end{equation}
	The Hubbard interaction in the above equation is more general than~\eqref{eq:Hubbard_int} since the latter is translationally invariant, while the former is not as we allow for an arbitrary dependence of the coupling constant $U_j$ on the site index $j$.
	The Hubbard interaction in~\eqref{eq:Hubbard_non_trans_symm} is obtained by choosing the interaction coefficients in~\eqref{eq:H_int_alt} as follows
	\begin{equation}
		V_{i\uparrow,j\downarrow} = V_{i\downarrow,j\uparrow} = -U_j\delta_{i,j}\,,\quad V_{i\sigma,j\sigma} = 0\,.	
	\end{equation}
	Then, the expectation value of the interaction term in~\eqref{eq:Hint_ev} becomes
	\begin{equation}
		\label{eq:interaction_matrix}
		\begin{split}
			&\ev*{\mathcal{\hat{H}}_{\rm int}}	=  \frac{1}{2}\pdv{\Omega_0}{\Delta^a}V^{ab}\pdv{\Omega_0}{\Delta^b} \\
			&= -\sum_jU_j\qty(\ev{\hat{n}_{j\uparrow}}\ev{\hat{n}_{j\downarrow}} + 
			\ev*{\hat{c}^\dg_{j\uparrow}\hat{c}^\dg_{j\downarrow}}
			\ev*{\hat{c}_{j\downarrow}\hat{c}_{j\uparrow}}) \\
			&=-\frac{1}{2}\sum_{j} 
			\pmqty{\expval*{\hat{n}_{j\uparrow }},\expval*{\hat{n}_{j\downarrow }},\expval*{\hat{c}_{j\uparrow}^\dg \hat{c}_{j\downarrow}^\dg},\expval*{\hat{c}_{j\downarrow}\hat{c}_{j\uparrow}}}\\
			&\qquad\times\pmqty{0 & U_j & 0 & 0 \\ 
				U_j & 0 & 0 & 0 \\
				0 & 0 & 0 & U_j \\
				0 & 0 & U_j & 0}
			\pmqty{\expval*{\hat{n}_{j\uparrow }} \\ \expval*{\hat{n}_{j\downarrow }} \\ \expval*{\hat{c}_{j\uparrow}^\dg \hat{c}_{j\downarrow}^\dg} \\ \expval*{\hat{c}_{j\downarrow}\hat{c}_{j\uparrow}}}   
			\,.
		\end{split}
	\end{equation}
	This gives the matrix $B$ in~\eqref{eq:B_alpha}, after translational invariance is enforced. Instead, the expressions for the matrices $A_{\alpha,\beta}$~\eqref{eq:A_alphabeta} and the vectors $\vb{v}_{l\alpha}$~\eqref{eq:vect_v_def} in terms of correlation functions of the form~\eqref{eq:corr_func_def} are obtained by using the results~\cite{Peotta2022}
	\begin{equation}
		\pdv{\Omega_0}{\Delta^a}{\Delta^b} = \qty(\pdv{\ham_0}{\Delta^a},\pdv{\ham_0}{\Delta^b})\,,
	\end{equation}  
	and
	\begin{equation}
		\pdv{\Omega_0}{A_l}{\Delta^a} = \qty(\pdv{\ham_0}{A_l},\pdv{\ham_0}{\Delta^a}) = -\qty(\hat{J}_l,\pdv{\ham_0}{\Delta^a})\,,	
	\end{equation}
	respectively. As an example, we have
	\begin{equation}
		\begin{split}
		&\pdv{\Omega_0}{A_l}{\Delta_{i,i}} = -\qty(\hat{J}_l,\pdv{\ham_0}{\Delta_{i,i}}) = 
		-\qty\big(\hat{J}_l,\hat{c}_{i\uparrow}^\dg\hat{c}_{i\downarrow}^\dg) \\
		&
		= -\qty\big(\hat{J}_l,\hat{c}_{\vb{i}\alpha\uparrow}^\dg\hat{c}_{\vb{i}\alpha\downarrow}^\dg) = -\qty(\hat{J}_l,\frac{1}{N_c}\sum_{\vb{i}}\hat{c}_{\vb{i}\alpha\uparrow}^\dg\hat{c}_{\vb{i}\alpha\downarrow}^\dg) \\
		& = - \qty\big(\hat{J}_l,\hat{D}_\alpha^\dg)
		\,.		
		\end{split}
	\end{equation}
	In the second line we have performed the substitution $i\to \vb{i}\alpha$ and used the fact that the correlation function $\qty\big(\hat{J}_l,\hat{c}_{\vb{i}\alpha\uparrow}^\dg\hat{c}_{\vb{i}\alpha\downarrow}^\dg)$ is independent of the unit cell index $\vb{i}$ since the operator $\hat{J}_l$ is translationally invariant.
 	This completes the derivation of the expression for the superfluid weight in the generalized random phase approximation given in~\eqref{eq:GRPA} in the main text.

 \section{Nambu formalism and evaluation of correlation functions}
  \label{app:corr_computation}
  In this section, it is explained how to evaluate the correlation function~\eqref{eq:corr_func_def} between translationally invariant quadratic operators, which are conveniently recast in the following Nambu form 
  
  \begin{gather}
  	\hat{A} = \sum_{\vb{k}}\hat{A}(\vb{k})\,,\\
  	\label{eq:quadratic_generic}
  	\begin{split}
  		&\hat{A}(\vb{k}) = \\
  		 &= \sum_{\alpha,\beta}\Big(\hat{c}_{\vb{k}\alpha\uparrow}^\dg [A(\vb{k})]_{\alpha,\beta}^{1,1} \hat{c}_{\vb{k}\beta\uparrow} +  \hat{c}_{-\vb{k}\alpha\downarrow} [A(\vb{k})]_{\alpha,\beta}^{2,2} \hat{c}^\dg_{-\vb{k}\beta\downarrow} 
  		\\&\qquad\,\,
  		+ \hat{c}_{\vb{k}\alpha\uparrow}^\dg [A(\vb{k})]_{\alpha,\beta}^{1,2}\hat{c}^\dg_{-\vb{k}\beta\downarrow} + \hat{c}_{-\vb{k}\alpha\downarrow} [A(\vb{k})]_{\alpha,\beta}^{2,1} \hat{c}_{\vb{k}\beta\uparrow}   \Big) \\
  		&=\hat{\vb{c}}_{\vb{k}}^\dg A(\vb{k}) \hat{\vb{c}}_{\vb{k}}\,,
  	\end{split} 
  \\
  \label{eq:Nambu_matrix}
  \qq{with} A(\vb{k}) = \pmqty{[A(\vb{k})]^{1,1} & [A(\vb{k})]^{1,2} \\
  [A(\vb{k})]^{2,1} & [A(\vb{k})]^{2,2}}\,,\\
  \label{eq:Nambu_spinor}
  \hat{\vb{c}}_{\vb{k}} = 	\pmqty{\hat{\vb{c}}_{\vb{k}\uparrow} \\
   (\hat{\vb{c}}^\dg_{-\vb{k}\downarrow})^T}\qq{and} \hat{\vb{c}}_{\vb{k}}^\dg = \pmqty{\hat{\vb{c}}_{\vb{k}\uparrow}^\dg & (\hat{\vb{c}}_{-\vb{k}\downarrow})^T}\,.
  \end{gather}
  Here $\hat{\vb{c}}_{\vb{k}\sigma}$ ($\hat{\vb{c}}_{\vb{k}\sigma}^\dg$) is the column (row) vector whose components are the field operators $\hat{c}_{\vb{k}\alpha\sigma}$ ($\hat{c}_{\vb{k}\alpha\sigma}^\dagger$) for $\alpha = 1,\dots,N_{\rm orb}$. The column vector $\hat{\vb{c}}_{\vb{k}}$ in~\eqref{eq:Nambu_spinor}, grouping together both creation and annihilation operators, is called a Nambu spinor. Consistently with the Nambu spinor structure, the quadratic operatic $\hat{A}(\vb{k})$ corresponds to the single-particle operator $A(\vb{k})$, a $2N_{\rm orb} \times 2N_{\rm orb}$ matrix consisting of four blocks of dimension $N_{\rm orb}$ denoted by $[A(\vb{k})]^{i,j}$, as shown in~\eqref{eq:Nambu_matrix}. We adopt the convention that the blocks of a single-particle operator in the Nambu representation are labeled by superscripts, while subscripts label the matrix elements in each block. This convention is used already in~\eqref{eq:quadratic_generic}. Note that, in order to bring a quadratic operator in Nambu form, it is necessary to rearrange the field operators, which  produces additional c-number terms due to the fermionic anticommutation relations. However, all of the c-number terms cancel out when the expectation value of the same operator is subtracted, namely $\hat{A}(\vb{k}) - \ev*{\hat{A}(\vb{k})}$. This combination is precisely the one appearing in~\eqref{eq:corr_func_def}, implying that, for the purpose of computing correlation functions, we are free to reorder the field operators and represent quadratic operators in Nambu form. The ultimate reason for using the Nambu formalism is that it allows to diagonalize in a convenient way quadratic operators that contain anomalous terms, such as $\hat{c}_{\vb{i}\alpha\downarrow}\hat{c}_{\vb{i}\alpha\uparrow}$ and $\hat{c}_{\vb{i}\alpha\uparrow}^\dg\hat{c}_{\vb{i}\alpha\downarrow}^\dg$. In our specific case, we need to diagonalize the variational Hamiltonian $\mathcal{\hat{H}}_0$ (see Sec.~\ref{sec:correction}).

In the following  we denote by $\ev*{\hat{\vb{c}}_{\vb{k}}(\tau)\hat{\vb{c}}_{\vb{k}'}}$
the set of expectation values obtained by replacing each of the Nambu spinors with any of their components, namely $\ev*{\hat{c}_{\vb{k}\alpha\uparrow}(\tau)\hat{c}_{\vb{k}'\beta\uparrow}}$, $\ev*{\hat{c}^\dg_{-\vb{k}\alpha\downarrow}(\tau)\hat{c}_{\vb{k}'\beta\uparrow}}$, $\ev*{\hat{c}_{\vb{k}\alpha\uparrow}(\tau)\hat{c}^\dg_{-\vb{k}'\beta\downarrow}}$ and $\ev*{\hat{c}^\dg_{-\vb{k}\alpha\downarrow}(\tau)\hat{c}^\dg_{-\vb{k}'\beta\downarrow}}$ for $\alpha,\beta = 1,\dots,N_{\rm orb}$. The same convention is used  for $\ev*{\hat{\vb{c}}^\dg_{\vb{k}}(\tau)\hat{\vb{c}}_{\vb{k}'}}$ and so on. Using this notation, we can express in a concise way the constraints imposed by the
conservation of spin $\hat{S}^z$ and momentum on the expectation values of products of two operators, which read
\begin{gather}
    \label{eq:spin_conservation}
    \ev*{\hat{\vb{c}}_{\vb{k}}^\dg(\tau)\hat{\vb{c}}_{\vb{k}'}^\dg} = \ev*{\hat{\vb{c}}_{\vb{k}}(\tau)\hat{\vb{c}}_{\vb{k}'}} = 0\,,	\\
    \label{eq:k_conservation_1}
    \ev*{\hat{\vb{c}}_{\vb{k}}^\dg(\tau)\hat{\vb{c}}_{\vb{k}'}} = \ev*{\hat{\vb{c}}_{\vb{k}}^\dg(\tau)\hat{\vb{c}}_{\vb{k}}}\delta_{\vb{k},\vb{k}'}\,, \\
    \label{eq:k_conservation_2}
    \ev*{\hat{\vb{c}}_{\vb{k}}(\tau)\hat{\vb{c}}^\dg_{\vb{k}'}} = 
		\ev*{\hat{\vb{c}}_{\vb{k}}(\tau)\hat{\vb{c}}^\dg_{\vb{k}}}\delta_{\vb{k},\vb{k}'}\,.
\end{gather}
These relations are used in the evaluation of the expectation value that appears under the integral sign in~\eqref{eq:corr_func_def}
\begin{equation}
		\label{eq:compact_correlation}
		\begin{split}
			&\ev*{(\hat{A}(\vb{k},\tau)-\ev*{\hat{A}(\vb{k})})(\hat{B}(\vb{k}')-\ev*{\hat{B}(\vb{k}')})} \\
            &= \ev*{\hat{A}(\vb{k},\tau)\hat{B}(\vb{k}')} - \ev*{\hat{A}(\vb{k})}\ev*{\hat{B}(\vb{k}')} 
			\\ 
			&= \ev*{\hat{\vb{c}}_{\vb{k}}^\dg(\tau) A(\vb{k}) \hat{\vb{c}}_{\vb{k}}(\tau) \hat{\vb{c}}_{\vb{k}'}^\dg B(\vb{k}') \hat{\vb{c}}_{\vb{k}'}} \\ 
			&\qquad- \ev*{\hat{\vb{c}}_{\vb{k}}^\dg A(\vb{k}) \hat{\vb{c}}_{\vb{k}}}\ev*{\hat{\vb{c}}_{\vb{k}'}^\dg B(\vb{k}') \hat{\vb{c}}_{\vb{k}'}} \\
			&= -\delta_{\vb{k},\vb{k}'} \Tr\big[A(\vb{k})\ev*{\mathcal{T}_\tau[ \hat{\vb{c}}_{\vb{k}}(\tau)\hat{\vb{c}}_{\vb{k}}^\dg(0)]} \\
			&\qquad \qquad\quad\times B(\vb{k})\ev*{\mathcal{T}_\tau[\hat{\vb{c}}_{\vb{k}}(0)\hat{\vb{c}}_{\vb{k}}^\dg(\tau)]}\big] 
			\\&= -\delta_{\vb{k},\vb{k}'}\Tr\big[A(\vb{k})\ev*{\mathcal{T}_\tau[ \hat{\vb{c}}_{\vb{k}}(\tau)\hat{\vb{c}}_{\vb{k}}^\dg(0)]}\\
			&\qquad \qquad\quad\times B(\vb{k})\ev*{\mathcal{T}_\tau[\hat{\vb{c}}_{\vb{k}}(-\tau)\hat{\vb{c}}_{\vb{k}}^\dg(0)]}\big] \\
			&= -\delta_{\vb{k},\vb{k}'}\Tr[A(\vb{k})\mathcal{G}(\vb{k},\tau)B(\vb{k})\mathcal{G}(\vb{k},-\tau)]\,.
		\end{split}
	\end{equation}	
Here $\mathcal{T}_\tau$ is the time-ordering symbol for imaginary time
	\begin{equation}
		\mathcal{T}_\tau[\hat{a}(\tau)\hat{b}(\tau')] = 
		\begin{cases}
			\hat{a}(\tau)\hat{b}(\tau') & \text{for } \tau > \tau'\,,\\
			-\hat{b}(\tau')\hat{a}(\tau) & \text{for } \tau < \tau'\,, 
		\end{cases}	
	\end{equation}
where $\hat{a}$ and $\hat{b}$ are anticommuting operators. Note that $0 < \tau < \beta$ in~\eqref{eq:corr_func_def} and the field operator are ordered accordingly in~\eqref{eq:compact_correlation}. 
In the third equality, we have used Wick's theorem~\cite{Fetter2003}, which holds since all expectation values are evaluated with respect to the quadratic Hamiltonian $\ham_0$, see~\eqref{eq:H-H0}.  
In the last line of~\eqref{eq:compact_correlation} we have introduced the standard definition of the imaginary-time Green's function

\begin{equation}
\label{eq:Green_imag_tau}
\mathcal{G}(\vb{k},\tau-\tau') = -\ev*{\mathcal{T}_\tau[\hat{\vb{c}}_{\vb{k}}(\tau)\hat{\vb{c}}_{\vb{k}}^\dg(\tau')]}\,.
\end{equation}
After expanding the Green's function using Matsubara frequencies $\omega_n=(2n+1)\pi/\beta$ in~\eqref{eq:compact_correlation}
	\begin{equation}
		\label{eq:Green_function_omega}
		\mathcal{G}(\vb{k},\tau) = \frac{1}{\beta}\sum_{\omega_n}\mathcal{G}(\vb{k},i\omega_n)e^{-i\omega_n\tau}\,, 
	\end{equation}
and performing the imaginary time integral in~\eqref{eq:corr_func_def}, one obtains
	\begin{equation}
		\label{eq:correlation_A_and_B}
			(\hat{A},\hat{B}) = \frac{1}{\beta}\sum_{\omega_n}\sum_{\vb{k}} \Tr[A(\vb{k})\mathcal{G}(\vb{k},i\omega_n)B(\vb{k})\mathcal{G}(\vb{k},i\omega_n)]\,.
	\end{equation}
The Nambu form $H_0(\vb{k})$ of the variational Hamiltonian is given  in~\eqref{eq:Bdg_Hamiltonian}. It is a standard result that the Matsubara Green's function can be written as~\cite{Fetter2003}
\begin{equation}
	\label{eq:Green_function_Bdg_Ham}
	\begin{split}
	&\mathcal{G}(\vb{k},i\omega_n) = \frac{1}{i\omega_n-H_0(\vb{k})} \\
	&= (U_{\vb{k}}\oplus U_{\vb{k}})W_{\vb{k}}\frac{1}{i\omega_n-E_{\vb{k}}}W_{\vb{k}}^\dagger
	(U_{\vb{k}}^\dagger\oplus U_{\vb{k}}^\dagger)\,.
	\end{split}
\end{equation}
The second line follows from~\eqref{eq:BdG_Ham_diagonalized}. To proceed, we also need  the operators $\hat{J}_l$, $\hat{N}_\alpha$ and $\hat{D}_\alpha$ in Nambu form, namely
\begin{gather}
\label{eq:current_op_Nambu}
J(\vb{k}) = \pmqty{\partial_{l}H_{\rm free}^\uparrow(\vb{k}) & 0 \\
	0 & -[\partial_{l}H_{\rm free}^{\downarrow}(-\vb{k})]^*}\,,\\
N_{\alpha\uparrow}(\vb{k}) = \pmqty{ [N_{\alpha\uparrow}(\vb{k})]^{1,1} & 0 \\ 0 & 0}\,,\\
N_{\alpha\downarrow}(\vb{k}) = \pmqty{ 0 & 0 \\ 0 & [N_{\alpha\downarrow}(\vb{k})]^{2,2}}\,,\\
 [N_{\alpha\uparrow}(\vb{k})]_{\beta,\gamma}^{1,1} = 
 - [N_{\alpha\downarrow}(\vb{k})]_{\beta,\gamma}^{2,2} =  \delta_{\alpha,\beta}\delta_{\alpha,\gamma}\,, \label{eq:minus_sign_N_down}\\
 \label{eq:D_alpha}
 D_{\alpha}(\vb{k}) = \pmqty{ 0 & 0 \\ [D_{\alpha}(\vb{k})]^{2,1}  & 0}\,,\\
 \label{eq:D_dg_alpha}
 D^\dg_{\alpha}(\vb{k}) = \pmqty{ 0 & [D^\dg_{\alpha}(\vb{k})]^{1,2}  \\ 0 & 0} \,,\\
 \label{eq:D_alpha_matr_elem}
 [D_{\alpha}(\vb{k})]^{2,1} = [D^\dg_{\alpha}(\vb{k})]^{1,2} = \delta_{\alpha,\beta}\delta_{\alpha,\gamma}\,.
\end{gather}
It is important to keep track of the minus sign associated with the down spin in~\eqref{eq:minus_sign_N_down}.

In order to perform the Matsubara frequency summation in~\eqref{eq:correlation_A_and_B}, it is convenient to introduce the following matrix
\begin{equation}
\label{eq:L_matrix_def}
\begin{split}
L(\vb{k}, i\omega) &= W_{\vb{k}}\frac{1}{i\omega-E_{\vb{k}}}W_{\vb{k}}^\dagger \\
&= \pmqty{[L(\vb{k},i\omega)]^{1,1} & [L(\vb{k},i\omega)]^{1,2} \\[0.3em] 
[L(\vb{k},i\omega)]^{2,1} & [L(\vb{k},i\omega)]^{2,2}}\,.
\end{split}
\end{equation}
This is simply the Green's function $\mathcal{G}(\vb{k},i\omega_n)$~\eqref{eq:Green_function_Bdg_Ham} from which the Bloch functions $U_{\vb{k}}$ have been removed. Note that each of the blocks of $L(\vb{k}, i\omega)$ is diagonal, namely $[L(\vb{k},i\omega)]^{i,j}_{m,n} =  [L(\vb{k},i\omega)]^{i,j}_{n,n} \delta_{m,n}$, and the diagonal elements are given by
\begin{gather}
	[L(\vb{k},i\omega)]^{1,1}_{n,n} = \frac{u^2_{n\vb{k}}}{i\omega-E_{n\vb{k}}} + 
	\frac{v^2_{n\vb{k}}}{i\omega+E_{n\vb{k}}} \,,\\
	[L(\vb{k},i\omega)]_{n,n}^{2,2} = 
	\frac{v^2_{n\vb{k}}}{i\omega-E_{n\vb{k}}} + 
	\frac{u^2_{n\vb{k}}}{i\omega+E_{n\vb{k}}} \,,\\
	[L(\vb{k},i\omega)]^{i,j}_{n,n} =  
	 \frac{u_{n\vb{k}}v_{n\vb{k}}}{i\omega-E_{n\vb{k}}} -
	\frac{u_{n\vb{k}}v_{n\vb{k}}}{i\omega+E_{n\vb{k}}}\,,
\end{gather}
where $i\neq j$ in the last equation. Using these definitions and the Matsubara frequency sum
\begin{equation}
\label{eq:Matsubara_sum_s}
\begin{split}
s(E_1,E_2) &= \frac{1}{\beta}\sum_{i\omega_n} \frac{1}{(i\omega_n-E_1)(i\omega_n-E_2)} \\
&= \frac{n_{\rm F}(E_1)-n_{\rm F}(E_2)}{E_1-E_2} \\
&= \frac{\partial n_{\rm F}(E)}{\partial E}\qq{for} E = E_1 = E_2\,,
\end{split}
\end{equation}
with $n_{\rm F}(E) = (e^{\beta E}+1)^{-1}$ the Fermi-Dirac distribution, we obtain the following useful results ($i\neq j$)
\begin{gather}
\label{eq:LL_sum_1}
\begin{split}
&\frac{1}{\beta} \sum_{\omega_l}[L(\vb{k},i\omega_l)]_{n,n}^{i,i} [L(\vb{k},i\omega_l)]_{m,m}^{i,i} 
\\
&= (u_{n\vb{k}}^2u_{m\vb{k}}^2+v_{n\vb{k}}^2v_{m\vb{k}}^2)
s(E_{n\vb{k}},E_{m\vb{k}}) 
\\
&\quad+ (u_{n\vb{k}}^2v_{m\vb{k}}^2+v_{n\vb{k}}^2u_{m\vb{k}}^2)
	s(E_{n\vb{k}},-E_{m\vb{k}})\,,
\end{split}
\\
\label{eq:LL_sum_2}
\begin{split}
&\frac{1}{\beta} \sum_{\omega_l}[L(\vb{k},i\omega_l)]_{n,n}^{i,i} [L(\vb{k},i\omega_l)]_{m,m}^{j,j} \\
&= (u_{n\vb{k}}^2v_{m\vb{k}}^2+v_{n\vb{k}}^2u_{m\vb{k}}^2)
s(E_{n\vb{k}},E_{m\vb{k}}) 
\\
&\quad+ (u_{n\vb{k}}^2u_{m\vb{k}}^2+v_{n\vb{k}}^2v_{m\vb{k}}^2)
s(E_{n\vb{k}},-E_{m\vb{k}})\,,
\end{split}
\\
\label{eq:LL_sum_3}
\begin{split}
	&\frac{1}{\beta} \sum_{\omega_l}[L(\vb{k},i\omega_l)]_{n,n}^{i,j} [L(\vb{k},i\omega_l)]_{m,m}^{j,i} \\
	&=\frac{1}{\beta} \sum_{\omega_l}[L(\vb{k},i\omega_l)]_{n,n}^{i,j} [L(\vb{k},i\omega_l)]_{m,m}^{i,j}\\
	&= 2u_{n\vb{k}}v_{n\vb{k}}u_{m\vb{k}}v_{m\vb{k}}
	\\
	&\quad\times\qty[
	s(E_{n\vb{k}},E_{m\vb{k}}) - s(E_{n\vb{k}},-E_{m\vb{k}})]\,,
\end{split}
\\
\label{eq:LL_sum_4}
\begin{split}
	&\frac{1}{\beta} \sum_{\omega_l}[L(\vb{k},i\omega_l)]_{n,n}^{i,i} [L(\vb{k},i\omega_l)]_{m,m}^{i,j} \\
	&= \frac{1}{\beta}\sum_{\omega_l}[L(\vb{k},i\omega_l)]_{n,n}^{i,i} [L(\vb{k},i\omega_l)]_{m,m}^{j,i} \\
	&= (-1)^{i-1}(u_{n\vb{k}}^2-v^2_{n\vb{k}})u_{m\vb{k}}v_{m\vb{k}}
	\\
	&\quad\times\qty[
	s(E_{n\vb{k}},E_{m\vb{k}}) - s(E_{n\vb{k}},-E_{m\vb{k}})]\,.
\end{split}
\end{gather}
As an example, it is shown how to compute the correlation function $(\hat{D}_\alpha,\hat{D}_\beta^\dg)$ using the above results. From~\eqref{eq:D_alpha}-\eqref{eq:D_alpha_matr_elem} and~\eqref{eq:correlation_A_and_B} 
\begin{equation}
	\label{eq:D_Ddg}
	\begin{split}
	&(\hat{D}_\alpha,\hat{D}_\beta^\dg) \\
	&= \frac{1}{\beta}\sum_{\omega_l}\sum_{\vb{k}}
	\Tr\qty\big[D_\alpha(\vb{k})\mathcal{G}(\vb{k},i\omega_n)D_\beta^\dg(\vb{k})\mathcal{G}(\vb{k},i\omega_n)] 
	\\
	&=\sum_{\vb{k}}\sum_{n,m}\sum_{\alpha,\beta} g_{n\vb{k}}(\alpha)g_{n\vb{k}}^*(\beta)g_{m\vb{k}}(\beta)g^*_{m\vb{k}}(\alpha)\\
	&\quad\times\frac{1}{\beta} \sum_{\omega_l}[L(\vb{k},i\omega_l)]_{n,n}^{1,1} [L(\vb{k},i\omega_l)]_{m,m}^{2,2}\,. \\
 	\end{split}
\end{equation}
Finally the Matsubara sum is evaluated with~\eqref{eq:LL_sum_2}.

From~\eqref{eq:LL_sum_1}-\eqref{eq:LL_sum_4}, it is apparent that several relations hold between the correlation functions $(\hat{A},\hat{B})$, where the operators $\hat{A}$ and $\hat{B}$ are taken from the set $\{ \hat{N}_{\alpha\sigma},\,\hat{D}_\alpha,\hat{D}_{\alpha}^\dg \}$. They are the following
\begin{gather}
	\label{eq:corr_func_rel_1}
	(\hat{N}_{\alpha\uparrow},\hat{N}_{\beta\uparrow}) = 
	(\hat{N}_{\alpha\downarrow},\hat{N}_{\beta\downarrow})\,, \\
	\label{eq:corr_func_rel_2}
	\begin{split}
	&(\hat{D}_\alpha,\hat{D}_\beta) = (\hat{D}_\alpha^\dg,\hat{D}_\beta^\dg) = -(\hat{N}_{\alpha\sigma},\hat{N}_{\beta\bar{\sigma}})\,, 
	\end{split}
	\\
	\label{eq:corr_func_rel_3}	
	\begin{split}
	&(\hat{N}_{\alpha\sigma},\hat{D}_\beta) = (\hat{N}_{\alpha\bar\sigma},\hat{D}^\dg_\beta) \\
	&\quad=(\hat{D}_\alpha^\dg,N_{\beta\sigma}) = (\hat{D}_\alpha,\hat{N}_{\beta\bar\sigma})\,,
	\end{split}
	\\
	\label{eq:corr_func_rel_4}
	(\hat{D}_\alpha,\hat{D}_\beta^\dg) = (\hat{D}_\alpha^\dg,\hat{D}_\beta)\,.
\end{gather}
In~\eqref{eq:corr_func_rel_2} and~\eqref{eq:corr_func_rel_3} $\bar\sigma$ denotes the spin opposite to $\sigma$. From the properties~\eqref{eq:corr_func_prop_1}-\eqref{eq:corr_func_prop_2}, we obtain also the additional relation 
\begin{equation}
\label{eq:corr_func_rel_5}
(\hat{N}_{\alpha \sigma},\hat{D}_\beta)^* = (\hat{N}_{\alpha \sigma},\hat{D}^\dg_\beta)\,.
\end{equation} 

If $\bar{n}$ denotes the only partially filled band, then in the isolated band limit $E_{\bar{n}\vb{k}} \sim U$ and $E_{n\vb{k}} \sim E_{\rm gap}$ for $n \neq \bar{n}$ and the leading order contribution $\sim U^{-1}$ to the matrix $A_{\alpha,\beta}$ is obtained by retaining only the term $n = m = \bar{n}$ in~\eqref{eq:D_Ddg} and in all of the other correlation functions appearing in~\eqref{eq:A_alphabeta}. Thus, the Matsubara sum~\eqref{eq:Matsubara_sum_s} becomes
\begin{gather}
	s(E_{\bar{n}\vb{k}},E_{\bar{n}\vb{k}}) = -\frac{\beta}{4\cosh[2](\frac{\beta E_{\bar{n} \bo k }}{2})}\,,\\
	s(E_{\bar{n}\vb{k}},-E_{\bar{n}\vb{k}}) = -\frac{1}{2E_{\bar{n}}}\tanh(\frac{\beta E_{\bar{n} \bo k }}{2})\,,\\
	s(E_{n\vb{k}},E_{m\vb{k}})=0\qq{if} n\neq \bar{n} \qq{or} m \neq \bar{n}\,.
\end{gather} 
Observe that, in the isolated band limit, all the matrix elements of $A_{\alpha,\beta}$ are real since the Matsubara sums~\eqref{eq:LL_sum_1}-\eqref{eq:LL_sum_4} are always real and the Bloch functions of the band $\bar{n}$ appear in the combination $|g_{\bar{n}\vb{k}}(\alpha)|^2|g_{\bar{n}\vb{k}}(\beta)|^2 = |\mel{\alpha}{P(\vb{k})}{\beta}|^2$, which is positive. As a consequence, the correlation functions that appear in~\eqref{eq:corr_func_rel_3} and~\eqref{eq:corr_func_rel_5} are all equal. With the notation and the results established so far, it is immediate to obtain the matrix elements of $A_{\alpha,\beta}$ in the isolated flat band limit, which are shown in~\eqref{eq:A_alphabeta_isolated_band}-\eqref{eq:C_matrix_last}.

The correlation functions that involve the current operator $\hat{J}_l$ require some care when taking the isolated flat band limit. To begin with, time-reversal symmetry implies that the current operator in Nambu form becomes
\begin{gather}
\label{eq:current_op_Nambu_TR}
J(\vb{k}) = \pmqty{\partial_{l}H_{\rm free}^\uparrow(\vb{k}) & 0 \\
	0 & \partial_lH_{\rm free}^\uparrow(\vb{k})} = \partial_lH_0(\vb{k})\tau^z\,,\\
\text{with} \quad \tau^z = \mqty(1 & 0 \\ 0 & -1)\,.
\end{gather}
To derive~\eqref{eq:current_op_Nambu_TR} from~\eqref{eq:current_op_Nambu}, we have used the identity
\begin{equation}
\partial_lH_{\rm free}^\uparrow(\vb{k}) = -[\partial_lH_{\rm free}^\downarrow(-\vb{k})]^*\,,	
\end{equation}
which is obtained by taking the partial derivative $\partial_l \equiv \pdv{}{k_l}$ on both sides of the relation expressing time-reversal symmetry $H_{\rm free}^\uparrow(\vb{k}) = [H_{\rm free}^\downarrow(-\vb{k})]^*$. 
An equivalent form of~\eqref{eq:current_op_Nambu_TR} is
\begin{equation}
J_l(\vb{k})\tau^z	=  -\partial_l\mathcal{G}^{-1}(\vb{k},i\omega_n)\,.
\end{equation}
Using this last result, one can show that the diamagnetic term in~\eqref{eq:zero_order} can written as a correlation function in the same way as the paramagnetic one, that is 
\begin{equation}
\label{eq:diagmagnetic_resp_corr_func}
\begin{split}
&\ev{\pdv{\ham_{\rm free}(\vb{A})}{A_l}{A_m}}_{\vb{A} = \vb{0}} = - \qty\big(\hat{J}_l\tau^z,\hat{J}_m\tau^z)\geq 0\,.
\end{split}
\end{equation}
The fact that the diamagnetic term is nonnegative is an immediate consequence of the general property~\eqref{eq:corr_func_prop_3}.
In order to compute $\qty\big(\hat{J}_l\tau^z,\hat{J}_m\tau^z)$ and other correlation functions that involve the current operator, it is useful to introduce the following operator in Nambu form
\begin{gather}
M_l^{\pm}(\vb{k}) = \frac{1}{\beta}\sum_{\omega_l} L(\vb{k},i\omega_l)N^{\pm}_l(\vb{k})L(\vb{k},i\omega_l)
\\
N_l^{\pm}(\vb{k})=\pmqty{\tilde{J}_l(\vb{k})  & 0 \\ 0 & \pm \tilde{J}_l(\vb{k})}
\,,\\
\label{eq:J_tilde_def}
\tilde{J}_l(\vb{k}) = U^\dg_{\vb{k}}\partial_lH_{\rm free}^\uparrow(\vb{k})U_{\vb{k}}\,.
\end{gather}
It is not difficult to show that this operator satisfies the following properties
\begin{gather}
[M_l^{\pm}(\vb{k})]^\dg = M_l^{\pm}(\vb{k})\,,\\
\tau^y M_l^{\pm}(\vb{k})\tau^y = \pm M_l^{\pm}(\vb{k})\,,\\
\text{with} \quad\tau^y = \pmqty{0 & -i \\ i & 0}\,.
\end{gather}
From these relations and the results in~\eqref{eq:LL_sum_1}-\eqref{eq:LL_sum_4}, one obtains the following expressions for the matrix elements of $M_l^{\pm}(\vb{k})$
\begin{gather}
\label{eq:Mpm_ii_first}
\begin{split}
&[M^{\pm}_l(\vb{k})]_{m,n}^{1,1} = \pm [M^{\pm}_l(\vb{k})]_{m,n}^{2,2} \\ &= [\tilde{J}_l(\vb{k})]_{m,n}
\big[s(E_{m\vb{k}},E_{n\vb{k}})(u_{m\vb{k}}u_{n\vb{k}}\pm v_{m\vb{k}}v_{n\vb{k}})^2 \\
&\hspace{1.3cm}+ s(E_{m\vb{k}},-E_{n\vb{k}})(u_{m\vb{k}}v_{n\vb{k}}\mp v_{m\vb{k}}u_{n\vb{k}})^2\big]\,,
\end{split}	
\\
\label{eq:Mpm_ij_first}
\begin{split}
	&[M^\pm_l(\vb{k})]^{1,2}_{m,n} = \qty([M_l^{\pm}(\vb{k})]_{n,m}^{2,1})^* 
	= \mp \qty([M_l^{\pm}(\vb{k})]_{n,m}^{1,2})^* 
	\\
	 &=
	[\tilde{J}_l(\vb{k})]_{m,n} 
	[s(E_{m\vb{k}},E_{n\vb{k}})-s(E_{m\vb{k}},-E_{n\vb{k}})] \\
	&\times \qty[(u_{m\vb{k}}^2-v_{m\vb{k}}^2)u_{n\vb{k}}v_{n\vb{k}}\mp u_{m\vb{k}}v_{m\vb{k}}(u^2_{n\vb{k}}-v^2_{n\vb{k}})]\,.
\end{split}
\end{gather}
Using the definitions in~\eqref{eq:def_unk}-\eqref{eq:def_Enk} and after some laborious algebra, one finds the following alternative expressions for the same matrix elements~\eqref{eq:Mpm_ii_first}-\eqref{eq:Mpm_ij_first}
\begin{gather}
\label{eq:Mm_ii_second}
\begin{split}
&[M^{-}(\vb{k})]^{1,1}_{m,n}	= \frac{1}{2}\frac{[\tilde{J}_l(\vb{k})]_{m,n}}{ \varepsilon_{m\vb{k}}-\varepsilon_{n\vb{k}}} \\
&\quad\times\qty[(\varepsilon_{n\vb{k}}-\mu)f(E_{n\vb{k}})-(\varepsilon_{m\vb{k}}-\mu)f(E_{m\vb{k}})]\,, \\
&\qq*{with} f(E) = \frac{1}{E}\tanh(\frac{\beta E}{2})\,,
\end{split}
\\
\label{eq:Mp_ii_second}
\begin{split}
&[M^{+}(\vb{k})]^{1,1}_{m,n} = [M^{-}(\vb{k})]^{1,1}_{m,n} \\
&\quad+\Delta^2\frac{[\tilde{J}_l(\vb{k})]_{m,n}}{ \qty(E^2_{m\vb{k}}-E^2_{n\vb{k}})}\qty[f(E_{n\vb{k}})-f(E_{m\vb{k}})]\,,
\end{split}
\\
\label{eq:Mpm_ij_second}
\begin{split}
&[M^\pm_l(\vb{k})]^{1,2}_{m,n} = \frac{\Delta}{2}\frac{[\tilde{J}_l(\vb{k})]_{m,n}}{E_{m\vb{k}}^2-E_{n\vb{k}}^2} \\
&\,\,\times \qty\big(f(E_{n\vb{k}})-f(E_{m\vb{k}}))
\qty\big[(\varepsilon_{m\vb{k}}-\mu)\mp(\varepsilon_{n\vb{k}}-\mu)]\,.
\end{split}
\end{gather}
These are useful to compute the components of the vector $\vb{v}_{l\alpha}$ in~\eqref{eq:vect_v_def}, which read
\begin{gather}
	\label{eq:JN_corr_M}
	\begin{split}
	&(\hat{J}_l, \hat{N}_{\alpha\uparrow}) = \sum_{m,n} \sum_{\vb{k}} \braket{\alpha}{g_{m\vb{k}}} 
	[M_l^+(\vb{k})]_{m,n}^{1,1} \braket{g_{n\vb{k}}}{\alpha} \\
	&= \sum_{m,n} \sum_{\vb{k}} \braket{\alpha}{g_{m\vb{k}}} 
	[M_l^+(\vb{k})]_{m,n}^{2,2} \braket{g_{n\vb{k}}}{\alpha} = -(\hat{J}_l, \hat{N}_{\alpha\downarrow})\,,
	\end{split}
	\\
	\label{eq:JD_corr_M}
	\begin{split}
		&\qty\big(\hat{J}_l,\hat{D}_\alpha)
		=\sum_{m,n,\vb{k}}\braket{\alpha}{g_{m\vb{k}}}[M^+_l(\vb{k})]_{m,n}^{1,2}\braket{g_{n\vb{k}}}{\alpha} \\
		&= - \sum_{m,n,\vb{k}}\braket{\alpha}{g_{m\vb{k}}}[M^+_l(\vb{k})]_{m,n}^{2,1}\braket{g_{n\vb{k}}}{\alpha} =  -\qty\big(\hat{J}_l,\hat{D}_\alpha^\dg)\,.
	\end{split}	
\end{gather}
Here, the relations $[M^{+}_l(\vb{k})]_{m,n}^{1,1} =  [M^{+}_l(\vb{k})]_{m,n}^{2,2}$~\eqref{eq:Mpm_ii_first} and $[M^{+}_l(\vb{k})]_{m,n}^{1,2} = - [M^{+}_l(\vb{k})]_{m,n}^{2,1}$~\eqref{eq:Mpm_ij_first} have been used in the first and second equation, respectively.
It is convenient to consider separately the terms with $m = n$ and $m\neq n$ in the double sum over the bands indices $\sum_{m,n}$ in~\eqref{eq:JN_corr_M}. For the terms with $m=n$ we use \eqref{eq:Mpm_ii_first}, which gives
\begin{equation}
\label{eq:M+_ii_eq_comp}
[M_l^+(\vb{k})]_{n,n}^{1,1} = -\frac{\beta}{4\cosh[2](\frac{\beta E_{n \bo k }}{2})}\partial_l\varepsilon_{n\vb{k}}\,.	
\end{equation}
Here we have also taken advantage of~\eqref{eq:Matsubara_sum_s} and of the identity
\begin{equation}
\label{eq:matr_elem_curr_op_identity}
	[\tilde{J}_l(\vb{k})]_{m,n} = 
	\begin{cases}
	\partial_l\varepsilon_{n\vb{k}}\,, & m = n\,,\\
	(\varepsilon_{m\vb{k}}-\varepsilon_{n\vb{k}})\braket{\partial_l g_{m\vb{k}}}{g_{n\vb{k}}}\,, & m \neq n\,,
	\end{cases}
\end{equation}
which is easily obtained by inserting $H_{\rm free}^\uparrow(\vb{k}) = U_{\vb{k}}\varepsilon_{\vb{k}}U_{\vb{k}}^\dg$ into~\eqref{eq:J_tilde_def} and is employed repeatedly in the following. The result for $[M_l^-(\vb{k})]_{n,n}^{1,1}$ is also useful and is computed in a similar way
\begin{equation}
	\label{eq:Mm_11_nn}
	\begin{split}
	&[M_l^-(\vb{k})]_{n,n}^{1,1} = -\bigg[
	\qty(\frac{\varepsilon_{n\vb{k}}-\mu}{E_{n\vb{k}}})^2
	\frac{\beta}{4\cosh[2](\frac{\beta E_{n\vb{k}}}{2})}
	\\
	&\hspace{0cm}+ \frac{\Delta^2}{2E_{n\vb{k}}^2}f( E_{n\vb{k}})
	\bigg]\partial_l \varepsilon_{n\vb{k}} = -
	\partial_l\qty[\frac{\varepsilon_{n\vb{k}}-\mu}{2}
	f(E_{n\vb{k}})]\,.
	\end{split} 
\end{equation}
The easiest way to obtain the second equality is to take the limit $\varepsilon_{m\vb{k}}-\varepsilon_{n\vb{k}} \to 0$ in~\eqref{eq:Mm_ii_second}.

In the case $m\neq n$, the term $\propto \Delta^2$ in~\eqref{eq:Mp_ii_second} can be ignored from the start, thus $[M^{+}(\vb{k})]^{1,1}_{m,n} \approx [M^{-}(\vb{k})]^{1,1}_{m,n}$. Indeed, this term gives a contribution of order $\Delta^2/E_{\rm gap}~\sim U^2_\alpha/E_{\rm gap}$, which vanishes in the isolated band limit. Then, using again~\eqref{eq:matr_elem_curr_op_identity}, we derive the following useful result
\begin{equation}
	\label{eq:M+_ii_neq_comp}
	\begin{split}
		&\sum_{\substack{m,n \\ m\neq n}} \sum_{\vb{k}} \braket{\alpha}{g_{m\vb{k}}} 
		[M_l^-(\vb{k})]_{m,n}^{1,1} \braket{g_{n\vb{k}}}{\beta} \\
		&=\frac{1}{2} \sum_{m,n,\vb{k}} \braket{\alpha}{g_{m\vb{k}}} \braket{\partial_l g_{m\vb{k}}}{g_{n\vb{k}}} \braket{g_{n\vb{k}}}{\beta}\\
		&\hspace{0.5cm}\times \qty[(\varepsilon_{n\vb{k}}-\mu)f(E_{n\vb{k}})-(\varepsilon_{m\vb{k}}-\mu)f(E_{m\vb{k}})] \\
		&= -\frac{1}{2} \sum_{n,\vb{k}}  \braket{\alpha}{\partial_lg_{n\vb{k}}} \braket{g_{n\vb{k}}}{\beta} (\varepsilon_{n\vb{k}}-\mu)f(E_{n\vb{k}}) \\
		&\quad-\frac{1}{2} \sum_{m,\vb{k}} \braket{\alpha}{g_{m\vb{k}}} \braket{\partial_l g_{m\vb{k}}}{\beta} (\varepsilon_{m\vb{k}}-\mu)f(E_{m\vb{k}}) \\ 
		&= -\frac{1}{2} \sum_{n,\vb{k}}  \partial_l \qty\big(\braket{\alpha}{g_{n\vb{k}}} \braket{g_{n\vb{k}}}{\beta}) (\varepsilon_{n\vb{k}}-\mu)f(E_{n\vb{k}})\\
		&=  \sum_{n,\vb{k}} \braket{\alpha}{g_{n\vb{k}}} \braket{g_{n\vb{k}}}{\beta}\partial_l\qty[\frac{\varepsilon_{n\vb{k}}-\mu}{2}f(E_{n\vb{k}})] \\
		&= -\sum_{n,\vb{k}} \braket{\alpha}{g_{n\vb{k}}} \braket{g_{n\vb{k}}}{\beta}[M_l^-(\vb{k})]_{n,n}^{1,1}
		\,.
	\end{split}
\end{equation}
Note that in the second line the sum over the band indices is unrestricted since for $m \neq n$ the term in square brackets vanishes. In the second equality, we have used the property $\braket{\partial_l g_{m\vb{k}}}{g_{n\vb{k}}} = -\braket{g_{m\vb{k}}}{\partial_lg_{n\vb{k}}}$, which is a consequence of the orthonormality of the Bloch wave functions $\braket{g_{m\vb{k}}}{g_{n\vb{k}}} = \delta_{m,n}$. Moreover, one sum of over the band indices has been carried out by using the completeness relation $\sum_n\dyad{g_{n\vb{k}}} = 1$. Then, an integration by part has been performed, which is allowed since in the thermodynamic limit the sum over wave vectors becomes an integral over the Brillouin zone $\sum_{\vb{k}} \to \frac{\mathcal{A}}{(2\pi)^2}\int \dd[2]{\vb{k}}$. It is understood that  in the following all sums over wave vectors represent Brillouin zone integrals. The last equality follows from~\eqref{eq:Mm_11_nn}.
 
Finally,~\eqref{eq:M+_ii_eq_comp},~\eqref{eq:Mm_11_nn} and~\eqref{eq:M+_ii_neq_comp}
are combined to give
\begin{equation}
	\label{eq:JN_final}
	\begin{split}
	&(\hat{J}_l, \hat{N}_{\alpha\uparrow}) = -(\hat{J}_l, \hat{N}_{\alpha\downarrow}) \\
	&\approx \sum_{n,\vb{k}} \qty|\braket{\alpha}{g_{n\vb{k}}}|^2
	\qty([M_l^+(\vb{k})]_{n,n}^{1,1}-[M_l^-(\vb{k})]_{n,n}^{1,1}) \\ &=\sum_{n,\vb{k}} \qty|\braket{\alpha}{g_{n\vb{k}}}|^2\frac{\Delta^2}{2E_{n\vb{k}}^2}
	\qty[f(E_{n\vb{k}})
	-\frac{\beta}{2\cosh[2](\frac{\beta E_{n\vb{k}}}{2})}]\partial_l \varepsilon_{n\vb{k}}\,. 
	\end{split}
\end{equation}
In the isolated band limit, only the term $n = \bar{n}$ corresponding to the partially filled band gives a nonzero contribution. Note that, to obtain the correct result for the correlation function~\eqref{eq:JN_final}, it is important to retain all of the matrix elements of the current operator $[\tilde{J}_l(\vb{k})]_{m,n}$, even if $n,m \neq \bar{n}$.  Indeed, retaining these matrix elements allowed us to use the completeness relation for the Bloch functions in~\eqref{eq:M+_ii_neq_comp}. The need to take into account also the interband matrix elements of the current operator even in the isolated flat band limit is a general phenomenon, as we will see in the following. For this reason, one has to be particularly careful when evaluating correlation functions that involve the current operator.

To compute the correlation function between the current operator and the pairing operator~\eqref{eq:JD_corr_M}, one can use the approximation
\begin{gather}
\begin{split}
&[M_l^+(\vb{k})]^{1,2}_{\bar{n},m}  = -\qty\big([M_l^+(\vb{k})]^{1,2}_{m,\bar{n}})^* \\
&\hspace{0.5cm}\approx \frac{\Delta}{2}f(E_{\bar{n}\vb{k}})\braket{\partial_lg_{\bar{n}\vb{k}}}{g_{m\vb{k}}}\,,
\end{split}
\\
[M_l^+(\vb{k})]^{1,2}_{m,n} \sim \frac{\Delta}{E_{\rm gap}} \sim 0\qq{for} m,n\neq \bar{n}\,,
\end{gather}
valid again in the isolated band limit. In addition, all the diagonal matrix elements vanish $[M_l^+(\vb{k})]^{1,2}_{n,n} = 0$, as one can see from~\eqref{eq:Mpm_ij_second}. Thus, for the correlation function~\eqref{eq:JD_corr_M} we have
\begin{equation}
	\label{eq:JD_final}
	\begin{split}
		&\qty\big(\hat{J}_l,\hat{D}_\alpha) = \sum_{m,n,\vb{k}}\braket{\alpha}{g_{m\vb{k}}}[M^+_l(\vb{k})]_{m,n}^{1,2}\braket{g_{n\vb{k}}}{\alpha} \\
		&\approx \frac{\Delta}{2}\sum_{\vb{k},m \neq \bar{n}}f(E_{\bar{n}\vb{k}}) \big(\braket{\alpha}{g_{\bar{n}\vb{k}}}\braket{\partial_lg_{\bar{n}\vb{k}}}{g_{m\vb{k}}}\braket{g_{m\vb{k}}}{\alpha} \\
		&\hspace{2.5cm}-  \braket{\alpha}{g_{m\vb{k}}}\braket{g_{m\vb{k}}}{\partial_lg_{\bar{n}\vb{k}}}\braket{g_{\bar{n}\vb{k}}}{\alpha}\big) \\
		&= \frac{\Delta}{2}\sum_{\vb{k}}f(E_{\bar{n}\vb{k}})\big[\braket{\alpha}{g_{\bar{n}\vb{k}}}\bra{\partial_lg_{\bar{n}\vb{k}}}\qty\big(1-P(\vb{k}))\ket{\alpha} \\
		&\hspace{2.2cm}-\bra{\alpha}\qty\big(1-P(\vb{k}))\ket{\partial_lg_{\bar{n}\vb{k}}}\braket{g_{\bar{n}\vb{k}}}{\alpha}\big]\\
		&=\frac{\Delta}{2}\sum_{\vb{k}}f(E_{\bar{n}\vb{k}})\big[\ev{\partial_lP(\vb{k})\qty\big(1-P(\vb{k}))}{\alpha} 
		\\&\hspace{2.5cm}- \ev{\qty\big(1-P(\vb{k}))\partial_lP(\vb{k})}{\alpha} \big]\\
		&= \frac{\Delta}{2}\sum_{\vb{k}}f(E_{\bar{n}\vb{k}})\ev{\qty[P(\vb{k}),\partial_l P(\vb{k})]}{\alpha}\,.
	\end{split} 
\end{equation} 
Again, the completeness relation in the form $1-P(\vb{k}) = \sum_{m\neq \bar{n}}\dyad{g_{m\vb{k}}}$ with $P(\vb{k}) = \dyad{g_{\bar{n}\vb{k}}}$ has been used.

The last two correlation functions needed in order to compute the superfluid weight are
\begin{gather}
\qty\big(\hat{J}_{l_1},\hat{J}_{l_2}) = 2\sum_{m,n}[M_{l_1}^+(\vb{k})]^{1,1}_{m,n}[\tilde{J}_{l_2}(\vb{k})]_{n,m}\,,\\
\qty\big(\hat{J}_{l_1}\tau^z,\hat{J}_{l_2}\tau^z) 
= 2\sum_{m,n}[M_{l_1}^-(\vb{k})]^{1,1}_{m,n}[\tilde{J}_{l_2}(\vb{k})]_{n,m}\,.
\end{gather}
In fact, according to~\eqref{eq:zero_order}, only their difference is required
\begin{equation}
\label{eq:Ds_0_full}
\begin{split}
&\qty\big(\hat{J}_{l_1},\hat{J}_{l_2})-\qty\big(\hat{J}_{l_1}\tau^z,\hat{J}_{l_2}\tau^z) \\ 
&= \sum_{n,\vb{k}} \frac{\Delta^2}{E_{n\vb{k}}^2}\qty[f(E_{n\vb{k}})-\frac{\beta}{2\cosh[2](\frac{\beta E_{n\vb{k}}}{2})}]\partial_{l_1}\varepsilon_{n\vb{k}}\partial_{l_2}\varepsilon_{n\vb{k}} \\
&\hspace{0.5cm}+2\Delta^2\sum_{\vb{k}}\sum_{\substack{m,n \\ m\neq n}}\frac{(\varepsilon_{m\vb{k}}-\varepsilon_{n\vb{k}})^2}{ E^2_{m\vb{k}}-E^2_{n\vb{k}}}\qty\big[f(E_{n\vb{k}})-f(E_{m\vb{k}})]\\
&\hspace{3cm}\times\braket{\partial_{l_1} g_{m\vb{k}}}{g_{n\vb{k}}}\braket{ g_{n\vb{k}}}{\partial_{l_2} g_{m\vb{k}}}\,.
\end{split}
\end{equation}
In the isolated band limit, the first sum gives the conventional contribution to the superfluid weight in~\eqref{eq:D0_sc}, while the second one with $m\neq n$ corresponds to the geometric contribution in~\eqref{eq:D0_sg}. Indeed, in the case of the geometric contribution, one can proceed as follows
\begin{equation}
\label{eq:quantum_metric_derivation}
\begin{split}
&\sum_{\substack{m,n \\ m\neq n}}\frac{(\varepsilon_{m\vb{k}}-\varepsilon_{n\vb{k}})^2}{ E^2_{m\vb{k}}-E^2_{n\vb{k}}}\qty\big[f(E_{n\vb{k}})-f(E_{m\vb{k}})]\\
&\hspace{1.5cm}\times\braket{\partial_{l_1} g_{m\vb{k}}}{g_{n\vb{k}}}\braket{ g_{n\vb{k}}}{\partial_{l_2} g_{m\vb{k}}} \\
&\approx f(E_{\bar{n}\vb{k}})\sum_{m\neq \bar{n}} \braket{\partial_{l_1} g_{\bar{n}\vb{k}}}{g_{m\vb{k}}}\braket{ g_{m\vb{k}}}{\partial_{l_2} g_{\bar{n}\vb{k}}} + (l_1\leftrightarrow l_2) \\
&= f(E_{\bar{n}\vb{k}}) \bra{\partial_{l_1} g_{\bar{n}\vb{k}}} \qty\big(1-P(\vb{k})) \ket{\partial_{l_2} g_{\bar{n}\vb{k}}} + (l_1\leftrightarrow l_2) \\
&= f(E_{\bar{n}\vb{k}})\Tr[\partial_{l_1}P(\vb{k})\partial_{l_2}P(\vb{k})]\,.
\end{split}	
\end{equation}
Thus, the Bloch function $\ket{g_{\bar{n}\vb{k}}}$ of the partially filled band  enters only through the quantum metric~\eqref{eq:quantum_metric}. 
Again, it is important to notice that the completeness relation has been used in the above derivation.


\section{Self-consistency equations of mean-field theory for the Hubbard interaction and uniform pairing assumption}
\label{app:uniform_pairing_condition}

In this section, we solve the self-consistency equations of mean-field theory in the case of the Hubbard interaction, thus justifying the uniform pairing condition~\eqref{eq:uniform_pairing}. In order to solve the self-consistency equations~\eqref{eq:Gamma_self_eq} and~\eqref{eq:Delta_self_eq}, it is necessary to compute the expectation values that appear on the right hand side.
These are obtained from the imaginary-time Green's function~\eqref{eq:Green_imag_tau} by taking the limit $\tau -\tau' \to 0^-$
\begin{gather}
\label{eq:nka_up}
n_{\vb{k}\alpha\uparrow} = \ev*{\hat{c}^\dg_{\vb{k}\alpha\uparrow}\hat{c}_{\vb{k}\alpha\uparrow}} = [\mathcal{G}(\vb{k},\tau = 0^-)]^{1,1}_{\alpha,\alpha}\,,\\
\label{eq:nka_do}
n_{\vb{k}\alpha\downarrow} = \ev*{\hat{c}^\dg_{\vb{k}\alpha\downarrow}\hat{c}_{\vb{k}\alpha\downarrow}} =
 1-[\mathcal{G}(-\vb{k},\tau = 0^-)]^{2,2}_{\alpha,\alpha}\,,\\
\label{eq:ev_ckaup_ckado}
\ev*{\hat{c}_{-\vb{k}\alpha\downarrow}\hat{c}_{\vb{k}\alpha\uparrow}} = [\mathcal{G}(\vb{k},\tau = 0^-)]^{1,2}_{\alpha,\alpha}\,,
\end{gather}
To evaluate the Green's function for $\tau =0^-$ one can use~\eqref{eq:Green_function_omega} together with the standard summation over Matsubara frequencies
 \begin{equation}
 	\frac{1}{\beta}\sum_{\omega_n} \frac{e^{i\omega_n\eta}}{i\omega_n-E} = \frac{1}{e^{\beta E}+1} = n_{\rm F}(E)\,,\quad \eta = 0^+\,.
 \end{equation}
Using the definition in~\eqref{eq:L_matrix_def}, we obtain
\begin{equation}
\label{eq:sum_omega_L}
\begin{split}
&\frac{1}{\beta}\sum_{\omega_n} L(\vb{k},i\omega_n)e^{i\omega_n\eta} = 
\frac{1}{\beta}\sum_{\omega_n}	W_{\vb{k}}\frac{e^{i\omega_n\eta}}{i\omega_n-E_{\vb{k}}}W_{\vb{k}}^\dagger \\
&=\frac{1}{2}\pmqty{1-\frac{\varepsilon_{\vb{k}}-\mu}{E^>_{\vb{k}}}\tanh\frac{\beta E^>_{\vb{k}}}{2} & - \frac{\Delta}{E^>_{\vb{k}}}\tanh\frac{\beta E^>_{\vb{k}}}{2} \\[0.3em]
-\frac{\Delta}{E^>_{\vb{k}}}\tanh\frac{\beta E^>_{\vb{k}}}{2} & 1+\frac{\varepsilon_{\vb{k}}-\mu}{E^>_{\vb{k}}}\tanh\frac{\beta E^>_{\vb{k}}}{2}}\,.
\end{split}
\end{equation}
Recall that $\varepsilon_{\vb{k}} = \mathrm{diag}(\varepsilon_{n\vb{k}})$ is a diagonal matrix containing the band dispersions $\varepsilon_{n\vb{k}}$, while $E_{\vb{k}}^> = \mathrm{diag}(E_{n\vb{k}})$ is also diagonal but contains the quasiparticle dispersions $E_{n\vb{k}}$ instead, that is the eigenvalues of the BdG Hamiltonian $H_0(\vb{k})$, see~\eqref{eq:BdG_Ham_diagonalized}. Note that the result in~\eqref{eq:sum_omega_L} is valid only under the uniform pairing condition~\eqref{eq:uniform_pairing}.
Then, the Green's function at $\tau=0^-$ is computed by combining~\eqref{eq:Green_function_Bdg_Ham} with~\eqref{eq:sum_omega_L}. Thus, from~\eqref{eq:nka_up}-\eqref{eq:ev_ckaup_ckado}, we can rewrite the self-consistency equations as
\begin{gather}
	\label{eq:self_const_exp_Gamma}
	\Gamma_\alpha^\sigma = -\frac{U_\alpha}{2N_{\rm c}} \sum_{n,\vb{k}} \abs{\braket{\alpha}{g_{n\vb{k}}}}^2\qty(1-\frac{\varepsilon_{n\vb{k}}-\mu}{E_{n\vb{k}}}\tanh\frac{\beta E_{n\vb{k}}}{2})\,,
	\\
	\label{eq:self_const_exp_Delta}
	\Delta_\alpha = \frac{U_{\alpha} }{2N_{\rm c}} \sum_{n,\vb{k}}\abs{\braket{\alpha}{g_{n\vb{k}}}}^2\frac{\Delta}{E_{n\vb{k}}}\tanh\qty(\frac{\beta E_{n\vb{k}}}{2})\,.
\end{gather}
 The parameters $\Delta_\alpha$, obtained from the second equation for a given value of $\Delta$, do not in general satisfy the uniform pairing condition. However, it is possible to adjust the relative strength of the coupling constants $U_\alpha$ so as to ensure that~\eqref{eq:uniform_pairing} is at least approximately satisfied.    
 
 In the isolated band limit, it is possible to derived an explicit condition on the coupling constants $U_\alpha$ that ensures uniform pairing. First, the self-consistency equation relative to the Hartree potential $\Gamma_\alpha^\sigma$ is neglected for simplicity and only the partially filled band $\bar{n}$ is retained in the sum over bands in~\eqref{eq:self_const_exp_Delta}. Indeed, the contribution of the terms $n\neq \bar{n}$ is negligible in the isolated band limit since $\Delta$ is of the same order of $U_\alpha$. If it is  assumed that the $\bar{n}$-th band is flat ($\varepsilon_{\bar{n}\vb{k}} = \varepsilon_{\bar{n}}$), then the quasiparticle dispersion $E_{\bar{n}\vb{k}} = E_{\bar{n}} = \sqrt{(\varepsilon_{\bar{n}}-\mu)^2+\Delta^2}$ is also flat and the uniform pairing conditions is equivalent to the following requirement
 \begin{equation}
 	\label{eq:uniform_pairing_2}
 	\begin{split}
 		\frac{U_{\alpha}}{N_c}\sum_{\bo k}\abs{\braket{\alpha}{g_{\bar n\vb{k}}}}^2 = \bar U >0\qquad \forall \alpha \, ,
 	\end{split}
 \end{equation} 
 namely that the quantity on the left hand side is independent of the orbital index $\alpha$, when it is not zero.~\eqref{eq:uniform_pairing_2} can always be satisfied by a suitable choice of the coupling constants $U_\alpha$.
 In this case, the self-consistent value of the pairing potential $\Delta$ (called also the order parameter) is obtained from
 \begin{equation}
 	\label{eq:self_cons_eqs}
 	\frac{\bar U}{2  E_{\bar{n}}} \tanh\qty(\frac{\beta E_{\bar{n}}}{2}) =1\,.
 \end{equation}
 This equation is obtained by combining~\eqref{eq:self_const_exp_Delta} and~\eqref{eq:uniform_pairing_2} and is identical to the self-consistency equation of the Weiss mean-field theory of ferromagnetism, in which the quasiparticle energy $E_{\bar{n}}$ plays the role of the magnetization. 
 If the quantity $\sum_{\vb{k}}\abs{\braket{\alpha}{g_{\bar n\vb{k}}}}^2$ is independent of the orbital index $\alpha$, for instance because of some lattice symmetry, then the uniform pairing condition follows from $U_\alpha = U_\beta = U = N_{\rm orb}\bar{U}$ for all $\alpha,\,\beta$, where $N_{\rm orb}$ is the number of orbitals in the unit cell. This is the case considered in Ref.~\cite{Peotta2015} and other subsequent works.

	\bibliography{mqp,mylibrary}
	
\end{document}